\documentclass[journal]{IEEEtran}
\usepackage{amsmath}
\usepackage{amsthm}
\usepackage{dsfont}
\usepackage{bbm}
\usepackage{amssymb}
\usepackage{graphicx}
\usepackage{graphics}

\ifCLASSINFOpdf
\else
\fi

\def\>{\rangle}
\def\<{\langle}

\def\id{\mathsf{id}}
\def\mB{\mathcal{B}}
\def\mE{\mathcal{E}}

\def\mN{\mathcal{N}}
\def\mC{\mathcal{C}}
\def\mL{\mathcal{L}}

\def\mD{\mathcal{D}}
\def\mT{\mathcal{T}}
\def\mV{\mathcal{V}}

\renewcommand{\geq}{\geqslant}
\renewcommand{\leq}{\leqslant}

\newtheorem{theorem}{Theorem}
\newtheorem{corollary}{Corollary}

\newtheorem{definition}{Definition}

\newtheorem{remark}{Remark}

\newcommand{\bea}{\begin{eqnarray}}
\newcommand{\eea}{\end{eqnarray}}
\newcommand{\be}{\begin{equation}}
\newcommand{\ee}{\end{equation}}
\newcommand{\ba}{\begin{equation}\begin{aligned}}
\newcommand{\ea}{\end{aligned}\end{equation}}

\newcommand{\J}{\mathbf{J}}

\newcommand{\1}{\mathds{1}}

\def\be{\begin{equation}}
\def\ee{\end{equation}}

\newcommand{\mU}{\mathcal{U}}

\newcommand{\mH}{\mathcal{H}}

\newcommand{\mK}{\mathcal{K}}

\newcommand{\lr}{\rangle\langle}
\newcommand{\la}{\langle}
\newcommand{\ra}{\rangle}
\newcommand{\tr}{{\rm Tr}}

\newcommand{\pree}{{\rm pre}}
\newcommand{\post}{{\rm post}}

\newcommand{\mbb}[1]{\mathbb{#1}}

\def\tA{\tilde{A}}
\def\tB{\tilde{B}}
\def\tC{\tilde{C}}

\def\u{\mathbf{u}}
\def\v{\mathbf{v}}

\newcommand*{\QED}{\hfill\ensuremath{\square}}

\begin{document}
%
\title{Comparison of Quantum Channels by Superchannels}
%
%
%

\author{Gilad~Gour
        \thanks{Department of Mathematics and Statistics, Institute for Quantum Science and Technology,
University of Calgary, AB, Canada T2N 1N4}}

%
%

\markboth{IEEE TRANSACTIONS ON INFORMATION THEORY}%
{Shell \MakeLowercase{\textit{et al.}}: Bare Demo of IEEEtran.cls for IEEE Journals}
%



\maketitle

\begin{abstract}
We extend the definition of the conditional min-entropy from bipartite quantum states to bipartite quantum channels. We show that many of the properties of the conditional min-entropy carry over to the extended version, including an operational interpretation as a guessing probability when one of the subsystems is classical. We then show that the extended conditional min-entropy can be used to fully characterize when two bipartite quantum channels are related to each other via a superchannel (also known as supermap or a comb) that is acting on one of the subsystems. This relation is a pre-order that extends the definition of ``quantum majorization" from bipartite states to bipartite channels, and can also be characterized with semidefinite programming. As a special case, our characterization provides necessary and sufficient conditions for when 
a set of quantum channels is related to another set of channels via a single superchannel. We discuss the applications of our results to channel discrimination, and to resource theories of quantum processes.
Along the way we study channel divergences, entropy functions of quantum channels, and noise models of superchannels, including random unitary superchannels, and doubly-stochastic superchannels. For the latter we give a physical meaning as being completely-uniformity preserving.
\end{abstract}


%
\IEEEpeerreviewmaketitle

\section{Introduction}
One of the most fundamental properties of quantum mechanics is that it is impossible to perfectly distinguish between distinct non-orthogonal states of a physical system.
This property, which lies at the heart of the uncertainty principle, may be perceived at first as a hindrance to the theory. In recent years, however, it was shown that it can be harnessed to the success of many interesting quantum information processing tasks, such as quantum cryptography~\cite{wilde2013,Nielsen-2000a,Watrous-2018a}. Consequently, different techniques have been developed to quantify the distinguishability of quantum states. For example, in quantum hypothesis testing (see, e.g.,~\cite{Audenaert2007,Brand2010,Mosonyi2015} and references therein), one uses as a measure of distinguishability the maximized success probability of guessing correctly the quantum state by a quantum measurement. This can be done either in the single-shot regime or asymptotically, and either in a symmetric or an asymmetric way. Alternatively, one may choose other figures of merit, such as the accessible information, to measure the distinguishability of quantum states.  

The variety of figures of merit indicates that the distinguishability of quantum states cannot be fully captured with a single function, but instead can be described by a pre-order. To understand this pre-order, suppose Alice holds one out of two quantum states $\rho_1$ and $\rho_2$, and does not know which state it is. Clearly, her ability to determine which state she holds cannot increase if she sends her state $\rho_x$ ($x=1,2$) through a quantum channel $\Phi$. Therefore, the pair $(\rho_1,\rho_2)$ is always more distinguishable than $(\Phi(\rho_1),\Phi(\rho_2))$ and any measure of distinguishability $D(\rho_1\|\rho_2)$ must behave monotonically under such a transformation; i.e.
$$
D\left(\Phi(\rho_1)\big\|\Phi(\rho_2)\right)\leq D(\rho_1\|\rho_2)\;.
$$

Given two pairs of quantum states $(\rho_1,\rho_2)$ and $(\sigma_1,\sigma_2)$, how can we determine if there exists a channel $\Phi$ such that $\Phi(\rho_x)=\sigma_x$ for both $x=1,2$? This question was answered already in 1953 by Blackwell~\cite{blackwell1953} for the classical case, and in 1980 by Alberti and Uhlmann~\cite{Alberti1980} for the qubit case. More recently, it was solved for pure states in~\cite{Winter2004}, characterized in~\cite{Bus12,B1,B2,B3}, and finally,  in~\cite{Gour2018} it was fully solved (for finite dimensions) with semidefinite programming. In~\cite{Gour2018,Bus18b} (see also references therein) it was also shown that this pre-order can be characterized completely in terms of a family of distinguishability measures that are given in terms of the conditional min-entropy~\cite{Renner-2005a,Konig2009}.

Quantum phenomena however are not static in general. They correspond to dynamical processes that characterize the evolution of a physical system and are described mathematically with completely positive and trace-preserving (CPTP) maps (also known as quantum channels). A quantum state can be viewed as a special type of a quantum channel with one dimensional input. In this operational view, a quantum state is a preparation process, and consequently, quantum channels can be viewed as the fundamental objects of quantum mechanics, describing both static and dynamical behaviours of physical systems. Much like quantum states, quantum channels can evolve and change over a period of time. The most general evolution of a quantum channel is described with a superchannel (introduced in~\cite{Pavia1} under the name of supermaps, or 2-comb in~\cite{Chir08}; see also~\cite{KU17,Karol}). A superchannel is a linear map that maps (even when act on subsystems) quantum channels to quantum channels and has a physical realization with a pre and post processing on the quantum channel upon which it acts (see Sec.~\ref{preliminary}, particularly, Fig.~\ref{superchannel}). In addition to being interesting mathematically, superchannels are expected to play an important role in quantum resource theories of processes (see for example the very recent works~\cite{FBB18,Diaz18,WW18}).
 
Discrimination of quantum channels can be defined similarly to its state (static) analog, although the theory is richer due to a variety of possible schemes (such as adaptive vs non-adaptive ones, single-shot vs asymptotic, etc) and recently, there has been an active research on the subject~\cite{Wil18b,Wil18,Puz17,Cooney2016,Har10,Dua09,Jen14,Jen12,Chir08}. Here, however, we will focus on the \emph{comparison} of channels in general, avoiding the optimization of the success probability of a specific scheme. More precisely,
suppose Alice holds at her disposal one out of two quantum channels $\Phi_1$ and $\Phi_2$, but she does not know which one. By ``sending" her channel $\Phi_x$ ($x=1,2$) through a superchannel $\Theta$, she ends up with the channel $\Theta[\Phi_x]$. Therefore, the pair $(\Phi_1,\Phi_2)$ must be more distinguishable than the pair of channels $(\Theta[\Phi_1],\Theta[\Phi_2])$, and any measure of distinguishability of quantum channels must behave monotonically under such a transformation.

In this paper we provide necessary and sufficient conditions for the existence of a superchannel $\Theta$ under which two sets of $n$ channels $(\Phi_1,...,\Phi_n)$ and $(\Psi_1,...,\Psi_n)$ are related by a superchannel $\Theta$ via $\Psi_x=\Theta[\Phi_x]$ for all $x=1,...,n$. 
Our conditions are given in terms of an SDP and therefore can be solved efficiently and algorithmically. Furthermore,
we show that the conditions
can be expressed in terms of a complete family of distinguishability measures given in terms of a function that we call the 
\emph{extended conditional min-entropy}.
The extended conditional min-entropy is an extension of the conditional min-entropy from bipartite states to bipartite channels. We show that the extended conditional min-entropy satisfies many properties similar to those satisfied by the conditional min-entropy. 
In addition, we develop an axiomatic approach for the entropy of a quantum channel and discuss the properties that the entropy of a quantum channel should satisfy. Particularly, we argue that entropy functions should behave monotonically under completely uniformity preserving superchannels (see Sec.~\ref{uniformitypreserving}). We show that the doubly stochastic superchannels (i.e. those for which both $\Theta$ and $\Theta^*$ are superchannels) have this property. Our work involves the extended conditional min-entropy, and not a von-Neumann version of it (see e.g.~\cite{GW18}), since we are studying here only the single-shot regime, while the i.i.d. version of our main result remains open.

This paper is organized as follows. In Sec.~\ref{preliminary}, we introduce our notations and discuss the properties of superchannels. In Sec.~\ref{noisy}, we introduce and study four different noise models for superchannels and discuss the relationships among them. In Sec.~\ref{entropy}, we define the entropy of a quantum channel and as an example define the extended min-entropy. In addition, we introduce the extended conditional min-entropy and study its properties and physical meaning.  In section~\ref{discrim}, we introduce the main result about discrimination of quantum channels in terms of both an SDP and the extended conditional min-entropy. In addition, we compare our work with the analog work on quantum states as given in~\cite{Gour2018} and characterize the pre-order that extends quantum majorization (as defined in~\cite{Gour2018}) from bipartite states to bipartite channels. Finally, we end in Sec.~\ref{outlook} with summary and conclusions.

\section{Notations and Preliminaries}\label{preliminary}

In this section, we introduce our notations and cover several topics that will be used in the subsequent sections.
While a significant part of the material presented in this section can be found (in some form) somewhere else, there are also new key observations that we will use extensively later on.

\subsection{The space of linear maps}

Let $\mB(\mH)$ denote the space of all (bounded) operators acting on a finite dimensional Hilbert space $\mH$, $\mB_h(\mH)$ the subset consisting of all Hermitian matrices in $\mB(\mH)$, $\mB_+(\mH)$ the subset of positive semidefinite matrices in $\mB(\mH)$, and $\mD(\mH)$ the subset of all density matrices in $\mB(\mH)$. We view $\mB(\mH)$ as an inner product space equipped with the Hilbert-Schmidt inner product: $\la X,Y\ra\equiv\tr\left[X^* Y\right]$ for all $X,Y\in\mB(\mH)$.  
Quite often we will consider quantum channels such that both their input and output systems are accessible to a single party. 
In this case, we will denote a channel in Alice's system by $\Psi^{A_0\to A_1}:\mB\left(\mH^{A_0}\right)\to\mB\left(\mH^{A_1}\right)$, where $A_0$ and $A_1$ are the input and output systems. Since we assume that both the input and output systems are held by Alice, we will denote by $A$ the joint system $A_0A_1$ and use the notation
$$
\Psi^A\equiv\Psi^{A_0\to A_1}\;.
$$
More generally, a multipartite quantum channel is a channel whose input and output spaces are composite quantum systems shared by several parties. If a party $B$ holds only an output subsystem (not an input one)  we will assign to it the trivial 1-dimensional system $B_0$. In this way, any party holds two subsystems, the input and output subsystems. For example, a channel shared by two parties will be denoted by
\be\label{ab}
\Psi^{AB}\equiv\Psi^{A_0B_0\to A_1B_1}:\mB(\mH^{A_0B_0})\to\mB(\mH^{A_1B_1})\;.
\ee
The structure of bipartite channels of the form above, has been studied extensively in~\cite{DW16} and also in~\cite{LM15}.

The space of all linear operators from the input space $\mB(\mH^{A_0})$ to the output space $\mB(\mH^{A_1})$ will be denoted by 
$$
\mL^A\equiv\left\{\Psi^A:\mB(\mH^{A_0})\to\mB(\mH^{A_{1}})\;\Big|\;\Psi^{A}\text{ is a linear map}\right\}
$$
Similarly, we will denote by $\mL^{AB}$  the space of all linear maps as in~\eqref{ab}.
We will view $\mL^A$ (and $\mL^{AB}$) as a vector space equipped with the following inner product. Let $\{X_{a}\}$ be an orthonormal basis of $\mB(\mH^{A_0})$ (i.e. $\tr[X_{a}^*X_{a'}]=\delta_{aa'}$). Then, the inner product between two elements $\Phi,\Psi\in\mL^A$ is defined by:
\ba
\left\la\Phi,\Psi\right\ra &\equiv \sum_a \left\la\Phi(X_{a}),\Psi(X_{a})\right\ra\\
&=\sum_a\tr\left[\left(\Phi(X_{a})\right)^{*}\Psi(X_{a})\right]\label{basic0}
\ea
Note that we used the symbol $\la\;,\;\ra$ to denote on the LHS the inner product in $\mL^A$, while on the RHS the Hilbert Schmidt inner product in $\mB(\mH^{B})$. 

The above inner product is independent of the choice of the orthonormal basis $\{X_{a}\}$ of $\mB(\mH^{A_0})$,
and can be expressed in terms of the Choi matrices. 
The Choi matrix of $\Psi^{A}$ is given by  
\be\label{choidef}
 J_{\Psi}^A\equiv J_{\Psi}^{A_0 A_1}\equiv\left(\id^{A_0}\otimes\Psi^{\tA_0 \to A_1}\right)\left(\phi_{+}^{A_0\tA_0}\right)
\ee
where the tilde symbol will always indicate an  identical copy of the system under it, and $\phi_{+}^{A_0\tA_0}\equiv|\phi_{+}\lr\phi_{+}|^{A_0\tA_0}$ is an unnormalized maximally entangled state $|\phi_{+}\ra^{A_0\tA_0}\equiv\sum_{i=1}^{d_{A_0}}|i\ra^{A_0}|i\ra^{\tA_0}$. 
It is straightforward to show that by taking a basis $X_{a\equiv(i,j)}=|i\lr j|^{A_0}$ in~\eqref{basic0},
the inner product between linear maps is equivalent to inner product between their corresponding Choi matrices:\begin{align}
\la\Phi,\Psi\ra=\la J_{\Phi}^{A},J_{\Psi}^{A}\ra=\tr\left[\left(J_{\Phi}^{ A}\right)^*J_{\Psi}^{A}\right].\label{equiva}
\end{align}

We now define the canonical orthonormal basis of $\mL^A$. Let $\{X_{a}\}$ and $\{Y_{b}\}$ be two orthonormal bases of $\mB(\mH^{A_0})$ and $\mB(\mH^{A_1})$, respectively. Then, one can construct an orthonormal basis, $\left\{\mE_{a_0a_1}^{A_0\to A_1}\right\}$ of $\mL^A$:
\be\label{ch-basis}
\mE_{a_0a_1}^{A_0\to A_1}(\rho)\equiv\tr[X_{a_0}^{* }\rho] Y_{a_1}\quad\forall\rho\in\mB(\mH^{A_0})\;.
\ee 
It is straight forward to check that $\left\{\mE_{a_0a_1}^{A_0\to A_1}\right\}$ is an orthonormal basis of $\mL^A$. The canonical orthonormal basis of $\mL^A$, is the one obtained from the above basis by taking $X_{a_0\equiv(i,j)}=|i\lr j|^{A_0}$ and $Y_{a_1\equiv(k,\ell)}=|k\lr \ell|^{A_1}$; i.e.
\be\label{elbasis}
\mE_{a_0a_1}^{A_0\to A_1}(\rho)\equiv{}^{A_0}\la i|\rho|j\ra^{A_0}\; |k\ra\la\ell|^{A_1}\;.
\ee 

\subsection{The space of supermaps}

We denote by $\mbb{L}^{AB}$ (with $A\equiv A_0A_1$ and $B\equiv B_0B_1$) the space of all linear maps $\Theta:\mL^A\to\mL^B$. This space is also a vector space, equipped with the following inner product: for all $\Theta_1,\Theta_2\in\mbb{L}^{AB}$ and an orthonormal basis $\{\mE_{a_0a_1}^{A_0\to A_1}\}$ of $\mL^A$:
\be
\la\Theta_1,\Theta_2\ra\equiv\sum_{a_0,a_1}\left\la\Theta_1\left[\mE_{a_0a_1}^{A_0\to A_1}\right],\Theta_2\left[\mE_{a_0a_1}^{A_0\to A_1}\right]\right\ra\;,
\ee 
where we used the symbol $\la\;,\;\ra$ to denote on the LHS the inner product in $\mbb{L}^{AB}$, while on the RHS the inner product in $\mL^B$ (with a definition of inner product as in~\eqref{basic0}). As before, this definition is also independent of the choice of the orthonormal basis $\{\mE_{a_0a_1}^{A_0\to A_1}\}$ of $\mL^A$.

In analogy with the space $\mL^A$, for any $\Theta\in\mbb{L}^{AB}$ we associate a Choi matrix $\mathbf{J}_{\Theta}^{AB}\in\mB(\mH^{AB})$ which is defined as follows. Let  $\{\mE_{a_0a_1}^{A_0\to A_1}\}$ be the canonical orthonormal basis of $\mL^A$ given in~\eqref{elbasis}, with indices $a\equiv(i,j)$ and $b\equiv(k,\ell)$. Then, (cf. Proposition 7 in~\cite{DW16})
\begin{align}\label{defichoi}
\J_{\Theta}^{AB}\equiv \sum_{a_0,a_1}J^{A}_{\mE_{a_0a_1}}\otimes J^{B}_{\Theta\left[\mE_{a_0a_1}\right]}\;,
\end{align}
where $J^{A}_{\mE_{a_0a_1}}$ and $J^{B}_{\Theta\left[\mE_{a_0a_1}\right]}$ are the Choi matrices of
the maps $\mE_{a_0a_1}^{A_{0}\to A_{1}}\in\mL^A$ and $\Theta\left[\mE_{a_0a_1}^{{A_{0}\to A_{1}}}\right]\in\mL^{B}$, respectively (recall that $\Theta$ transforms linear maps from $A_0$ to $A_1$ to linear maps from $B_0$ to $B_1$). In the following we provide several motivations for this definition. We start with the preservation of inner products.

Consider two maps $\Theta_1,\Theta_2\in\mbb{L}^{AB}$ and their corresponding Choi matrices $\J_{\Theta_1}^{AB}$ and $\J_{\Theta_2}^{AB}$. The Hilbert Schmidt inner product
between $\J_{\Theta_1}^{AB}$ and $\J_{\Theta_2}^{AB}$ can be expressed as follows:
\ba
&\left\la \J_{\Theta_1}^{AB},\J_{\Theta_2}^{AB}\right\ra\\
&=\sum_{a_0,a_1,a_0',a_1'}\left\la J^{A}_{\mE_{a_0a_1}}\otimes J^{B}_{\Theta_1\left[\mE_{a_0a_1}\right]},J^{A}_{\mE_{a_0'a_1'}}\otimes J^{B}_{\Theta_2\left[\mE_{a_0'a_1'}\right]}\right\ra\\
&=\sum_{a_0,a_1,a_0',a_1'}\left\la J^{A}_{\mE_{a_0a_1}},J^{A}_{\mE_{a_0'a_1'}}\right\ra\left\la J^{B}_{\Theta_1\left[\mE_{a_0a_1}\right]}, J^{B}_{\Theta_2\left[\mE_{a_0'a_1'}\right]}\right\ra\\
&=\sum_{a_0,a_1,a_0',a_1'}\left\la \mE_{a_0a_1},\mE_{a_0'a_1'}\right\ra\left\la {\Theta_1\left[\mE_{a_0a_1}\right]}, {\Theta_2\left[\mE_{a_0'a_1'}\right]}\right\ra\\
&=\sum_{a_0,a_1,a_0',a_1'}\delta_{a_0a_0'}\delta_{a_1a_1'}\left\la {\Theta_1\left[\mE_{a_0a_1}\right]}, {\Theta_2\left[\mE_{a_0'a_1'}\right]}\right\ra\\
&=\sum_{a_0,a_1}\left\la {\Theta_1\left[\mE_{a_0a_1}\right]}, {\Theta_2\left[\mE_{a_0a_1}\right]}\right\ra=\la\Theta_1,\Theta_2\ra\;,
\ea
where in the third equality we used~\eqref{equiva}.

The second motivation for the definition~\eqref{defichoi} is the following. Consider a linear map $\Theta\in\mbb{L}^{AB}$, a linear map $\Psi^A\in\mL^A$, and define $\Phi^B\equiv\Theta\left[\Psi^A\right]\in\mL^B$. Then, the Choi matrices of $J^{A}_{\Psi}$ and $J^{B}_{\Phi}$ of $\Psi^{A}$ and $\Phi^{B}$, respectively,  are related via (see also~\cite{Pavia1,LM15})
\be\label{trans}
J^{B}_{\Phi}=\tr_{A}\left[\J^{AB}_{\Theta}\left(\left(J^{A}_{\Psi}\right)^T\otimes I^{B}\right)\right]\;.
\ee
That is, $\J^{AB}_{\Theta}$ can be interpreted as the Choi matrix of the linear map $\Delta^{A\to B}_{\Theta}$ that converts $J^{A}_{\Psi}$ to $J^{B}_{\Phi}$. It is defined by
\be
\Delta^{A\to B}_{\Theta}\left(J^A_\Psi\right)\equiv J^{B}_{\Theta[\Psi]}
\ee

To see why~\eqref{trans} holds, note that the Choi matrix $J^{A}_{\Psi}$ is linear in $\Psi$, 
and from~\eqref{elbasis} $J_{\mE_{a_0a_1}}^{A}$ is a real matrix for the canonical basis $\{\mE_{a_0a_1}^{A_0\to A_1}\}$, so that
\ba
&\tr_{A}\left[\J^{AB}_{\Theta}\left(\left(J^{A}_{\Psi}\right)^T\otimes I^{B}\right)\right]\\
&=\sum_{a_0,a_1}\tr\left[J^{A}_{\mE_{a_0a_1}}\left(J^{A}_{\Psi}\right)^T\right] J^{B}_{\Theta\left[\mE_{a_0a_1}\right]}\\
&=\sum_{a_0,a_1}\tr\left[\left(J^{A}_{\mE_{a_0a_1}}\right)^{*}J^{A}_{\Psi}\right] J^{B}_{\Theta\left[\mE_{a_0a_1}\right]}\\
&=\sum_{a_0,a_1}\la \mE_{a_0a_1},\Psi\ra J^{B}_{\Theta\left[\mE_{a_0a_1}\right]}\\
&=J^{B}_{\Theta\left[\sum_{a_0,a_1}\la \mE_{a_0a_1},\Psi\ra\;\mE_{a_0a_1}\right]}\\
&=J^{B}_{\Theta\left[\Psi\right]}=J^{B}_{\Phi}\;.\nonumber\ea

The matrix $\J^{AB}_{\Theta}$ can also be expressed as the Choi matrix of the linear map 
$\Lambda^{AB}_{\Theta}: \mB(\mH^{A_0B_0})\to\mB(\mH^{A_1B_1})$ defined by the relation:
\be\label{lambdatheta}
\J^{AB}_{\Theta}=\id^{A_0B_0}\otimes\Lambda^{AB}_{\Theta}\left(\phi^{A_0\tA_0}_{+}\otimes\phi_{+}^{B_0\tB_0}\right)\;.
\ee
From~\eqref{defichoi} it follows that $\Lambda^{AB}_\Theta$ can be expressed as:
\ba\label{reltheta}
\Lambda^{AB}_\Theta&=\sum_{a_0,a_1}\mE_{a_0a_1}^{A_0\to A_1}
\otimes \Theta\left[\mE_{a_0a_1}^{A_0\to A_1}\right]\\
&\equiv(\1^{A}\otimes\Theta)\left[\Upsilon^{A\tA}\right]\;,
\ea
where we denote by $\1^{A}:\mL^A\to\mL^A$ the identity map (whereas the symbol $\id$ is reserved for the identity map from $\mB(\mH)\to\mB(\mH)$; see for example its use in Eq.~\eqref{choidef}), and
$$
\Upsilon^{A\tA}\equiv \sum_{a_0,a_1}\mE_{a_0a_1}^{A_0\to A_1}\otimes \mE_{a_0a_1}^{\tA_0\to \tA_1}\;.
$$
Note that the mapping $\Theta\mapsto\Lambda^{AB}_\Theta=(\1^{A}\otimes\Theta)\left[\Upsilon^{A\tA}\right]$ defines an isomorphism between $\Theta$ and $\Lambda_{\Theta}^{AB}$.
The map $\Upsilon^{A\tA}$ is completely positive, and it is the CP map analog of the maximally entangled state. Its action on matrices $\rho^{A_0\tA_0}\in\mB(\mH^{A_0\tA_0})$ is given by: 
$$
\Upsilon^{A\tA}\left(\rho^{A_0\tA_0}\right)=\tr\left[\phi^{A_0\tA_0}_{+}\rho^{A_0\tA_0}\right]\phi^{A_1\tA_1}_{+}\;.
$$
Similar to the property of the  maximally entangled state, $\Upsilon^{A\tA}$ satisfies for any $\Theta\in\mbb{L}^{AB}$ the relation
\be
\1^A\otimes \Theta[\Upsilon^{A\tA}]=\Theta^T\otimes\1^B[\Upsilon^{\tB B}]\label{swapp}
\ee
where $\Theta^T:\mL^{\tB}\to\mL^A$ is the transposition of $\Theta$ which is defined by its components 
$$
\left\la\mE_{a_0a_1}^{A},\Theta^T\left[\mE_{a_0'a_1'}^{B}\right]\right\ra\equiv \left\la\mE_{a_0'a_1'}^{B},\Theta\left[\mE_{a_0a_1}^{A}\right]\right\ra\quad\forall\;a_0,a_1,a_0',a_1'
$$
where $\{\mE_{a_0a_1}^A\}$ and $\{\mE_{a_0'a_1'}^B\}$ are the canonical orthonormal bases of $\mL^A$ and $\mL^B$.

Both the maps $\Lambda^{AB}_\Theta$ and $\Delta^{A\to B}_{\Theta}$ correspond to the same map $\Theta\in\mbb{L}^{AB}$ and have the \emph{same} Choi matrix $\J^{AB}_\Theta$. 
That is,
\ba
\J^{AB}_\Theta&=\id^{A_0B_0}\otimes\Lambda^{\tA_0\tB_0\to A_1B_1}_\Theta\left(\phi_{+}^{A_0\tA_0}\otimes\phi_{+}^{B_0\tB_0}\right)\nonumber\\
&=\id^{A_0A_1}\otimes\Delta^{\tA_0\tA_1\to B_0B_1}_{\Theta}\left(\phi_{+}^{A_0\tA_0}\otimes\phi_{+}^{A_1\tA_1}\right)
\ea
Note that~\eqref{swapp} implies that 
$
\Lambda^{BA}_{\Theta^T}=\1^B\otimes\Theta^T[\Upsilon^{B \tB}]=\Theta\otimes\1^A[\Upsilon^{\tA A}]\;.
$
Therefore, the Choi matrix \be \J_{\Theta^T}^{BA}=\text{swap}(\J^{AB}_\Theta)\label{z1}\ee where the swap operator is between $A$ and $B$.
Moreover, 
the mapping $\Theta\mapsto\Lambda_\Theta^{AB}$ is an isomorphism map between $\mbb{L}^{AB}$ and the space of bipartite maps $\mL^{AB}$, and similarly the mapping $\Theta\mapsto\Delta^{A\to B}_{\Theta}$ is an isomorphism as well.

The dual of a linear map $\Theta\in\mbb{L}^{AB}$ is a linear map $\Theta^{*}\in\mbb{L}^{BA}$ with the property that for all $\Phi^A\in\mL^A$ and for all $\Psi^B\in\mL^B$
\be\label{dualtheta}
\left\la\Phi^A,\Theta^*[\Psi^B]\right\ra=\left\la\Theta[\Phi^A],\Psi^B\right\ra\;.
\ee
Note that the LHS of the equation above can be expressed as:
\begin{align}\label{a1}
\left\la\Phi^{A}\;,\;\Theta^{*}\left[\Psi^{B}\right]\right\ra
&=\left\la J_{\Phi}^{A}\;,\;J_{\Theta^{*}[\Psi]}^{A}\right\ra\nonumber\\
&=\left\la J_{\Phi}^{A}\;,\;\Delta_{\Theta^{*}}^{B\to A}\left(J_{\Psi}^{B}\right)\right\ra
\end{align}
and the RHS of~\eqref{dualtheta} can be expressed as:
\begin{align}\label{a2}
\left\la\Theta[\Phi^A],\Psi^B\right\ra&=\left\la J_{\Theta[\Phi]}^{B}\;,\;J_{\Psi}^{B}\right\ra\nonumber\\
&=\left\la \Delta_{\Theta}^{A\to B}\left(J_{\Phi}^{A}\right)\;,\;J_{\Psi}^{B}\right\ra\nonumber\\
&=\left\la J_{\Phi}^{A}\;,\;\Delta_{\Theta}^{*B\to A}\left(J_{\Psi}^{B}\right)\right\ra
\end{align}
Comparing~\eqref{a1} and~\eqref{a2} we conclude that
$$
\Delta_{\Theta^{*}}^{B\to A}=\Delta_{\Theta}^{*B\to A}\;.
$$
Consequently, this also implies that
$$
\J_{\Theta^*}^{BA}=\overline{\J_{\Theta^T}^{BA}}\;.
$$

Finally, we point out that for two supermaps $\Theta_1\in\mbb{L}^{AB}$ and $\Theta_2\in\mbb{L}^{CD}$ we have
\ba
& \Delta_{\Theta_2\circ\Theta_1}^{A\to D}=\Delta_{\Theta_2}^{C\to D}\circ\Delta_{\Theta_1}^{A\to B}\quad\quad(\text{here }B\cong C)\nonumber\\
& \Delta_{\Theta_1\otimes\Theta_2}^{AC\to BD}=\Delta_{\Theta_1}^{A\to B}\otimes\Delta_{\Theta_2}^{C\to D}
\ea

\subsection{Completely Positive Preserving (CPP) Maps and Superchannels}

We will denote by $\mL^{A}_{+}$ (and similarly $\mL^B_+$) the convex subset of $\mL^A$ consisting of all completely positive (CP) maps in $\mL^A$. We also denote by $\mC^A$ all the elements in $\mL^A_{+}$ 
that are also trace-preserving (TP); i.e. $\mC^A$ is the convex set of all quantum channels in $\mL^A$.

\begin{definition}
Let $\Theta\in\mbb{L}^{AB}$ be a linear map. We say that:
\begin{enumerate}
\item $\Theta$ is \emph{CP preserving (CPP)}  if $\Theta[\Psi^A]\in\mL^{B}_{+}$  for any CP map $\Psi^A\in\mL^{A}_+$.
\item $\Theta$ is \emph{completely CPP}  if  $\1^{C}\otimes\Theta$ is positive for all dimensions of system $C\equiv(C_0,C_1)$.
\item $\Theta$ is \emph{TP preserving (TPP)} if $\Theta[\Psi^A]$ is a TP map for any TP map $\Psi^A\in\mL^A$.
\item $\Theta$ is  a \emph{superchannel} if it is completely CPP and TPP.
\end{enumerate} 
\end{definition}

The following theorem provides the characterization and realization of a superchannel.

\begin{figure}[t]
    \includegraphics[width=0.4\textwidth]{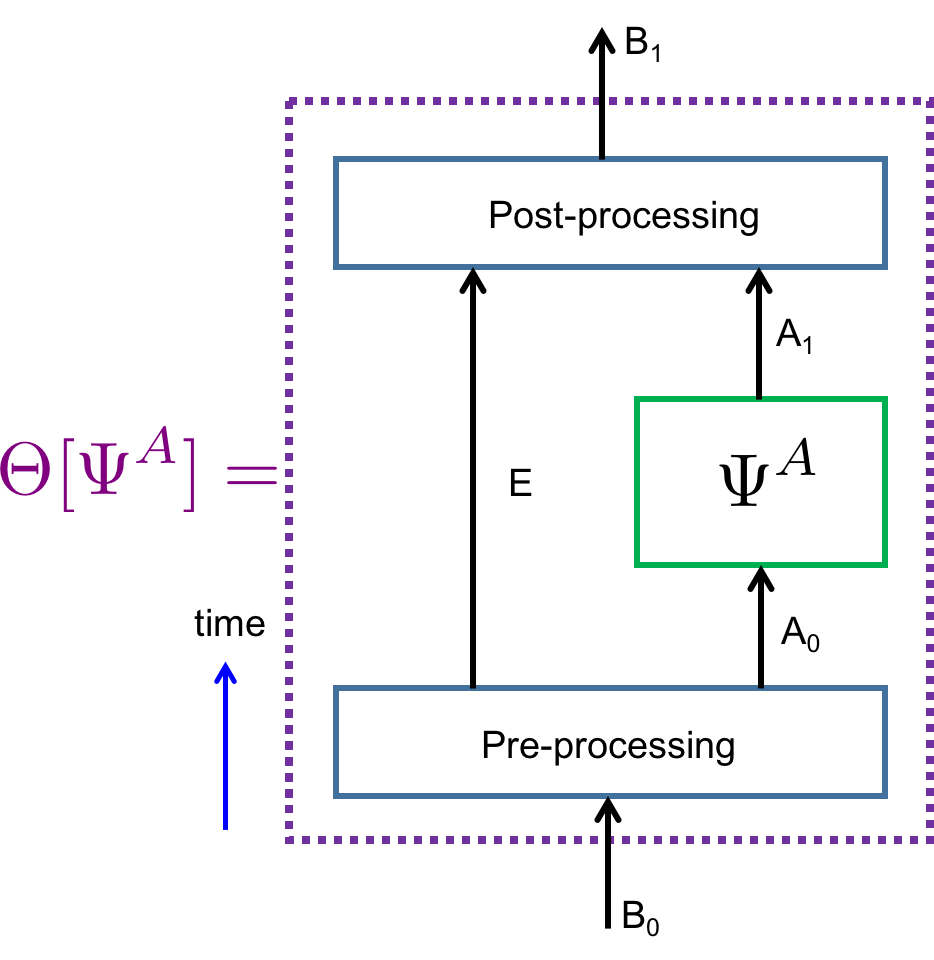}
  \caption{\linespread{1}\selectfont{\small Realization of a Superchannel}}
  \label{superchannel}
\end{figure}

\begin{theorem}\label{premain}{\rm \cite{Pavia1}}
Let $\Theta\in\mbb{L}^{AB}$. The following are equivalent.
\begin{enumerate}
\item $\Theta$ is a superchannel.
\item The Choi matrix $\J^{AB}_{\Theta}\geq 0$ with marginals
\be\label{marginals}
\J^{A_1 B_0}_{\Theta}=I^{A_1 B_0}\quad;\quad \J_{\Theta}^{AB_0}=\J^{A_0 B_0}_{\Theta}\otimes u^{A_1}
\ee
where $u^{A_1}\equiv\frac{1}{d_{A_1}}I^{A_1}$ is the maximally mixed state (i.e. the uniform state) on system $A_1$.
\item The map $\Delta_{\Theta}^{A\to B}$ is CP, and there exists a unital CP map $\Delta_{\Theta}^{A_0\to B_0}$ such that the map $\Delta_{\Theta}^{A\to B_0}\equiv\tr_{B_1}\circ\Delta_{\Theta}^{A\to B}$ satisfies
\be\label{formap}
\Delta_{\Theta}^{A\to B_0}=\Delta_{\Theta}^{A_0\to B_0}\circ\tr_{A_1}
\ee
\item There exists a Hilbert space $\mH^E$, with $d_{E}\leq d_{A_0}d_{B_0}$, and two CPTP maps $\Gamma_{\rm pre}^{B_0\to A_0 E}:\mB(\mH^{B_0})\to\mB(\mH^{A_{0}E})$ and $\Gamma_{\rm post}^{A_{1}E\to B_{1}}:\mB(\mH^{A_{1}E})\to\mB(\mH^{B_1})$ such that for all $\Psi^{A}\in\mL^A$
\ba\label{realization}
&\Theta\left[\Psi^{A}\right]=\\
&\Gamma_{\rm post}^{A_{1}E\to B_{1}}\circ\left(\Psi^{A_0\to A_1}\otimes\id^E\right)\circ \Gamma_{\rm pre}^{B_0\to A_0 E}
\ea
(see Fig.~\ref{superchannel}).
\end{enumerate}
\end{theorem}

The original proof of this theorem can be found in~\cite{Pavia1}. However, for the purpose of being self contained, we provide here an alternative proof that is based on similar ideas as given in~\cite{Pia2005} for the characterization of semi-causal maps~\cite{Beck2001,Wer2002,Pia2005,Dar2011}.

{{\it{Proof.}}}
We start by proving ${\it 1}\Rightarrow{\it 2}$. Let $Z^{B_0}\in\mB(\mH^{B_0})$ be an arbitrary element on the bounded operators on $B_0$.
Multiplying both sides of~\eqref{trans} by $Z^{B_0}\otimes I^{B_1}$  and taking the trace we get that  
\be\label{11}
\tr\left[\J^{AB_0}_{\Theta}\left(J_{\Psi}^{A}\otimes Z^{B_0}\right)^T\right]=\tr[Z^{B_0}]\;.
\ee
The above equation holds for all $Z^{B_0}\in\mB(\mH^{B_0})$ and all trace preserving maps $\Psi^{A}\in\mL^{A}$ for which $J_{\Psi}^{A_0}=I^{A_0}$. 
Since the above equation holds for all such $J_\Psi^{A}$, it also holds for $\frac{1}{d_{A_1}}I^{A}$, so that 
\be
\tr\left[\J^{AB_0}_{\Theta}\left(Y^{A}\otimes Z^{B_0}\right)^T\right]=0
\ee
for any matrix $Y^{A}\equiv J_{\Psi}^{A}-\frac{1}{d_{A_1}}I^{A}$ whose marginal $Y^{A_0}=0$ and any $Z^{B_0}$. Therefore, $\J_{\Theta}^{AB_0}$ must have the form $\J^{A_0B_0}_{\Theta}\otimes u^{A_1}$. Finally, substituting this form into~\eqref{11} and taking $J_{\Psi}^{A}=\frac{1}{d_{A_1}}I^{A}$ gives for all $Z^{B_0}\in\mB(\mH^{B_0})$
\be
\tr\left[\J^{B_0}_{\Theta}(Z^{B_0})^T\right]=d_{A_1}\tr[Z^{B_0}]\;.
\ee
Therefore, $\J^{B_0}_{\Theta}=d_{A_1}I^{B_0}$ so that $\J^{A_1B_0}_\Theta=I^{A_1B_0}$. The converse follows from the form of the Choi matrix. This completes the proof that $1\Rightarrow 2$.

We now prove that $2\Rightarrow 4$. Let $\phi^{ABC}$ be purification of $\J_{\Theta}^{AB}$, and let
$\psi^{A_0B_0E}$ be a purification $\frac{1}{d_{A_1}}\J^{A_0 B_0}_{\Theta}$. The latter always exists with $d_{E}\leq d_{A_0}d_{B_0}$. Then, from the relation
$\J_{\Theta}^{AB_0}=\J^{A_0 B_0}_{\Theta}\otimes u^{A_1}$ we conclude that $\psi^{A_0B_0E}\otimes\phi_{+}^{A_1\tA_1}$ is a purification of  $\J_{\Theta}^{AB_0}$. Therefore, since $\phi^{ABC}$ is also a purification of $\J_{\Theta}^{AB_0}$ there exists an isometric channel $\mV^{\tA_1E\to B_1C}$ such that
$$
\phi^{ABC}=\id^{AB_0}\otimes\mV^{\tA_1E\to B_1C}\left(\psi^{A_0B_0E}\otimes\phi_{+}^{A_1\tA_1}\right)
$$
Tracing out system $C$ on both sides, and denoting $\Gamma^{\tA_1E\to B_1}_{\post}\equiv\tr_{C}\circ\mV^{\tA_1E\to B_1C}$ gives
\be\label{choiform9}
\J_{\Theta}^{AB}=\id^{AB_0}\otimes\Gamma^{\tA_1E\to B_1}_{\post}\left(\psi^{A_0B_0E}\otimes\phi_{+}^{A_1\tA_1}\right)
\ee
From its definition, $\tr_{A_0E}[\psi^{A_0B_0E}]=I^{B_0}$. Hence, there exists a CPTP map (in fact an isometry) $\Gamma^{\tB_0\to A_0E}_{\pree}$ such that $\psi^{A_0B_0E}=\id^{B_0}\otimes \Gamma^{\tB_0\to A_0E}_{\pree}(\phi_{+}^{B_0\tB_0})$. Hence,
\ba
&\J_{\Theta}^{AB}=\left(\id^{A_1B_0}\otimes\Gamma^{\tA_1\tB_0\to A_0B_1}_\Theta\right)\left(\phi_{+}^{A_1\tA_1}\otimes\phi_{+}^{B_0\tB_0}\right)\\
&\Gamma^{\tA_1\tB_0\to A_0B_1}_\Theta\equiv\\
&\left(\id^{A_0}\otimes\Gamma^{\tA_1E\to B_1}_{\post}\right)\circ\left(\id^{\tA_1}\otimes\Gamma^{\tB_0\to A_0E}_{\pree}\right)\label{gammapre}
\ea
To finish the proof, denote $\Phi^B\equiv\Theta[\Psi^A]$ and from~\eqref{trans} we get for arbitrary $\rho\in\mB(\mH^{B_0})$ 
\ba
\Phi(\rho)&=\tr_{B_0}\left[J^{B}_{\Phi}\left(\rho^T\otimes I^{B_1}\right)\right]\\
&=\tr_{AB_0}\left[\J^{AB}_{\Theta}\left(\left(J^{A}_{\Psi}\right)^T\otimes \rho^T\otimes I^{B_1}\right)\right]\;.
\ea
Next, we substitute the above expression for $\J^{AB}_{\Theta}$ to get
\ba
\Phi(\rho)&=\tr_{A}\Big[\left(\left(J^{A}_{\Psi}\right)^T \otimes I^{B_1}\right)\times\\
&\left(\id^{A}\otimes\Gamma^{\tA_1E\to B_1}_{\post}\left(\phi_{+}^{A_1\tA_1}\otimes\Gamma^{\tB_0\to A_0E}_{\pree}(\rho)\right)\right)\Big]\label{formx}
\ea
where we traced out system $B_0$ after using the relation 
$$\left(\rho^T\otimes I^{\tB_0}\right)
\phi_{+}^{B_0\tB_0}=\left(I^{B_0}\otimes \rho\right)
\phi_{+}^{B_0\tB_0}\;.
$$
Finally, note that 
$$
\left(J^{A}_{\Psi}\right)^T=\sum_{i,j}|j\lr i|^{A_0}\otimes\left(\Psi^{\tA_0\to A_1}\left[|i\lr j|^{\tA_0}\right]\right)^T\;,
$$
so that 
\ba
&\tr_{A_1}\left[\left(\left(J^{A}_{\Psi}\right)^T\otimes I^{\tA_1}\right)\left( I^{A_0}\otimes \phi_{+}^{A_1\tA_1}\right)\right]\\
&=\sum_{i,j}|j\lr i|^{A_0}\otimes \Psi^{\tA_0\to \tA_1}\left[|i\lr j|^{\tA_0}\right]\;.\nonumber
\ea
Substituting this into~\eqref{formx} we conclude that
\ba
&\Phi(\rho)=\sum_{i,j}\Gamma^{\tA_1E\to B_1}_{\post}\\
&\left(\Psi^{\tA_0\to \tA_1}\left[|i\lr j|^{\tA_0}\right]\otimes \left\la i^{A_0}\left|\Gamma^{\tB_0\to A_0E}_{\pree}(\rho) \right|j^{A_0}\right\ra\right)\\
&=\Gamma_{\rm post}^{A_{1}E\to B_{1}}\left[(\Psi^{A_0\to A_1}\otimes\id^E)\left( \Gamma_{\rm pre}^{B_0\to A_0 E}(\rho)\right)\right]\;.\nonumber
\ea
This completes the proof that $2\Rightarrow 4$. The proof that $4\Rightarrow 1$ is trivial, so we have $1\Rightarrow 2\Rightarrow 4\Rightarrow 1$. Hence, we proved that points $1$, $2$, and $4$, are all equivalent.
To complete the proof, we show now that $2$ and $3$ are equivalent.

Suppose $\Delta^{A\to B}_\Theta$ has the form~\eqref{formap}. Then, since $\J^{AB}_\Theta$ is its Choi matrix we get
\begin{align*}
\J^{AB_0}_\Theta&=\Delta^{\tA\to B_0}_\Theta\left(\phi_{+}^{A_0\tA_0}\otimes\phi_{+}^{A_1\tA_1}\right)\nonumber\\
&=\Delta^{\tA_0\to B_0}_\Theta\left(\phi_{+}^{A_0\tA_0}\right)\otimes I^{A_1}\;,
\end{align*}
where we used the form~\eqref{formap}. Hence, $\J^{AB_0}_\Theta=\J^{A_0B_0}_\Theta\otimes u^{A_1}$, and since $\Delta^{\tA\to B_0}_\Theta$ is a unital CP map we conclude that $\J^{B_0}_\Theta=d_{A_1}I^{B_0}$ so that $\J^{A_1B_0}_\Theta=I^{A_1B_0}$.

Conversely, suppose the Choi matrix of $\Theta$ is positive semidefinite and has marginals as in~\eqref{marginals}. Define the map $\Delta^{A_0\to B_0}_\Theta$ to be the (unique) map satisfying $$\frac{1}{d_{A_1}}\J^{A_0B_0}_\Theta=\Delta^{\tA_0\to B_0}_\Theta\left(\phi_{+}^{A_0\tA_0}\right)\;.$$
Therefore, $\Delta^{A_0\to B_0}_\Theta$ is a unital CP map since $\J^{B_0}_\Theta=d_{A_1}I^{B_0}$. Moreover, from the relation $\J^{AB_0}_\Theta=\J^{A_0B_0}_\Theta\otimes u^{A_1}$ we get that the two maps 
$
\Delta^{A\to B_0}_\Theta
$
and $\Delta^{A_0\to B_0}_\Theta\circ\tr_{A_1}$ have the same Choi matrix and therefore they must be the same. This completes the proof of the equivalence between 2 and 3.
\QED

\begin{figure}[t]
    \includegraphics[width=0.4\textwidth]{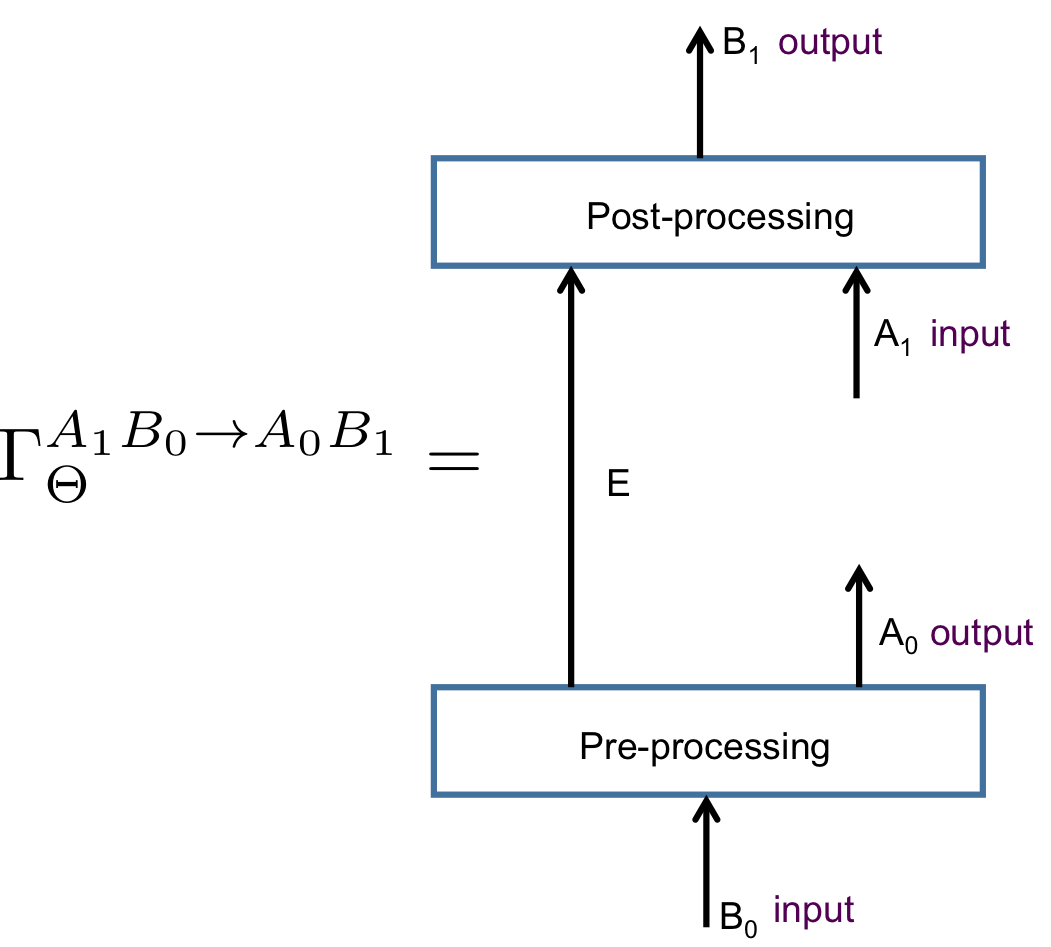}
  \caption{\linespread{1}\selectfont{\small Realization of the channel $\Gamma_{\Theta}^{A_1B_0\to A_0B_1}$}}
  \label{bigam}
\end{figure}

From the theorem above, it follows that $\Theta$ is a superchannel if and only if the CPTP map $\Gamma^{A_1B_0\to A_0B_1}_{\Theta}$ which corresponds to the Choi matrix 
\be\label{cgam}
\J^{AB}_\Theta=\id^{A_1B_0}\otimes\Gamma^{\tA_1\tB_0\to A_0B_1}_{\Theta}\left(\phi_{+}^{A_1\tA_1}\otimes\phi_{+}^{B_0\tB_0}\right)
\ee
can be expressed as 
\be\label{regam}
\Gamma^{A_1B_0\to A_0B_1}_{\Theta}=\left(\id^{A_0}\otimes\Gamma^{A_1E\to B_1}_{\post}\right)\circ\left(\id^{A_1}\otimes\Gamma^{B_0\to A_0E}_{\pree}\right)\;.
\ee
Therefore, $\Gamma^{A_1B_0\to A_0B_1}_{\Theta}$ is the CPTP map obtained from $\Theta$ by taking $A_0$ and $B_1$ to be the outputs and $A_1$ and $B_0$ to be the inputs as described in Fig.~\ref{bigam}.

Note that the CPTP map $\Gamma^{A_1B_0\to A_0B_1}_{\Theta}$ has the form~\eqref{regam} if and only if its marginal map $\Gamma^{A_1B_0\to A_0}_{\Theta}\equiv\tr_{B_1}\circ\Gamma^{A_1B_0\to A_0B_1}_{\Theta}$ has the form: 
\be\label{int}
\Gamma^{A_1B_0\to A_0}_{\Theta}=\Gamma^{B_0\to A_0}_\Theta\circ\tr_{A_1}\;,
\ee
where $\Gamma^{B_0\to A_0}_\Theta\equiv\tr_E\circ\Gamma_{\pree}^{B_0\to A_0E}$ is some CPTP map. This condition is somewhat similar to the condition in~\eqref{formap}.

\begin{figure}[h]
    \includegraphics[width=0.3\textwidth]{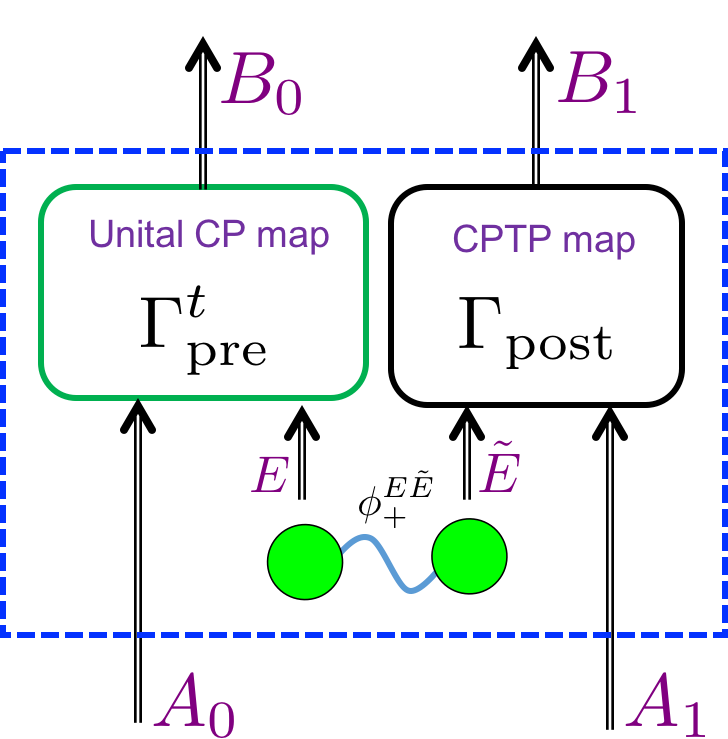}
  \caption{\linespread{1}\selectfont{\small Realization of the CP map $\Delta_{\Theta}^{A\to B}$}}
  \label{delta}
\end{figure}

The maps $\Delta^{A\to B}_{\Theta}$ and $\Gamma^{A_1B_0\to A_0B_1}_{\Theta}$ are closely related as they correspond to the same Choi matrix $\J^{AB}_{\Theta}$. In particular, the CPTP map $\Gamma^{B_0\to A_0}_\Theta$ of~\eqref{int} is related to the unital CP map $\Delta_\Theta^{A_0\to B_0}$ of~\eqref{formap} via
\be\label{rel}
\Gamma^{B_0\to A_0}_\Theta(\rho^{B_0})= \left[\Delta^{*B_0\to A_0}_{\Theta}\left(\left(\rho^{B_0}\right)^T\right)\right]^T
\ee
for all $\rho^{B_0}\in\mB(\mH^{B_0})$. More generally, it can be shown that (see also Fig.~\ref{delta})
\begin{align}
&\Delta_{\Theta}^{A\to B}\left(\sigma^{A}\right)=\nonumber\\
&\left(\Gamma_{\text{pre}}^{tA_0E\to B_0}\otimes\id^{B_1}\right)\circ\left(\id^{A_0E}\otimes\Gamma_{\text{post}}^{A_1\tilde{E}\to B_1}\right)\left(\sigma^{A}\otimes \phi_{+}^{E\tilde{E}}\right)\;,\nonumber
\end{align}
where $\Gamma_{\text{pre}}^{tA_0E\to B_0}$ is the unital CP map obtained from the CPTP map $\Gamma_{\text{pre}}^{B_0\to A_0E}$ by replacing all of its Kraus operators (in the operator sum representation) with their transpose.

\subsection{Entropies}\label{pre-entropies}
 Entropy functions, such as the family of R\'enyi entropies, measure how noisy a quantum state is. Therefore, they do not change under unitary channels, meaning that the entropy of a quantum state $\rho\in\mB(\mH)$ is the same as that of $\mU(\rho)\equiv U\rho U^*$, for any unitary matrix $U$ acting on $\mH$. Now, suppose that $U$ is chosen at random from some ensemble $\{p_x,\;U_x\}_{x=1}^{m}$ of unitary matrices. Still the entropy of 
$\mU_x(\rho)\equiv U_x\rho U_x^*$ and $\rho$ are the same for each $x$. If the information about $x$ is lost, the state of the system becomes:
$$
\sum_{x=1}^{m}p_x\;\mU_x(\rho) .
$$
Since ``losing information" cannot decrease noise, the entropy of the state above can only be larger than the entropy of $\rho$. That is, entropy functions must behave monotonically under random unitary channels. 

Since we identify the maximally mixed state as the state with the most noise, quantum channels with the same input and output dimensions, and that do not decrease noise, must 
preserve the maximally mixed state. Indeed, random unitary channels have this property.
The set of all channels with the same input and output dimensions that preserve the maximally mixed state are called unital CPTP maps or doubly stochastic channels. While they form a strictly larger set of channels than the set of random unitary channels, 
if a state $\rho$ can be converted into $\sigma$ via a doubly stochastic channel, then this transformation can also be achieved by a random unitary channel (see for example Lemma 10 in Ref.~\cite{Gour15}). 

From the discussion above it follows that all entropy functions cannot decrease under doubly stochastic channels. In addition, entropy functions are also additive under tensor products. To summarize, an entropy function $f:\mD(\mH)\to\mbb{R}$ satisfies the following 3 conditions:
\begin{enumerate}
\item {\it Monotonicity}: For any random unitary channel $\Phi:\mB(\mH)\to\mB(\mH)$, 
$$
f(\Phi(\rho))\geq f(\rho)\;.
$$
\item {\it Additivity}: For any two quantum states $\rho\in\mB(\mH^A)$ and $\sigma\in\mB(\mH^B)$ we have
$$
f(\rho\otimes\sigma)=f(\rho)+f(\sigma)\;.
$$
\item  {\it Normalization}: on maximally mixed state $f(\frac{1}{d}I)=\log(d)$, and on pure state $f(|\psi\lr \psi|)=0$.
\end{enumerate}

One may add other conditions such as concavity, or subadditivity, but they are not as fundamental as the above three (for example, not all the R\'enyi entropies satisfy them).

\subsubsection{The conditional min-entropy}

The min-entropy plays an important role in quantum information. It is the smallest entropy in the family of R\'enyi entropies, and in this sense provides the most conservative way to quantify uncertainty.
The min-entropy of a density matrix $\rho\in\mB(\mH^{A_1})$ is defined by:
$$
H_{\min}(A_1)_\rho\equiv -\log\min\{t\in\mbb{R}\;:\;t I\geq\rho\}=-\log\lambda_{\max}(\rho)
$$
where $\lambda_{\max}(\rho)$ is the maximum eigenvalue of $\rho$. Like the R\'enyi entropies, the min-entropy has many interesting properties including additivity under tensor products and monotonicity under unital CPTP maps.

The conditional min-entropy of a density matrix $\rho^{A}\in\mB(\mH^{A_0A_1})$ is defined by~\cite{Renner-2005a}:
\ba
H_{\min}(A_1|A_0)_\rho&\equiv  -\log\min\left\{\tr[\sigma^{A_0}]\;:\; \sigma^{A_0}\otimes I^{A_1}\geq\rho^{A}\right\}
\ea
Note that the condition $\sigma^{A_0}\otimes I^{A_1}\geq\rho^{A}$ implies that $\sigma^{A_0}\geq 0$. The conditional min-entropy can be calculated using semi-definite programming (SDP). Any SDP optimization problem has a dual. For the conditional min-entropy, the dual is given by~\cite{Konig2009}:
\be\label{chminmax}
2^{-H_{\min}(A_1|A_0)_\rho}=\max_{\Lambda\in\mC^{A}}\tr\left[\phi_{+}^{\tA_1 A_1}\Big(\Lambda\otimes\id^{A_1}\left(\rho^{A_0A_1 }\right)\Big)\right]
\ee
where the maximum is over all CPTP maps $\Lambda:\mB(\mH^{A_0})\to\mB(\mH^{\tA_1})$. 
We use the notation $\phi_{+}^{\tA_1 A_1}\equiv|\phi_{+}^{\tA_1 A_1}\lr\phi_{+}^{\tA_1 A_1}|$, where $|\phi_{+}^{\tA_1 A_1}\ra=\sum_{x=1}^{d_{A_1}}|x\ra^{\tA_1}|x\ra^{A_1}$ is an unnormalized maximally entangled state.
If system $A_1$ is classical then the above expression reduces to the optimal
guessing  probability (that is, the optimal probability to guess the classical value of $A_1$ after measuring the quantum system $A_0$).

The conditional min-entropy has many properties reminiscent of the conditional von-Neumann entropy. First, it is indeed a generalization of the min-entropy. Particularly, if $\dim(\mH^{A_0})=1$ we get that $H_{\min}(A_1|A_0)=H_{\min}(A_1)$. Second, if $\rho^{A}=\rho^{A_0}\otimes\rho^{A_1}$ then $H_{\min}(A_1|A_0)=H_{\min}(A_1)$. In addition, conditioning can only reduce the min-entropy; i.e. for any density matrix $\rho^{AB}\in\mB(\mH^{AB})$ we have
$H_{\min}(A_1|A_0B)\leq H_{\min}(A_1|A_0)$. Finally, its smoothed version satisfies the fully quantum asymptotic equipartition property, which states that in the limit of many copies of a state, the smoothed conditional min-entropy approaches the conditional von-Neuman entropy~\cite{Tom2009}.

\subsubsection{Entropies and support functions of convex sets}

The support function of a convex set is one of the most central and basic concepts in convex geometry. The support function $f_A:\mbb{R}^n\to\mbb{R}$ of a non-empty closed convex set $\mC$ in $\mbb{R}^n$ is defined by:
$$
f_{\mC}(\v)=\sup\left\{\u\cdot\v\;:\;\u\in\mC\right\}\quad\forall\;\v\in\mbb{R}^n\;.
$$
Consider the real vector space, $\mB_h(\mH)$, of all Hermitian matrices in $\mB(\mH)$. Note that the set of all density matrices in this space, $\mD(\mH)$, is closed and convex. The support function of $\mD(\mH)$ is given by:
$$
f_{\mD(\mH)}(\rho)=\sup\left\{\tr[\rho\sigma]\;:\;\sigma\in\mD(\mH)\right\}\quad\forall\;\rho\in\mB_h(\mH)\;.
$$
This support function can be expressed as:
\be
f_{\mD(\mH)}(\rho)=\lambda_{\max}(\rho)= 2^{-H_{\min}(\rho)}\;.
\ee
That is, the min-entropy is simply $-\log_2$ of the support function of the set of density matrices $\mD(\mH)$ in $\mB_h(\mH)$.

Similarly, consider the space of all linear maps from $\mB(\mH^{A_0})$ to $\mB(\mH^{A_1})$, which we denoted by $\mL^A$. Since this is a complex vector space, consider its subspace $\mL_{\text{HP}}^{A}$ of all Hermitian preserving linear maps in $\mL^A$. Clearly, this is a real vector space equipped with the inner product~\eqref{basic0}.
Recall our notation $\mC^{A}\subset\mL_{\text{HP}}^{A}$ for the set of all CPTP maps in the space $\mL_{\text{HP}}^{A}$.
Since $\mC^{A}$ is a closed convex subset of $\mL_{\text{HP}}^{A}$, its support function is well defined: for all $\Psi^{A_0\to A_1}\in\mL_{\text{HP}}^{A}$
$$
f_{\mC^A}(\Psi^{A_0\to A_1})=\sup\left\{\la \Lambda,\Psi\ra\;:\;\Lambda\in\mC^{A}\right\}.\nonumber
$$

As we show now, this expression is closely related to the conditional min-entropy. Using the relation~\eqref{equiva} we get that
\ba
f_{\mC^{A}}(\Psi^{{A_0}\to {A_1}})&=\sup_{\Lambda\in\mC^{A}}\tr\left[J_{\Lambda}^{A }J_{\Psi}^{A }\right]
\ea
Recall that 
$$
J_{\Lambda}^{A }=\id^{A_0}\otimes\Lambda\left(\phi_{+}^{A_0\tA_0}\right)=\Lambda^T\otimes\id^{A_1}\left(\phi_{+}^{\tA_1 A_1}\right)
$$
where $\Lambda^T$ is a CP unital map obtained from $\Lambda^{A_0\to A_1}$ by taking the transpose on the Kraus operators in an operator sum representation of $\Lambda^{A_0\to A_1}$. Therefore, its dual $\left(\Lambda^T\right)^{*}\equiv\bar{\Lambda}$ is a CPTP map. With these notations we get
\ba
f_{\mC^{A}}(\Psi^{A_0\to {A_1}})&=\sup_{\Lambda\in\mC^{A}}\tr\left[\Lambda^T\otimes\id^{{A_1}}\left(\phi_{+}^{\tA_1 {A_1}}\right)J_{\Psi}^{A }\right]\\
&=\sup_{\Lambda\in\mC^{A}}\tr\left[\phi_{+}^{\tA_1 A_1}\Big(\bar{\Lambda}\otimes\id^{A_1}\left(J_{\Psi}^{A }\right)\Big)\right]\\
&=\sup_{\Lambda\in\mC^{A}}\tr\left[\phi_{+}^{\tA_1 A_1}\Big(\Lambda\otimes\id^{A_1}\left(J_{\Psi}^{A }\right)\Big)\right]\\
&= 2^{-H_{\min}(A_1|A_0)_{J_{\Psi}}}\;,
\ea
where the third equality follows from the fact that optimization over all $\bar{\Lambda}^{A_0\to\tA_1}$ is equivalent to optimization over all $\Lambda^{A_0\to\tA_1}$, and the last equality follows from~\eqref{chminmax}. We therefore conclude that the conditional min-entropy can be viewed as the support function of the set of quantum channels, while the min-entropy can be viewed as the support function of the set of quantum states. Note that $J^{A}_{\Psi}$ in the above equation is not normalized since its marginal $J^{A_0}_{\Psi}=I^{A_0}$.   

\section{Noisy Superchannels}\label{noisy}

We study here different types of superchannels that correspond to noisy processes. We will use these models in the next section for the definition of the entropy of a quantum channel, and particularly for the definition of the extended conditional min-entropy of a bipartite channel.
Similar to noisy channels, for which we defined in the previous section both random unitaries and doubly stochastic (unital) channels, we define here different types of noisy superchannels, including random-unitary superchannels, completely uniformity preserving superchannels, completely unital-channel preserving, and doubly stochastic superchannels. 


\subsection{Random Unitary Superchannels}

If $\Theta\in\mbb{L}^{AB}$ is a reversible map, with $A\cong B$, i.e. $d_{A_0}=d_{B_0}$ and $d_{A_1}=d_{B_1}$, then the entropy of a quantum channel must be defined in such a way that any channel $\Phi^{A}$ has the same entropy as $\Theta[\Phi^A]$. Such a reversible transformation $\Theta$
has the form 
\be\label{reversible}
\Theta[\Phi^A]=\mU^{A_1\to B_1}_{\post}\circ\Phi^{A_0\to A_1}\circ\mU^{B_0\to A_0}_{\pree}\;,
\ee 
where $\mU^{B_0\to A_0}_{\pree}$ and $\mU^{A_1\to B_1}_{\post}$ are unitary CPTP maps (i.e. acting by a conjugation with a unitary matrix).
If the dimensions of systems $A$ and $B$ are not the same, then one can replace $\mU^{B_0\to A_0}_{\pree}$ and $\mU^{A_1\to B_1}_{\post}$ above with isometric channels (i.e. channels acting by a conjugation with an isometry). Therefore, similar to the arguments given in the previous section, 
a convex combination of reversible superchannels is an entropy non-decreasing map since it corresponds to implementing a reversible transformation and then ``forgetting" the information about which reversible transformation has been applied. 
We call such a convex combination of reversible superchannels, \emph{a random unitary superchannel}. It can be expressed as a linear map $\Theta\in\mbb{L}^{AB}$ (with $d_{A_0}=d_{B_0}$ and $d_{A_1}=d_{B_1}$)
given by
\be\label{ru}
\Theta[\Phi^A]=\sum_{x=1}^{m}p_x\;\mU^{A_1\to B_1}_{\post,x}\circ\Phi^{A_0\to A_1}\circ\mU^{B_0\to A_0}_{\pree,x}\;,
\ee 
where $\{p_x\}_{x=1}^{m}$ is a probability distribution. 

From its definition in~\eqref{defichoi}, the Choi matrix of the above map is given by
$$
\J_{\Theta}^{AB}=\sum_{x=1}^{m}p_x \sum_{a_0,a_1}J^{A}_{\mE_{a_0a_1}}\otimes J^{B}_{\mU_{\post,x}\circ\mE_{a_0a_1}\circ\mU_{\pree,x}}
$$
where $a=(i,j)$, $b=(k,\ell)$, $J^{A}_{\mE_{a_0a_1}}=|i\lr j|^{A_0}\otimes|k\lr\ell|^{A_1}$ and
\begin{align}
&J^{B}_{\mU_{\post,x}\circ\mE_{a_0a_1}\circ\mU_{\pree,x}}=\nonumber\\
& \sum_{i',j'}\left\la i\left |\mU_{\pree,x}\left(|i'\lr j'|^{B_0}\right)\right|j\right\ra
 |i'\lr j'|^{B_0}\otimes\mU_{\post,x}\left(|k\lr\ell|^{A_1}\right)\nonumber\\
 &=\mU_{\pree,x}^t\left(|i\lr j|^{B_0}\right)\otimes\mU_{\post,x}\left(|k\lr\ell|^{A_1}\right)\;,
\end{align}
where $\mU_{\pree,x}^t(\rho)\equiv \left(\mU_{\pree,x}^*(\rho^T)\right)^T=U_{\pree,x}^T\rho (U_{\pree,x}^T)^*$.
Combining all this, we conclude that
\be\label{ruchoi}
\J_{\Theta}^{AB}=\sum_{x=1}^{m}p_x |\alpha_x\lr\alpha_x|^{A_0B_0}\otimes|\beta_x\lr\beta_x|^{A_1B_1}
\ee
where 
\begin{align}
& |\alpha_x\ra^{A_0B_0}\equiv I^{A_0}\otimes U_{\pree,x}^T \left|\phi_{+}\right\ra^{A_0B_0}\nonumber\\
& |\beta_x\ra^{A_1B_1}\equiv I^{A_1}\otimes U_{\post,x} |\phi_{+}\ra^{A_1B_1}
\end{align}
are all (unnormalized) maximally entangled states. In particular,  the marginals of the above Choi matrix 
satisfy
\begin{align}\label{4marginals}
&\J^{A_1B}_{\Theta}=u^{B_0}\otimes \J^{A_1B_1}_{\Theta}\\
&\J^{AB_1}_{\Theta}=u^{A_0}\otimes \J^{A_1B_1}_{\Theta}\label{s2}\\
&\J^{A_0B}_{\Theta}=\J^{A_0B_0}_{\Theta}\otimes u^{B_1}\label{t3}\\
&\J^{AB_0}_{\Theta}=\J^{A_0B_0}_{\Theta}\otimes u^{A_1}\label{f4}
\end{align}
along with
$$
\J^{A_1B_0}_{\Theta}=I^{A_1B_0}\quad\text{and}\quad\J^{A_0B_1}_{\Theta}=I^{A_0B_1}\;.
$$
Note that the last condition in~\eqref{4marginals} is satisfied for \emph{any} superchannel. Moreover, even if a Choi matrix satisfies the 4 conditions above, it is still not sufficient to guarantee the random unitary form of~\eqref{ruchoi}. For example, if we replace the unitary maps $\mU_{\pree,x}^{B_0\to A_0}$ and $\mU_{\post,x}^{A_1\to B_1}$ with unital CPTP maps, the conditions in~(\ref{4marginals}-\ref{f4})
 will still hold! In the following subsections we give physical meanings to the different conditions above. In particular, we will see that~\eqref{t3} ensures that the dual supermap $\Theta^*$ is also a superchannel, while the condition~\eqref{s2} is satisfied by superchannels with the property that they take unital channels to unital channels.

\subsection{Doubly Stochastic Superchannels}\label{uniformitypreserving}

A CPTP map $\Phi:\mB(\mH)\to\mB(\mH')$ is doubly stochastic if both $\Phi$ and $\Phi^*$ are CPTP maps. This is possible only if $\dim(\mH)=\dim(\mH')$. For superchannels, we introduce a similar definition.
\begin{definition}
A linear map $\Theta\in\mbb{L}^{AB}$ is said to be doubly stochastic if both $\Theta$ and its dual $\Theta^{*}\in\mbb{L}^{BA}$ are superchannels.
\end{definition}
\begin{remark}
The input and output dimensions of a doubly stochastic superchannel must satisfy
\be\label{dim}
\frac{d_{A_0}}{d_{A_1}}=\frac{d_{B_0}}{d_{B_1}}\;.
\ee
This follows from the fact that the traces of the Choi matrices of both $\Theta$ and $\Theta^*$ are the same (see the theorem below).
\end{remark}

Random unitary superchannels are doubly stochastic and in particular satisfy  trivially the above condition with $d_{A_0}=d_{B_0}$ and $d_{A_1}=d_{B_1}$. However, doubly stochastic superchannels form a much larger set of operations as can be seen in the following theorem.

\begin{theorem}\label{dsthm}
Let $\Theta$ be a superchannel in $\mbb{L}^{AB}$ with dimensions as in~\eqref{dim}.
The following are equivalent.
\begin{enumerate}
\item The dual map $\Theta^{*}\in\mbb{L}^{BA}$ is also a superchannel (i.e. $\Theta$ is doubly stochastic).
\item In addition to the conditions given in~\eqref{marginals}, the Choi matrix $\J^{AB}_{\Theta}\geq 0$ satisfies
\begin{align}
& \J_{\Theta}^{A_0B_1}=I^{A_0B_1}\quad;\quad\J^{A_0B}_{\Theta}=\J^{A_0B_0}_{\Theta}\otimes u^{B_1}\;.\label{dss}
\end{align}
\item The superchannel $\Theta$ can be realized as in~\eqref{realization} with the the quantum channels $\Gamma^{B_0\to A_0E}_{\pree}$ and $\Gamma_{\post}^{A_1E\to B_1}$ satisfying the following property. For any matrix $\rho^{E}\in\mB(\mH^E)$
\begin{align}
& \Gamma_{\post}^{A_1E\to B_1}\left(u^{A_1}\otimes\rho^{E}\right)=\tr\left[\rho^E\right]u^{B_1}\label{gpost0}\\
& \Gamma^{B_0\to A_0}_{\pree}(u^{B_0})=u^{A_0}
\end{align}
where $\Gamma^{B_0\to A_0}_{\pree}$ is the map $\tr_E\circ\Gamma^{B_0\to A_0E}_{\pree}$.
\end{enumerate}
\end{theorem}
\begin{remark}
The conditions in~\eqref{marginals}  ensure that $\Theta$ is a superchannel. The additional conditions are those in~\eqref{dss}. Moreover, from the first condition in~\eqref{marginals} we have $\tr[\J^{AB}_{\Theta}]=d_{A_1}d_{B_0}$ while from the first condition in~\eqref{dss} we have  $\tr[\J^{AB}_{\Theta}]=d_{A_0}d_{B_1}$. Hence, the dimensions of doubly stochastic superchannels must satisfy~\eqref{dim}.
\end{remark}
{\it{Proof.}}
We first prove the equivalence of 1 and 2. From Theorem~\ref{premain} the map $\Theta^*$ is a superchannel if and only if $\J^{BA}_{\Theta^*}\geq 0$ and it has marginals
\be\label{x}
\J^{B_1 A_0}_{\Theta^*}=I^{B_1 A_0}\quad;\quad \J_{\Theta^*}^{BA_0}=\J^{B_0 A_0}_{\Theta^*}\otimes u^{B_1}
\ee
Now, since $\Delta_{\Theta^*}^{\tB\to A}=\Delta_{\Theta}^{*\tB\to A}$ (see the discussion in the preliminary section) we conclude
that the Choi matrix of $\Theta^{*}\in\mbb{L}^{BA}$ is given by
\begin{align}\label{z2}
\J^{BA}_{\Theta^*}&=\id^{B}\otimes \Delta_{\Theta^*}^{\tB\to A}\left(\phi_{+}^{B_0\tB_0}\otimes\phi_{+}^{B_1\tB_1}\right)\nonumber\\
&=\id^{B}\otimes \Delta_{\Theta}^{*\tB\to A}\left(\phi_{+}^{B_0\tB_0}\otimes\phi_{+}^{B_1\tB_1}\right)\nonumber\\
&=\overline{\Delta_{\Theta}^{\tA\to B}\otimes \id^{A} \left(\phi_{+}^{\tA_0A_0}\otimes\phi_{+}^{\tA_1A_1}\right)}=\overline{\J^{BA}_{\Theta}}
\end{align}
where the bar above matrices indicates complex conjugation on the components of the matrix (without performing the transpose). 
Here $\J^{BA}_{\Theta}=\text{swap}(\J^{AB}_{\Theta})$ is obtained from $\J^{AB}_{\Theta}$ by swapping between systems $A$ and $B$.
Combining this with~\eqref{x} we get the marginals in~\eqref{dss}. This completes the proof of the equivalence between 1 and 2.

To prove part~3, note that if $\Gamma_{\post}^{A_1E\to B_1}$ has the form~\eqref{gpost0} then from~\eqref{choiform9} it follows that
$$
\J_{\Theta}^{A_0B}=\Gamma^{\tA_1E\to B_1}_{\post}\left(\psi^{A_0B_0E}\otimes I^{\tA_1}\right)=\J_{\Theta}^{A_0B_0}\otimes u^{B_1}
$$
since $\psi^{A_0B_0E}$ is a purification of $\frac{1}{d_{A_1}}\J_{\Theta}^{A_0B_0}$. Conversely, suppose $\J_{\Theta}^{A_0B}=\J_{\Theta}^{A_0B_0}\otimes u^{B_1}$, and let $d_{E}=\text{Rank}(\J_{\Theta}^{A_0B_0})$. Then, we can write $|\psi\ra^{A_0B_0E}=\sum_{j=1}^{d_E}r_j|u_j\ra^{A_0B_0}|j\ra^E$ with $r_j>0$ and $\left\{|u_j\ra^{A_0B_0}\right\}$ being an orthonormal set of vectors in $\mH^{A_0B_0}$. From its definition, the marginal state 
$\sum_{j=1}^{d_E}r_{j}^2|u_j\lr u_j|^{A_0B_0}=\frac{1}{d_{A_1}}\J_{\Theta}^{A_0B_0}$. Therefore, from our assumption on the form of $\J_{\Theta}^{A_0B}$,
it follows that 
\begin{align*}
&\sum_{j=1}^{d_E}r_{j}^2|u_j\lr u_j|^{A_0B_0}\otimes u^{B_1}\\
&=\sum_{j,k=1}^{d_E}r_{j}r_k|u_j\lr u_k|^{A_0B_0}\otimes\Gamma^{\tA_1E\to B_1}_{\post}\left(|j\lr k|^{E}\otimes u^{\tA_1}\right)\;. 
\end{align*}
Finally, from the independence of $\left\{|u_j\lr u_k|^{A_0B_0}\right\}$ it follows that
$$
\Gamma^{\tA_1E\to B_1}_{\post}\left(|j\lr k|^{E}\otimes u^{\tA_1}\right)=\delta_{jk}u^{B_{1}}\;.
$$
This proves the equivalence of $\J^{A_0B}_{\Theta}=\J^{A_0B_0}_{\Theta}\otimes u^{B_1}$ with the condition that $\Gamma_{\post}^{A_1E\to B_1}\left(u^{A_1}\otimes\rho^{E}\right)=\tr\left[\rho^E\right]u^{B_1}$ for all density matrices $\rho^E$. To complete the proof of the equivalence between 2 and 3,  note that the condition 
$\J^{B_1 A_0}_{\Theta^*}=I^{B_1 A_0}$ is equivalent to $\Gamma_{\Theta}^{A_1B_0\to A_0B_1}$ being a unital channel. Hence, (recall, $d_{A_0}d_{B_1}=d_{A_1}d_{B_0}$)
\ba
u^{A_0B_1}&=\Gamma_{\Theta}^{A_1B_0\to A_0B_1}(u^{A_1B_0})\\
&=\left(\id^{A_0}\otimes\Gamma^{A_1E\to B_1}_{\post}\right)\Big(u^{A_1}\otimes\Gamma^{B_0\to A_0E}_{\pree}(u^{B_0})\Big)\\
&=u^{B_1}\otimes \Gamma^{B_0\to A_0}_{\pree}(u^{B_0})\;,
\ea
where in the last equality we used the property that $\Gamma_{\post}^{A_1E\to B_1}\left(u^{A_1}\otimes\rho^{E}\right)=\tr\left[\rho^E\right]u^{B_1}$. That is,
$\Gamma^{B_0\to A_0}_{\pree}(u^{B_0})=u^{A_0}$. This completes the proof.
\QED

Note that a doubly stochastic superchannel satisfies the last two conditions in~\eqref{4marginals} but not necessarily the first two conditions nor does it necessarily have the form~\eqref{ruchoi}.
We now discuss how the condition~\eqref{dss} is related to the fact that doubly stochastic superchannels do not decrease noise.

\subsection{Completely Uniformity Preserving Superchannel}\label{unipre}

The noisiest quantum channel in $\mL^A$, which we denote by $\mN^A$ and call the uniform channel, is given by
\be\label{mn}
\mN^A(X)\equiv\tr[X]u^{A_1}\quad\forall\;X\in\mB(\mH^{A_0})\;.
\ee
This channel is also known in the literature as the completely depolarizing channel or the replacer channel.
That is, irrespective of the input state, the output state of a uniform channel is always maximally mixed (i.e. uniform).

With this in mind, a superchannel $\Theta\in\mbb{L}^{AB}$ with dimensions $d_{A_0}=d_{B_0}$ and $d_{A_1}=d_{B_1}$ (i.e. $A\cong B$), that does not decrease noise must satisfy
$$
\Theta\left[\mN^A\right]=\mN^B\;.
$$
That is, $\Theta$ is a uniformity preserving superchannel. The above condition is analogous to the condition that noisy channel $\mE\in\mC^A$ must preserve the uniform (maximally mixed) state $u^A\equiv\frac{1}{d_A}I^A$, since otherwise, the output state $\mE(u^A)$ will be less noisy than the input state (the maximally mixed state).
 
The Choi matrix of $\mN^A$ and $\mN^B$ are given by $I^{A_0}\otimes u^{A_1}$ and $I^{B_0}\otimes u^{B_1}$, respectively. Hence, from~\eqref{trans} the equation above becomes equivalent to
$$
\J^{B}_{\Theta}=I^B\;.
$$
Unlike the parallel discussion on noisy channels, for noisy superchannels, there is a stronger condition than the one above, that one can expect from any superchannel that does not decrease noise. 

\begin{figure}[t]
    \includegraphics[width=0.4\textwidth]{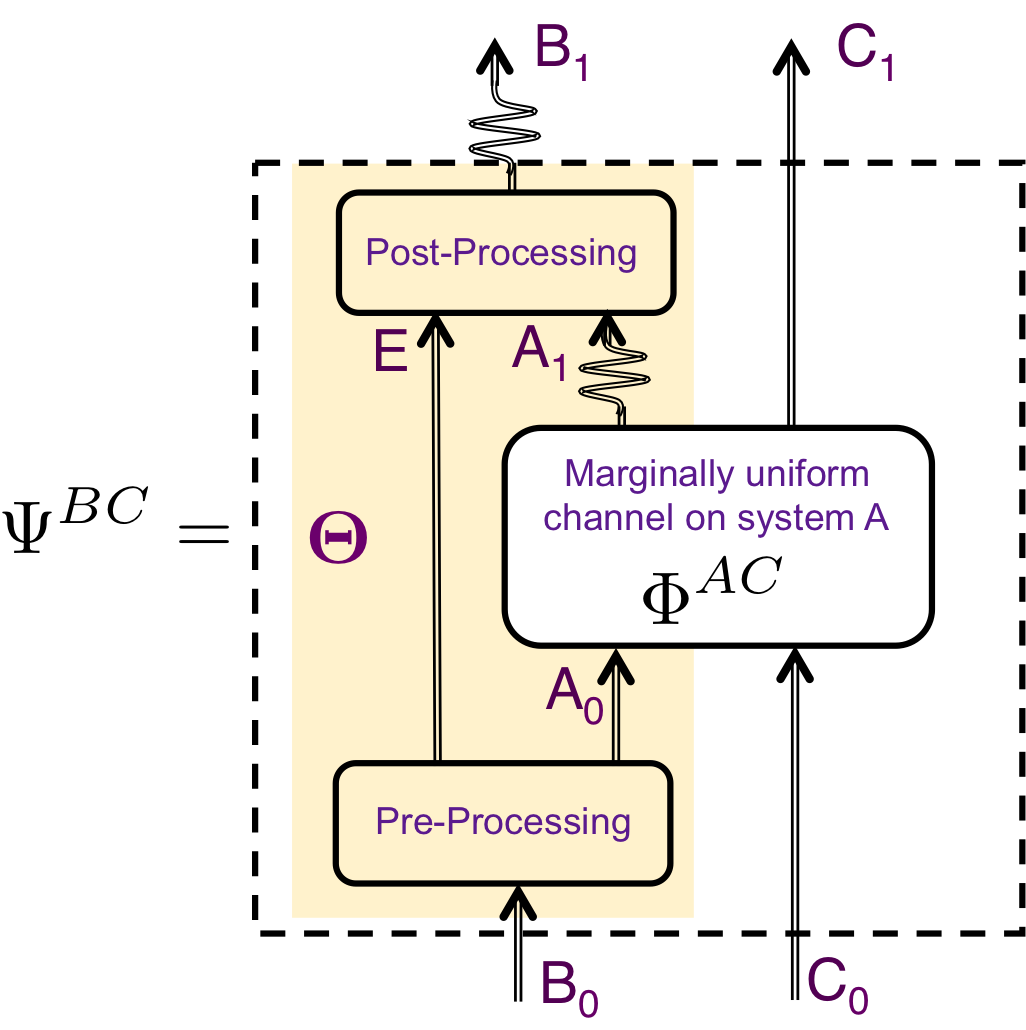}
  \caption{\linespread{1}\selectfont{\small The action of a superchannel $\Theta\in\mbb{L}^{AB}$ on a marginally uniform channel $\Phi^{AC}$. If $\Theta$ is completely uniformity preserving then $\Psi^{BC}$ is marginally uniform on $B$. The curly arrows indicate maximally mixed states.}}
  \label{uniformity}
\end{figure}

Consider a bipartite channel $\Phi^{AC}$ shared between two parties, and suppose that it has the property 
that the output state on subsystem $A_1$ is always the maximally mixed state. That is, $\Phi^{AB}$ satisfies for all $\rho^{A_0C_0}\in\mB(\mH^{A_0C_0})$
$$
\Phi^{AC}(\rho^{A_0C_0})= u^{A_1}\otimes\tr_{A_1}\left[\Phi^{AC}(\rho^{A_0C_0})\right]\;,
$$
where $\tr_{A_1}\left[\Phi^{AC}(\rho^{A_0C_0})\right]$ is a matrix in $\mB(\mH^{C_1})$. We say that such a bipartite channel is \emph{marginally uniform} on $A$. Therefore, if $\Theta\in\mbb{L}^{AB}$ (with dimensions $d_{A_0}=d_{B_0}$ and $d_{A_1}=d_{B_1}$) does not decrease noise,
and $\Phi^{AC}$ is marginally uniform on $A$, then  $\Psi^{BC}\equiv\Theta\otimes\1^C\left[\Phi^{AC}\right]$ should also be marginally uniform (on $B$); see Fig.~\eqref{uniformity}. We call such a superchannel a  
\emph{completely uniformity preserving} superchannel. Note that this condition is somewhat similar to the condition that physical operations are not only positive but  completely positive.

\begin{theorem}\label{cup}
Let $\Theta$ be a superchannel in $\mbb{L}^{AB}$ with dimensions $d_{A_0}=d_{B_0}$ and $d_{A_1}=d_{B_1}$. The following are equivalent.
\begin{enumerate}
\item $\Theta$ is a completely uniformity preserving superchannel.
\item In addition to~\eqref{marginals}, the Choi matrix of $\Theta$ satisfies
\be
\J^{A_0B}_{\Theta}=\J^{A_0B_0}_{\Theta}\otimes u^{B_1}\;.
\ee
\item The map $\Theta$ can be realized as in~\eqref{realization} with the additional condition that the quantum channel $\Gamma_{\post}^{A_1E\to B_1}$ satisfies for all $\rho^{E}\in\mB(\mH^E)$
\be\label{gpost}
\Gamma_{\post}^{A_1E\to B_1}\left(u^{A_1}\otimes\rho^{E}\right)=\tr[\rho^E]u^{B_1}\;.
\ee
\end{enumerate}
\end{theorem}
{\it{Proof.}}  
Suppose $\Theta$ is a completely uniformity preserving superchannel. Let $\Phi^{AC}$ be a marginally uniform channel on $A$ and define 
$$
\Psi^{BC}\equiv\Theta\otimes\1^C\left[\Phi^{AC}\right]\;.
$$
From our assumptions both $\Phi^{AC}$ and $\Psi^{BC}$ are marginally uniform, so that their Choi matrices are given by
$$
J^{AC}_{\Phi}=J^{A_0C}_{\Phi}\otimes u^{A_1}\quad\text{and}\quad J^{BC}_{\Psi}=J^{B_0C}_{\Psi}\otimes u^{B_1}\;,
$$
respectively. Taking $\Theta$ in~\eqref{trans} to be $\Theta\otimes\1^C$ we have
$$
J^{B_0C}_{\Psi}\otimes u^{B_1}=\tr_{A\tC}\left[\J^{ABC\tC}_{\Theta\otimes\1^C}\left(\left(J^{A_0\tC}_{\Phi}\right)^T\otimes u^{A_1}\otimes I^{B}\right)\right]
$$
From its definition,
$$
\J^{ABC\tC}_{\Theta\otimes\1^C}=\J^{AB}_{\Theta}\otimes\phi^{C\tC}_{+}\;.
$$
Moreover, note that
$$
\tr_{\tC}\left[\left(I^{A_0}\otimes\phi^{C\tC}_{+}\right)\left(\left(J^{A_0\tC}_{\Phi}\right)^T\otimes I^C\right)\right]=\left(J^{A_0C}_{\Phi}\right)^{T_{A_0}}
$$
where $T_{A_0}$ is the partial transpose on system $A_0$. We therefore get that
\ba
&J^{B_0C}_{\Psi}\otimes u^{B_1}\\
&=\tr_{A}\left[\left(\J^{AB}_{\Theta}\otimes I^C\right)\left(\left(J^{A_0C}_{\Phi}\right)^{T_{A_0}}\otimes u^{A_1}\otimes I^{B}\right)\right]\\
&=\frac{1}{d_{A_1}}\tr_{A_0}\left[\left(\J^{A_0B}_{\Theta}\otimes I^C\right)\left(\left(J^{A_0C}_{\Phi}\right)^{T_{A_0}}\otimes I^{B}\right)\right]\;.\nonumber
\ea
By multiplying both sides of the equation above by a traceless matrix $Z^{B_1}$ and taking the trace over $B_1$ we get that
\be\label{z}
\tr_{A_0B_1}\left[\left(\J^{A_0B}_{\Theta}\otimes I^C\right)\left(\left(J^{A_0C}_{\Phi}\right)^{T_{A_0}}\otimes I^{B_0}\otimes Z^{B_1}\right)\right]=0\;.
\ee
We show now that if the condition above holds for all traceless matrices $Z^{B_1}$, all systems $C$, and all positive semidefinite matrices $M^{A_0C}\equiv J^{A_0C}_{\Phi}$ with the property that $M^{A_0C_0}=I^{A_0C_0}$, then $\J^{A_0B}_{\Theta}=\J^{A_0B_0}_{\Theta}\otimes u^{B_1}$. To do that, let $\{X_j^{B_1}\}_{j=0}^{d_{B_1}^2-1}$ be an orthogonal basis of $\mB(\mH^{B_1})$ such that $X_0^{B_1}=u^{B_1}$ and $\tr[X_j^{B_1}]=0$ for $j>0$. With this basis we can express $\J^{A_0B}_{\Theta}$ as
$$
\J^{A_0B}_{\Theta}=\J^{A_0B_0}_{\Theta}\otimes u^{B_1}+\sum_{j=1}^{d_{B_1}^2-1}N_{j}^{A_0B_0}\otimes X_j^{B_1}\;,
$$
where $N_{j}^{A_0B_0}$ are some matrices in $\mB(\mH^{A_0B_0})$. Substituting this expression into~\eqref{z} and taking $Z^{B_1}=X_j^{B_1}$ for some $j>0$ we conclude that
$$
\tr_{A_0}\left[\left(N_{j}^{A_0B_0}\otimes I^C\right)\left(\left(M^{A_0C}\right)^{T_{A_0}}\otimes I^{B_0}\right)\right]=0
$$
for all positive semidefinite matrices with $M^{A_0C}$ with marginal $M^{A_0C_0}=I^{A_0C_0}$. Note also that the above equation has to hold for all dimensions of system $C$. If we take the dimensions of system $C$ to be $d_{C_0}=1$ and $d_{C}=d_{C_1}=d_{A_0}$, then we can choose $M^{A_0C}=\phi_{+}^{A_0C}$. With this choice the equation above become
$$
0=\tr_{A_0}\left[\left(N_{j}^{A_0B_0}\otimes I^C\right)\left(\left(\phi_{+}^{A_0C}\right)^{T_{A_0}}\otimes I^{B_0}\right)\right]=N_{j}^{CB_0}
$$
where $N_{j}^{CB_0}$ is a copy of $N_{j}^{A_0B_0}$ in $\mB(\mH^{CB_0})$ (recall that $d_C=d_{A_0}$). We therefore conclude that $N_{j}^{A_0B_0}=0$ for all $j>0$ so that $\J^{A_0B}_{\Theta}=\J^{A_0B_0}_{\Theta}\otimes u^{B_1}$. Note that the converse of this argument also holds. That is, following the above lines backwards we conclude that if $\J^{A_0B}_{\Theta}=\J^{A_0B_0}_{\Theta}\otimes u^{B_1}$ then $\Theta$ is a completely uniformity preserving superchannel.

In Theorem~\ref{dsthm} we proved the equivalence of $\J^{A_0B}_{\Theta}=\J^{A_0B_0}_{\Theta}\otimes u^{B_1}$ with the condition that $\Gamma_{\post}^{A_1E\to B_1}\left(u^{A_1}\otimes\rho^{E}\right)=u^{B_1}$ for all density matrices $\rho^E$.
This also provides the proof of Part 3 here.
\QED

A comparison between Theorem~\ref{dsthm} and Theorem~\ref{cup} demonstrates that doubly stochastic superchannels are completely uniformity preserving. However, the converse is not true in general since  completely uniformity preserving superchannels do not require that $\J^{A_0B_1}_{\Theta}=I^{A_0B_1}$. The latter is equivalent to the condition that $\Gamma_{\Theta}^{A_1B_0\to A_0B_1}$ is unital, and is related to another physical condition that we discuss in the following subsection.
 
\subsection{Completely Unital-Channel Preserving Superchannels}

In the preliminary section we discussed that random unitary channels can be viewed as noisy channels; i.e. channels that always increase noise no matter what the input state is. On the other hand, in any noise model, a superchannel that increases noise should not generate a non-noisy channel from a noisy one.  Therefore, in addition to being completely uniformity preserving, noisy superchannels should at least not convert random unitary channels to non-unital channels. Here we study superchannels with this property, and particularly those that are completely unital-channel preserving. Since we will consider unital channels we will assume here that $d_{A_0}=d_{A_1}$ and $d_{B_0}=d_{B_1}$.

We first show that if a superchannel takes random unitary channels to unital channels then it also takes unital channels to unital channels. To see why, note that from~\eqref{trans}, the superchannel $\Theta$ converts random unitary channels to unital channels if and only if
\be\label{up}
I^{B_1}=\tr_{AB_0}\left[\J^{AB}_{\Theta}\left(\left(J^{A}_{\Psi}\right)^T\otimes I^{B}\right)\right]
\ee
for any random unitary channel $\Psi^A$. Now, suppose $\Psi^A$ is a unital channel. In~\cite{Mendl2009} it was shown that it can be expressed as an affine linear combination of unitary channels; i.e. $\Psi^A=\sum_{j}r_j\mU_j^A$, where $r_j\in\mbb{R}$ and $\sum_jr_j=1$. Since for each $\mU_j$ the above equation holds, from its linearity it also holds for $\Psi^A$. We therefore conclude that the above equation holds for any unital channel which implies that $\Theta$ is unital-channel preserving. 

The condition in~\eqref{up} holds in particular for the completely dephasing channel whose Choi matrix is given by  $I^{A_0}\otimes u^{A_1}$. From the linearity, any matrix of the form $X^A\equiv J^{A}_{\Psi}-I^{A_0}\otimes u^{A_1}$ with $\Psi^A$ being unital satisfies
$$
\tr_{A}\left[\J^{AB_1}_{\Theta}\left(\left(X^{A}\right)^T\otimes I^{B_1}\right)\right]=0\;.
$$
Note that $X^A$ is any Hermitian matrix in $\mB(\mH^{A_0A_1})$ with zero marginals $X^{A_0}=0$ and $X^{A_1}=0$. This means that $\J^{AB_1}_{\Theta}$ is orthogonal (in the Hilbert Schmidt inner product) to any Hermitian matrix in $\mB(\mH^{AB_1})$ of the form $T^{A_0}_1\otimes T_{2}^{A_1}\otimes Y^{B_1}$, where
$T_1^{A_0}$ and $T_2^{A_1}$ are arbitrary traceless Hermitian matrices, and $Y^B$ is an arbitrary Hermitian matrix. There are exactly two types of matrices that are orthogonal to the subspace spanned by matrices of the form $T^{A_0}_1\otimes T_{2}^{A_1}\otimes Y^{B_1}$. These are either matrices of the form $I^{A_0}\otimes\alpha^{A_1B_1}$ or of the form
$I^{A_1}\otimes\beta^{A_0B_1}$, where $\alpha^{A_1B_1}$ and $\beta^{A_0B_1}$ are Hermitian matrices.
Therefore, $\J^{AB_1}_{\Theta}$ must be a linear combination of matrices of the form $I^{A_0}\otimes\alpha^{A_1B_1}$ and $I^{A_1}\otimes\beta^{A_0B_1}$, so that the superchannel $\Theta$ is a unital-channel preserving if and only if it's Choi matrix $\J^{AB}_{\Theta}$  satisfies the superchannel condition~\eqref{marginals} and in addition
$$
\J^{AB_1}_{\Theta}=I^{A_0}\otimes\alpha^{A_1B_1}+I^{A_1}\otimes\beta^{A_0B_1}\;.
$$
This last condition is somewhat cumbersome, but in the following theorem we show that $\beta^{A_0B_1}$ must be zero if $\Theta$ is a completely unital-channel preserving superchannel. That is, $\Theta$ is a superchannel such that for any system $C$, the superchannel $\Theta\otimes \1^C$ is unital-channel preserving.

\begin{theorem}\label{thmucp}
Let $\Theta\in\mbb{L}^{AB}$ be a superchannel with $d_{A_0}=d_{A_1}$ and $d_{B_0}=d_{B_1}$. Then, the following are equivalent.
\begin{enumerate}
\item $\Theta$ is completely unital-channel preserving.
\item The Choi matrix $\J^{AB}_{\Theta}$ has marginals (in addition to those in Eq.~\eqref{marginals})
$$
\J^{AB_1}_{\Theta}=u^{A_0}\otimes\J^{A_1B_1}_{\Theta}\quad\text{and}\quad\J^{A_0B_1}_{\Theta}=I^{A_0B_1}\;.
$$
\item The CPTP map $\Gamma_{\Theta}^{A_1B_0\to A_0B_1}$ is unital, and in addition to~\eqref{int} it satisfies for any density matrix $\rho^{A_1}\in\mD(\mH^{A_1})$
\ba\label{part3}
&\Gamma_{\Theta}^{A_1B_0\to A_0B_1}\left(\rho^{A_1}\otimes u^{B_0}\right)\\
&=u^{A_0}\otimes \Gamma_{\Theta}^{A_1B_0\to B_1}\left(\rho^{A_1}\otimes u^{B_0}\right)\;.
\ea
\end{enumerate}
\end{theorem}
{\it{Proof.}}
We first prove the equivalence of 1 and 2. Suppose $\Theta$ is completely unital-channel preserving. Then, from~\eqref{trans} if follows that the relation $\Phi^{BC}=\Theta\otimes\1^C[\Psi^{AC}]$ can be expressed in the Choi form as
\be\label{gggg}
J_{\Phi}^{BC}=\tr_{A}\left[\left(\J_{\Theta}^{AB}\otimes I^C\right)\left(\left(J^{AC}_{\Psi}\right)^{T_A}\otimes I^B\right)\right]
\ee
Suppose now that $\Psi^{AC}$ is a bipartite unital channel. Then, from our assumption, $\Phi^{BC}$ is unital as well. Hence,
$$
I^{B_1C_1}=\tr_{A}\left[\left(\J_{\Theta}^{AB_1}\otimes I^{C_1}\right)\left(\left(J^{AC_1}_{\Psi}\right)^{T_A}\otimes I^{B_1}\right)\right]
$$
Note that the equation above holds for $J^{AC_1}_{\Psi}=u^{A_0}\otimes I^{A_1C_1}$, which corresponds to the marginal of the Choi matrix of the completely dephasing map (which is unital). Hence, for this choice of $J^{AC_1}_{\Psi}$ we get the condition that $\J^{B_1}_{\Theta}=d_{A_0}I^{B_1}$. 

Next, note that for any unital bipartite channel $\Psi^{AC}$, the matrix $X^{AC_1}= \left(J_\Psi^{AC_1}\right)^{T_A}-u^{A_0}\otimes I^{A_1C_1}$ has the property that $X^{A_1C_1}=0$ and $X^{A_0}=0$. From the linearity of the equation above we conclude that for any such matrix 
$$
\tr_{A}\left[\left(\J_{\Theta}^{AB_1}\otimes I^{C_1}\right)\left(X^{AC_1}\otimes I^{B_1}\right)\right]=0
$$
Let $Z^{C_1}$ be some fixed normalized (in the Hilbert-Schmidt inner product) \emph{traceless} Hermitian matrix in $\mB_h(\mH^{C_1})$, and let $Y^{B_1}$ be arbitrary Hermitian matrix in $\mB_h(\mH^{B_1})$. Then, the equation above implies that
$$
\left\la \J_{\Theta}^{AB_1}\otimes Z^{C_1}\;,\;X^{AC_1}\otimes Y^{B_1}\right\ra=0
$$
for all such $Y^{B_1}$ and all $X^{AC_1}$ with zero marginals as above.
This implies that $\J_{\Theta}^{AB_1}\otimes Z^{C_1}$ is orthogonal to any matrix of the form
$T_{1}^{A_0}\otimes T^{A_1C_1}_{2}\otimes Y^{B_1}$ where $T_1^{A_0}$ and $T_2^{R_1B_1}$ are arbitrary Hermitian traceless matrices. Therefore, there must exist $\alpha^{A_0B_1}$ and $\beta^{A_1B_1C_1}$ such that
$$
\J_{\Theta}^{AB_1}\otimes Z^{C_1}=\alpha^{A_0B_1}\otimes I^{A_1C_1}+I^{A_0}\otimes\beta^{A_1B_1C_1}\;.
$$
Finally, by multiplying both sides of the equation above by $I^{AB_1}\otimes Z^{C_1}$ and taking the partial trace over system $C_1$ we conclude that
$$
\J_{\Theta}^{AB_1}=I^{A_0}\otimes\tr_{C_1}\left[\left(I^{AB_1}\otimes Z^{C_1}\right)\beta^{A_1B_1C_1}\right]\;,
$$
where we used the fact that $Z^C$ is a normalized traceless matrix. The equation above implies that 
$$
\J_{\Theta}^{AB_1}=u^{A_0}\otimes\J_{\Theta}^{A_1B_1}\;.
$$
Conversely, suppose the equation above holds, and in addition $\J_{\Theta}^{B_1}=d_{A_0}I^{B_1}$. Then, for any bipartite unital channel $\Psi^{AC}$ we get
\ba
&\tr_{A}\left[\left(\J_{\Theta}^{AB_1}\otimes I^{C_1}\right)\left(\left(J^{AC_1}_{\Psi}\right)^{T_A}\otimes I^{B_1}\right)\right]\nonumber\\
&=\frac{1}{d_{A_0}}\tr_{A_1}\left[\left(\J_{\Theta}^{A_1B_1}\otimes I^{C_1}\right)\left(\left(J^{A_1C_1}_{\Psi}\right)^{T_A}\otimes I^{B_1}\right)\right]\nonumber\\
&=\frac{1}{d_{A_0}}\J_{\Theta}^{B_1}\otimes I^{C_1}=I^{B_1C_1}\;.\nonumber
\ea
This completes the proof of the equivalence between 1 and 2. We now prove the equivalence between 2 and 3.

From~\eqref{cgam} it follows that
$$
\J^{AB_1}_\Theta=\Gamma^{\tA_1\tB_0\to A_0B_1}_{\Theta}\left(\phi_{+}^{A_1\tA_1}\otimes I^{\tB_0}\right)\;.
$$
In particular, the marginal is given by $$\J^{A_1B_1}_\Theta=\Gamma^{\tA_1\tB_0\to B_1}_{\Theta}\left(\phi_{+}^{A_1\tA_1}\otimes I^{\tB_0}\right)\;.$$
Hence, $\J_{\Theta}^{AB_1}=u^{A_0}\otimes\J_{\Theta}^{A_1B_1}$ if and only if~\eqref{part3} holds for all density matrices $\rho^{A_1}$. The converse follows trivially from the fact that the Choi matrix of $\Theta$ equals the Choi matrix of $\Gamma_{\Theta}^{A_1B_0\to A_0B_1}$. This completes the proof.
\QED

\section{The Entropy of a Quantum Channel}\label{entropy}

We now extend the definition of entropies, and particularly the conditional min-entropy, from  states to  channels. Since the entropy of a quantum channel measures how noisy the channel is, it must behave monotonically under noisy operations. Below we give an axiomatic and minimalistic approach for the definition of entropy. 

We call a function $f:\mL^A_+\to\mbb{R}$ an entropy if it satisfies the following conditions (see the analogous conditions on entropy of states in Sec.~\ref{pre-entropies})
\begin{enumerate}
\item {\it Monotonicity}: For any random unitary superchannel $\Theta:\mL^A\to\mL^A$, 
$$
f(\Theta[\Phi])\geq f(\Phi)\;,
$$
for all channels $\Phi\in\mC^A$.
\item {\it Additivity}: For any two quantum channels $\Phi^A\in\mC^A$ and $\Psi^B\in\mC^B$ we have
$$
f(\Phi^A\otimes\Psi^B)=f(\Phi^A)+f(\Psi^B)\;.
$$
\item  {\it Normalization}: on a uniform channel, $\mN^A$, as in~\eqref{mn}, $f(\mN^A)=\log(d_{A_1})$, and on any replacement map, $\Phi\in\mL^A$, of the form $\Phi(X)=\tr[X]|\psi\lr\psi|$ with $|\psi\ra$ being some fixed pure state, 
$f(\Phi^A)=0$.
\end{enumerate}

The last condition is motivated by the fact that replacement maps can be viewed as quantum states and consequently the entropy of these channels should reduce to the entropy of states. For the monotonicity, we only require monotonicity under random unitary superchannels; however, we expect many entropy functions to be monotonic under a larger set of superchannels such as, for example, doubly stochastic superchannels.
Regarding the additivity requirement, while it is a natural condition (since entropies of states are required to be additive), some natural candidates, as we discuss now, fail to satisfy this requirement.

The von-Neumann entropy of states is defined by $S(\rho)\equiv-\tr\left[\rho\log\rho\right]$ for any density matrix $\rho\in\mB(\mH)$. For quantum channels, one can propose a natural generalization, $\tilde{S}$, given by the minimum entropy output
$$
\tilde{S}[\Phi^A]\equiv\min_{\rho^{A_0}\in\mD(\mH^{A_0})}S\left(\Phi^A\big(\rho^{A_0}\big)\right)
$$
where the minimum is over all input density matrices $\rho^{A_0}$. This candidate for an entropy of channel is monotonic under random unitary superchannels. Indeed, let $\Theta$ be a random unitary superchannel as in~\eqref{ru}. Then,
\ba
&\tilde{S}\left[\Theta[\Phi^A]\right]\\
&=\min_{\rho^{A_0}}S\left(\sum_{x=1}^{m}p_x\;\mU^{A_1\to B_1}_{\post,x}\circ\Phi^{A_0\to A_1}\circ\mU^{B_0\to A_0}_{\pree,x}(\rho^{A_0})\right)\\
&\geq\min_{\rho^{A_0}}\sum_{x=1}^{m}p_xS\left(\mU^{A_1\to B_1}_{\post,x}\circ\Phi^{A_0\to A_1}\circ\mU^{B_0\to A_0}_{\pree,x}(\rho^{A_0})\right)\\
&=\min_{\rho^{A_0}}\sum_{x=1}^{m}p_xS\left(\Phi^{A_0\to A_1}\circ\mU^{B_0\to A_0}_{\pree,x}(\rho^{A_0})\right)\\
&\geq \sum_{x=1}^{m}p_x\min_{\rho^{A_0}}S\left(\Phi^{A_0\to A_1}\circ\mU^{B_0\to A_0}_{\pree,x}(\rho^{A_0})\right)
=\tilde{S}[\Phi^A]\nonumber\;,
\ea
where in the first inequality we used the concavity of the von-Neumann entropy. Since the R\'enyi entropies with parameter $\alpha\in[0,1]$ are also concave, their extension to channels as above will also be monotonic under random unitary superchannels. 

While the minimum entropy output of a quantum channel is monotonic under random unitary superchannels, it fails to satisfy the additivity property~\cite{Hastings2009}, and therefore, according to our definition above it is not an entropy function. One may choose to replace the additivity condition with a weaker one, in which $f$ is only required to be additive under tensor product of two replacement maps (i.e. states). With this modification, the minimum entropy output of a quantum channel is an entropy function.
However, we include the full additivity property in the definition of an entropy function, as the functions that we will consider here will be fully additive.
We now give an example of such an entropy function that is based on the min-entropy.

\begin{definition}
Let $\Phi^A\in\mL^A$ be a quantum channel. The \emph{extended min-entropy} of $\Phi^A$ is the function
$$
H_{\min}^{\rm ext}(A)_{\Phi}\equiv H_{\min}(A_1|A_0)_{J_{\Phi}/d_{A_0}}\;,
$$
where $J_{\Phi}^{A_0A_1}/d_{A_0}$ is the normalized Choi matrix of $\Phi^{A_0\to A_1}$. 
\end{definition}

\begin{remark}
The extended min-entropy was shown in~\cite{DW16} to have an operational interpretation. In
particular, $d_{A_0}H_{\min}^{\rm ext}(A)_{\Phi}$ was shown to be the zero-error classical simulation cost of the quantum channel $\Phi^A$. Furthermore, its smooth version was recently introduced in~\cite{FWTB18}. 
\end{remark}

The \emph{extended min-entropy} is an entropy function that satisfies all the 3 conditions above of monotonicity, additivity, and normalization. The additivity and the normalization follow immediately from the properties of the conditional min-entropy. To show monotonicity, we show now
that the extended min-entropy behaves monotonically not only under random unitary superchannels, 
but also under the much larger set of doubly stochastic superchannels. 

Let $\Theta\in\mbb{L}^{AB}$ be a doubly stochastic superchannel and suppose $d_{A_0}=d_{B_0}$ and $d_{A_1}=d_{B_1}$. Then, for any quantum channel $\Phi^A\in\mL^A$ we have
\ba
2^{-H_{\min}^{\rm ext}(B)_{\Theta[\Phi]}}&=2^{-H_{\min}(B_1|B_0)_{J_{\Theta[\Phi]}/d_{B_0}}}\\
&=\frac{1}{d_{B_0}}\max_{\Psi\in\rm \mC^B}\left\la \Psi^B,\Theta[\Phi^{A}]\right\ra\\
&=\frac{1}{d_{A_0}}\max_{\Psi\in\rm \mC^{B}}\left\la \Theta^*[\Psi^B],\Phi^{A}\right\ra\\
&\leq \frac{1}{d_{A_0}}\max_{\Lambda\in\rm \mC^{A}}\left\la \Lambda^A,\Phi^{A}\right\ra\\
&=2^{-H_{\min}^{\rm ext}(A)_{\Phi}}\;.
\ea
That is, for any doubly stochastic superchannel $\Theta\in\mbb{L}^{AB}$ as above (with $A\cong B$), and any CPTP map $\Phi^A\in\mC^A$,
$$
H_{\min}^{\rm ext}(B)_{\Theta[\Phi]}\geq H_{\min}^{\rm ext}(A)_{\Phi}\;.
$$
This completes the proof that the extended min-entropy is an entropy function.

\subsection{The extended conditional min-entropy}\label{ecme}

We extend here the definition of the conditional min-entropy to quantum channels. This function will play a key role in our results on the comparison of quantum channels and we therefore devote the rest of this section to study it along with its properties, and its operational interpretations.

We consider here a bipartite quantum channel $\Omega^{AB}\equiv\Omega^{A_0B_0\to A_1B_1}:\mB(\mH^{A_0B_0})\to\mB(\mH^{A_1B_1})$ and denote by
$$
\omega^{AB}\equiv\frac{1}{d_{A_0}d_{B_0}}\id^{A_0B_0}\otimes \Omega^{\tA_0\tB_0\to A_1B_1}\left(\phi_{+}^{A_0\tA_0}\otimes\phi_{+}^{B_0\tB_0}\right)
$$
its (normalized) Choi matrix, where $\phi_{+}^{A_0\tA_0}$ and $\phi_{+}^{B_0\tB_0}$ are unnormalized maximally entangled states.

\begin{definition}\label{emin}
The extended conditional min-entropy of a bipartite channel $\Omega^{AB}$ as above, is the function
\begin{align}
H_{\min}^{\rm ext}\left(B|A\right)_{\Omega}&\equiv-\log_2\min\tr[\gamma^{AB_0}]\nonumber\\
\text{subject to: }\; &{\it 1.}\;\;\gamma^{AB_0}\otimes I^{B_1}\geq\omega^{AB}\nonumber\\
&{\it 2.}\;\;\gamma^{A_0B_0}=u^{A_0}\otimes \gamma^{B_0}
\label{maind}
\end{align}
where $u^{A_0}$ is the maximally mixed state on system $A_0$.
\end{definition}

The above optimization problem is SDP, and therefore can be solved efficiently and algorithmically using standard techniques.
Furthermore, since any SDP problem has a dual problem, the extended conditional min-entropy can be expressed as (see Appendix~\ref{AppA} for details)
\begin{align}
2^{-H_{\min}^{\rm ext}\left(B|A\right)_{\Omega}}&= d_{A_0}\max\tr\left[\alpha^{AB}\omega^{AB}\right]\nonumber\\
\text{subject to:}\quad &{\it 1.}\;\;\alpha^{AB_0}=\alpha^{A_0B_0}\otimes u^{A_1}\nonumber\\
&{\it 2.}\;\;\alpha^{A_1B_0}=I^{A_1B_0}\nonumber\\
&{\it 3.}\;\;\alpha^{AB}\geq 0\;.
\label{maind2}
\end{align}
Note that the matrix $\alpha^{AB}$ above satisfies precisely the conditions that a Choi matrix of a superchannel satisfies. We can therefore identify each such $\alpha^{AB}$ with a Choi matrix $\J^{AB}_{\Theta}$ of some superchannel $\Theta$. Denoting the set of superchannels by $\mbb{S}^{AB}\subset\mbb{L}^{AB}$ we get that
\be\label{sf}
d_{B_0}2^{-H_{\min}^{\rm ext}\left(B|A\right)_{\Omega}}= \max_{\Theta\in\mbb{S}^{AB}}\tr\left[\J^{AB}_{\Theta}J_{\Omega}^{AB}\right]
\ee
Note that the trace on the RHS can be expressed as $\tr\left[\J^{AB}_{\Theta}J_{\Omega}^{AB}\right]=\la \Theta,\Theta'\ra$, where $\Theta'$ is a CPP map in $\mbb{L}^{AB}$ (but not necessarily a superchannel) that corresponds to the bipartite channel $\Omega^{AB}$ via the relation $\Omega^{AB}=\1^A\otimes\Theta'\left[\Upsilon^{A\tA}\right]$ (see~\eqref{lambdatheta} for a discussion on this relation). Therefore, the RHS of~\eqref{sf} can be viewed as the support function of superchannels.

A similar definition to~\eqref{sf} was studied in~\cite{Chir2016} for min/max entropies of more general objects known as quantum combs. However, in~\cite{Chir2016} the entropies were viewed as a function of general operators like the comb itself, and here we define the extended-conditional min-entropy as a function of bipartite channels, and not as a function of superchannels. Even the normalization (see the factor of $d_{B_0}$ in~\eqref{sf}) is not arbitrary. This distinction will become clearer when we study the properties of the extended conditional min-entropy. 

Moreover, recall that $\J^{AB}_{\Theta}$ can be viewed as the Choi matrix of the bipartite channel $\Lambda_{\Theta}^{AB}$ as defined in~\eqref{lambdatheta}, and note that $\Lambda_{\Theta}^{AB}$ is a channel if $\Theta$ is a superchannel. Therefore, alternatively,
\be\label{inch}
d_{B_0}2^{-H_{\min}^{\rm ext}\left(B|A\right)_{\Omega}}=\max_{\Theta\in\mbb{S}^{AB}}\left\la \Lambda_{\Theta}^{AB},\Omega^{AB}\right\ra\;,
\ee
where we replaced the inner product between Choi matrices to inner product between the corresponding channels as defined in~\eqref{basic0}.
Now, recall that $\Theta\mapsto\Lambda^{AB}_\Theta=(\1^{A}\otimes\Theta)\left[\Upsilon^{A\tA}\right]$ (see~\eqref{reltheta}) is an isomorphism. Substituting this expression of $\Lambda_{\Theta}^{AB}$ into~\eqref{inch},  we get that the extended conditional min-entropy can be expressed in the following form.
\ba\label{dualtheta2}
&d_{B_0}2^{-H_{\min}^{\rm ext}\left(B|A\right)_{\Omega}}\\
&=\max_{\Theta\in\mbb{S}^{AB}}\left\la (\1^{A}\otimes\Theta)\left[\Upsilon^{A\tA}\right],\Omega^{AB}\right\ra\\
&=\max_{\Theta\in\mbb{S}^{A\tB}}\left\la (\Theta^T\otimes\1^B)\left[\Upsilon^{\tB B}\right],\Omega^{AB}\right\ra\\
&=\max_{\Theta\in\mbb{S}^{A\tB}}\left\la \Upsilon^{\tB B},\left((\Theta^T)^*\otimes\1^B\right)\left[\Omega^{AB}\right]\right\ra\\
&=\max_{\Theta\in\mbb{S}^{A\tB}}\left\la \Upsilon^{\tB B},\left(\Theta\otimes\1^B\right)\left[\Omega^{AB}\right]\right\ra\\
&=\max_{\Theta\in\mbb{S}^{A\tB}}
\left\la\phi_{+}^{\tB_1B_1}\left|\left(\Theta\otimes\1^{B}\right)\left[\Omega^{AB}\right](\phi_{+}^{\tB_0B_0})\right|\phi_{+}^{\tB_1B_1}\right\ra
\ea
where $\left(\Theta\otimes\1^{B}\right)\left[\Omega^{AB}\right]$ is a quantum channel from $\mB(\mH^{\tB_0B_0})$ to $\mB(\mH^{\tB_1B_1})$.
We used the fact that $\Theta$ is a superchannel iff $(\Theta^T)^*$ is a superchannel. To see why, note that from~\eqref{z1} and~\eqref{z2} it follows that
$\J^{AB}_{(\Theta^T)^*}=\overline{\J^{AB}_{\Theta}}$, and therefore $\J^{AB}_{(\Theta^T)^*}$ satisfies the conditions~\eqref{marginals} of a superchannel. For the last equality we used the definition~\eqref{basic0}
of an inner product between channels, and took an orthonormal basis $\{X_a\}_{a=1}^{d_{B_0}^4}$ of $\mB(\mH^{\tB_0B_0})$, whose first element $X_1=\frac{1}{d_{B_0}}\phi_{+}^{B_0\tB_0}$ so that $\Upsilon^{\tB B}(X_a)=0$ unless $a=1$. 
Note that the expression above for the extended conditional 
min-entropy is reminiscent to the one given in~\eqref{chminmax} for the conditional min-entropy.

\subsection{Properties of the extended conditional min-entropy}

The extended conditional min-entropy provides a generalization for the conditional min-entropy. The following theorem demonstrates it by showing that many of the properties of the conditional min-entropy carry over to the extended conditional min-entropy.
\begin{theorem}\label{properties}
Consider a quantum channel $\Omega^{AB}:\mB(\mH^{A_0B_0})\to\mB(\mH^{A_1B_1})$, and denote its normalized Choi matrix by $\omega^{AB}$ as in Definition~\ref{emin}. 
\begin{enumerate}
\item {\it Generalization of conditional min-entropy:} If $\Omega^{AB}$ is a replacement channel
(i.e. $\Omega^{AB}(\rho^{A_0B_0})=\omega^{A_1B_1}$ for all density matrices $\rho^{A_0B_0}$), then $$H_{\min}^{\rm ext}\left(B|A\right)_{\Omega}=H_{\min}(B_1|A_1)_{\omega}\;.$$
\item {\it Independence:} If 
\be\label{tensor}
\Omega^{AB}=\Psi^{A}\otimes\Phi^{B}\;,
\ee
where $\Psi^{A}$ and $\Phi^{B}$ are local channels, then
$$
H_{\min}^{\rm ext}(B|A)_{\Omega}= H_{\min}^{\rm ext}(B)_{\Phi}
$$
is independent of $\Psi^{A}$. 
\item {\it Additivity}: Consider a second quantum channel $\Gamma^{A'B'}:\mB(\mH^{A_0'B_0'})\to\mB(\mH^{A_1'B_1'})$. Then,
$$
H_{\min}^{\rm ext}(BB'|AA')_{\Omega\otimes\Gamma}=H_{\min}^{\rm ext}(B|A)_{\Omega}+H_{\min}^{\rm ext}(B'|A')_{\Gamma}
$$
\item {\it Monotonicity:} For any superchannel $\Theta':\mL^A\to\mL^C$
\be\label{noisyevolution}
H_{\min}^{\rm ext}\left(B|A\right)_{\Omega^{AB}}\leq H_{\min}^{\rm ext}\left(B|C\right)_{\left(\Theta'\otimes\1^{B}\right)\left[\Omega^{AB}\right]}
\ee
\item {\it Conditioning}: Consider a tripartite quantum channel $\Omega^{ABC}:\mB(\mH^{A_0B_0C_0})\to\mB(\mH^{A_1B_1C_1})$. Then,
\be\label{mono}
H_{\min}^{\rm ext}(B|AC)_{\Omega}\leq H_{\min}^{\rm ext}(B|A)_{\Omega}
\ee

\item {\it Bounds:} Upper bound:
\be
H_{\min}^{\rm ext}\left(B|A\right)_{\Omega}\leq H_{\min}(B_1|A_1)_\omega\;.
\ee
Lower bound:
\be
H_{\min}^{\rm ext}\left(B|A\right)_{\Omega}\geq H_{\min}(AB_1|B_0)_\omega-\log(d_{A})
\ee
where $d_{A}\equiv d_{A_0}d_{A_1}$.
\end{enumerate}
\end{theorem}

\begin{remark}
The conditioning property involves the marginal bipartite channel $\Omega^{AB}$ on the RHS of~\eqref{mono}.
This marginal channel is obtained from the tripartite channel $\Omega^{ABC}$ by inputing a fixed state $\gamma^{C_0}$ into the input of system C and tracing out system $C_1$. That is,
$$
\Omega^{AB}(\rho^{A_0B_0})\equiv\tr_{C_1}\left[\Omega^{ABC}\left(\rho^{A_0B_0}\otimes\gamma^{C_0}\right)\right]
$$
The theorem above states that the inequality in~\eqref{mono} holds for all density matrices $\gamma^{C_0}$.
\end{remark}

{\it{Proof.}}
{\it Part 1.} The condition $\gamma^{AB_0}\otimes I^{B_1}\geq\omega^{AB}$ in~\eqref{maind} implies that
$\gamma^{A_1}\otimes I^{B_1}\geq\omega^{A_1B_1}$. Therefore, since 
$\tr[\gamma^{AB_0}]=\tr[\gamma^{A_1}]$ we always have
\be\label{ub}
2^{-H_{\min}^{\rm ext}\left(B|A\right)_{\Omega}}\geq 2^{-H_{\min}\left(B_1|A_1\right)_{\omega}}\;.
\ee
On the other hand, since $\Omega$ is a replacement map, its normalized Choi matrix is given by $\omega^{AB}=u^{A_0B_0}\otimes\omega^{A_1B_1}$. Let $\gamma^{A_1}$ be an optimal positive semidefinite matrix that satisfies $\gamma^{A_1}\otimes I^{B_1}\geq\omega^{A_1B_1}$ such that
$\tr[\gamma^{A_1}]=2^{-H_{\min}^{\rm ext}\left(B_1|A_1\right)_{\omega}}$. Define $\gamma^{AB_0}\equiv u^{A_0B_0}\otimes\gamma^{A_1}$. It is easy to check that this $\gamma^{AB_0}$ satisfies the two conditions in~\eqref{maind}. We therefore get that 
$$
2^{-H_{\min}^{\rm ext}\left(B|A\right)_{\Omega}}\leq 2^{-H_{\min}\left(B_1|A_1\right)_{\omega}}\;.
$$
Hence, we must have $H_{\min}^{\rm ext}\left(B|A\right)_{\Omega}=H_{\min}\left(B_1|A_1\right)_{\omega}$.
 
{\it Part 2.} From~\eqref{tensor} it follows that the normalized Choi matrix of $\Omega^{AB}$ can be decomposed as $\omega^{AB}=\omega^{A}\otimes\omega^{B}$, where $\omega^A$ and $\omega^B$ are the normalized Choi matrices of $\Psi^{A}$ and $\Phi^{B}$, respectively. 
In this case, the two conditions of~\eqref{maind} take the form
\begin{align}
&\gamma^{AB_0}\otimes I^{B_1}\geq \omega^A\otimes\omega^B\label{c1}\\
&\gamma^{A_0B_0}=u^{A_0}\otimes \gamma^{B_0}\label{c2}
\end{align}
Tracing out system $A_1$ on both sides of~\eqref{c1} gives 
$$
\gamma^{A_0B_0}\otimes I^{B_1}\geq u^{A_0}\otimes \omega^B\;,
$$
and when combined with~\eqref{c2} yields
\be\label{zzz}
\gamma^{B_0}\otimes I^{B_1}\geq  \omega^B\;.
\ee
Note that $\tr[\gamma^{AB_0}]=\tr[\gamma^{B_0}]$ so that we must have
$$
2^{-H_{\min}^{\rm ext}(B|A)}\geq 2^{-H_{\min}(B_1|B_0)}\;.
$$
On the other hand, for the choice $\gamma^{AB_0}=\omega^A\otimes\gamma^{B_0}$ with optimal $\gamma^{B_0}$ (i.e. $\gamma^{B_0}$ satisfies~\eqref{zzz} and $\tr[\gamma^{B_0}]=2^{-H_{\min}(B_1|B_0)_{\omega}}$)
we obtain the other side of the inequality. We therefore conclude that
$$
H_{\min}^{\rm ext}(B|A)_{\Omega}= H_{\min}(B_1|B_0)_{\omega}\equiv H_{\min}^{\rm ext}(B)_{\Phi}\;.
$$
{\it Part 3.} Denote the Choi matrix of $\Gamma^{A'B'}$ by $\omega^{A'B'}$. From Definition~\ref{emin}:
\begin{align}
2^{-H_{\min}^{\rm ext}\left(BB'|AA'\right)_{\Omega\otimes\Gamma}}&=\min\tr[\gamma^{AA'B_0B_0'}]\nonumber\\
\text{subject to: }\; &{\it 1.}\;\;\gamma^{AA'B_0B_0'}\otimes I^{B_1B_1'}\geq\omega^{AB}\otimes\omega^{A'B'}\nonumber\\
&{\it 2.}\;\;\gamma^{A_0A_0'B_0B_0'}=u^{A_0A_0'}\otimes \gamma^{B_0B_0'}\nonumber
\end{align}
Note that the above two conditions follow from the following 5 conditions:
\ba\label{5}
&\gamma^{AA'B_0B_0'}=\gamma^{AB_0}\otimes\gamma^{A'B_{0}'}\\
&\gamma^{AB_0}\otimes I^{B_1}\geq\omega^{AB}\;\;,\;\;\gamma^{A'B_0'}\otimes I^{B_1'}\geq\omega^{A'B'}\\
&\gamma^{A_0B_0}=u^{A_0}\otimes \gamma^{B_0}\;\;,\;\;\gamma^{A_0B_0}=u^{A_0}\otimes \gamma^{B_0}
\ea
Therefore, $2^{-H_{\min}^{\rm ext}\left(BB'|AA'\right)_{\Omega\otimes\Gamma}}\leq \min\tr[\gamma^{AA'B_0B_0'}]$, where the minimization is over all $\gamma^{AA'B_0B_0'}$ that satisfies the 5 conditions in~\eqref{5}.
That is,
$$
2^{-H_{\min}^{\rm ext}\left(BB'|AA'\right)_{\Omega\otimes\Gamma}}\leq
2^{-H_{\min}^{\rm ext}\left(B|A\right)_{\Omega}}2^{-H_{\min}^{\rm ext}\left(B'|A'\right)_{\Gamma}}\;.
$$
To prove the converse, consider the dual expression~\eqref{maind2} for the extended conditional min-entropy. 
We get that
\begin{align}\label{3}
2^{-H_{\min}^{\rm ext}\left(BB'|AA'\right)_{\Omega\otimes\Gamma}}&=d_{A_0}d_{A_0'}\max\tr\left[\alpha^{ABA'B'}\omega^{AB}\otimes\omega^{A'B'}\right]\nonumber\\
\text{subject to:}\quad &{\it 1.}\;\;\alpha^{AA'B_0B_0'}=\alpha^{A_0A_0'B_0B_0'}\otimes u^{A_1A_1'}\nonumber\\
&{\it 2.}\;\;\alpha^{A_1A_1'B_0B_0'}=I^{A_1A_1'B_0B_0'}\nonumber\\
&{\it 3.}\;\;\alpha^{AA'BB'}\geq 0\;.
\end{align}
Similarly to the previous argument, the 3 conditions above follow from the following conditions:
\ba
&\alpha^{ABA'B'}=\alpha_1^{AB}\otimes\alpha^{A'B'}_2\quad,\quad \alpha^{AB}_1\geq 0\quad,\quad\alpha^{A'B'}_2\geq 0\\
& \alpha^{AB_0}_1=\alpha^{A_0B_0}\otimes u^{A_1}\quad,\quad\alpha^{A'B_0'}_2=\alpha^{A_0'B_0'}\otimes u^{A_1'}\\
& \alpha_1^{A_1B_0}=I^{A_1B_0}\quad,\quad\alpha^{A_1'B_0'}_2=I^{A_1'B_0'}\nonumber
\ea
Therefore, if we replace the 3 conditions in~\eqref{3} with the above conditions we get that
$$
2^{-H_{\min}^{\rm ext}\left(BB'|AA'\right)_{\Omega\otimes\Gamma}}\geq
2^{-H_{\min}^{\rm ext}\left(B|A\right)_{\Omega}}2^{-H_{\min}^{\rm ext}\left(B'|A'\right)_{\Gamma}}\;.
$$
This completes the proof of Part 3.

{\it Part 4.} From~\eqref{dualtheta2} we get that the extended conditional min-entropy can be expressed in the following form: 
$$
d_{B_0}2^{-H_{\min}^{\rm ext}\left(B|A\right)_{\Omega}}=\max_{\Theta\in\mbb{S}^{A\tB}}\left\la \Upsilon^{\tB B},\left(\Theta\otimes\1^B\right)\left[\Omega^{AB}\right]\right\ra
$$
where $\left(\Theta\otimes\1^{B}\right)\left[\Omega^{AB}\right]$ is a quantum channel from $\mB(\mH^{\tB_0B_0})$ to $\mB(\mH^{\tB_1B_1})$.
Let  $\Theta':\mL^A\to\mL^C$ be a superchannel. Then,
\begin{align*}
&d_{B_0}2^{-H_{\min}^{\rm ext}\left(B|C\right)_{\Theta'\otimes\1^B[\Omega^{AB}]}}\\
&=\max_{\Theta\in\mbb{S}^{C\tB}}\left\la \Upsilon^{\tB B},\left(\Theta\circ\Theta'\otimes\1^B\right)\left[\Omega^{AB}\right]\right\ra\\
&\leq \max_{\Theta''\in\mbb{S}^{A\tB}}\left\la \Upsilon^{\tB B},\left(\Theta''\otimes\1^B\right)\left[\Omega^{AB}\right]\right\ra\\
&=d_{B_0}2^{-H_{\min}^{\rm ext}\left(B|A\right)_{\Omega^{AB}}}\;.
\end{align*}
Therefore,
\be
H_{\min}^{\rm ext}\left(B|A\right)_{\Omega^{AB}}\leq H_{\min}^{\rm ext}\left(B|C\right)_{\left(\Theta'\otimes\1^{B}\right)\left[\Omega^{AB}\right]}
\ee
for any superchannel $\Theta':\mL^A\to\mL^C$.

{\it Part 5.} Let $\Theta_\gamma:\mL^{AC}\to\mL^{A}$ be the superchannel defined by: for all $\Phi^{AC}\in\mL^{AC}$ and $\rho^{A_0}\in\mB(\mH^{A_0})$
$$
\Theta_\gamma[\Phi^{AC}](\rho^{A_0})\equiv\tr_{C_1}\left[\Phi^{AC}(\rho^{A_0}\otimes\gamma^{C_0})\right]\;.
$$
It is straightforward to see that $\Theta_\gamma$ is a superchannel if $\gamma^{C_0}$ is a density matrix. Let 
$$
\Omega^{AB}_\gamma\equiv\left(\Theta_\gamma\otimes\1^{B}\right)\left[\Omega^{ABC}\right]\;.
$$
In particular, $$\Omega^{AB}_\gamma(\rho^{A_0B_0})=\tr_{C_1}\left[\Omega^{ABC}(\rho^{A_0B_0}\otimes\gamma^{C_0})\right]\;.$$
Hence, from Part 4 we get that for any density matrix $\gamma^{C_0}\in\mB(\mH^{C_0})$
\ba
H_{\min}^{\rm ext}\left(B|AC\right)_{\Omega^{ABC}}&\leq 
H_{\min}^{\rm ext}\left(B|A\right)_{\left(\Theta_\gamma\otimes\1^{B}\right)\left[\Omega^{ABC}\right]}\\
&=H_{\min}^{\rm ext}\left(B|A\right)_{\Omega^{AB}_\gamma}\nonumber\;.
\ea
This completes the proof of Part 5.

{\it Part 6.}
The upper bound follows from~\eqref{ub}.
For the lower bound, if we add to the two conditions in~\eqref{maind} a third condition that 
$\gamma^{AB_0}=\gamma^{A_0B_0}\otimes u^{A_1}$ we get that
\ba
2^{-H_{\min}^{\rm ext}\left(B|A\right)_{\Omega}}&\leq \min\left\{\tr\left[\gamma^{B_0}\right]\;:\;I^{AB_1}\otimes\gamma^{B_0}\geq d_{A}\omega^{AB}\right\}\\
&=d_A2^{-H_{\min}\left(AB_1|B_0\right)_{\omega}}\;.
\ea
This completes the proof.
\QED

The properties above demonstrate that the extended conditional min-entropy indeed quantifies the uncertainty about one dynamical system conditioned on another. Particularly,
note that property 5 is consistent with the intuition that the uncertainty (i.e. entropy) about system $B$ increases if the system one has access to (i.e. system $A$) undergoes a physical evolution.

\subsection{Operational interpretation as a guessing probability} 

The conditional min-entropy $H(A_1|A_0)$ has an operational interpretation as a guessing probability when system $A_1$ is classical. 
Here we show that a similar interpretation can be made for the extended conditional min-entropy if system $B$ is classical. 
Since system $B_1$ is classical, for all $\rho^{A_0B_0}\in\mB(\mH^{A_0B_0})$
\be\label{om1}
\Omega^{A_0B_0\to A_1B_1}(\rho^{A_0B_0})=\sum_{x=1}^{d_{B_1}}\Omega_{x}^{A_0B_0\to A_1}(\rho^{A_0B_0})\otimes |x\lr x|^{B_1}\;,
\ee
where $\{\Omega_{x}^{A_0B_0\to A_1}\}$ form a quantum instrument. Moreover, since system $B_0$ is classical, we denote 
\be\label{om2}
\Omega_{x|y}^{A_0\to A_1}(\rho)\equiv\Omega_{x}^{A_0B_0\to A_1}\left(\rho\otimes|y\lr y|\right)\quad\forall\rho\in\mB(\mH^{A_0})
\ee
where for each $y$ the set $\{\Omega_{x|y}^{A_0\to A_1}\}_{x=1}^{d_{B_1}}$ form a quantum instrument.

\begin{figure}[t]
    \includegraphics[width=0.3\textwidth]{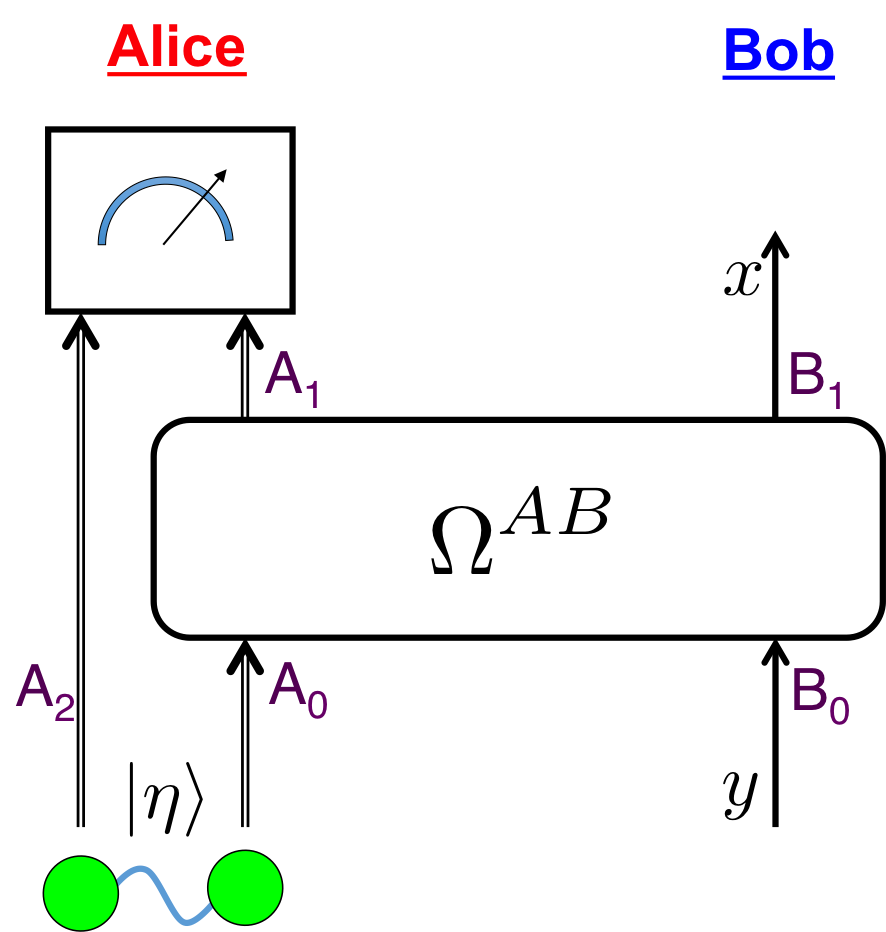}
  \caption{\linespread{1}\selectfont{\small A strategy for Alice to guess Bob's output.}}
  \label{guessing}
\end{figure}

In Fig.~\ref{guessing} we describe a strategy for Alice to guess Bob's outcome $x$ if Bob's input is $y$.
In this strategy, Alice sends through her share of the channel $\Omega^{AB}$ one part of a possibly entangled state $|\eta^{A_0A_2}\ra$. At her output of the channel she measures the joint system $A_1A_2$. 
The outcome of the measurement is Alice's guess of Bob's output value. The maximum probability that Alice guesses Bob's outcome correctly, given that Bob's input is $y$ can be expressed as
\begin{align}\label{guessy}
&P_{\text{guess}}^{(y)}(\Omega^{AB})\equiv\nonumber\\
&\max\sum_{x=1}^{d_{B_1}}
\tr\left[P_{x}\left(\Omega_{x|y}^{A_0\to A_1}\otimes\id^{A_2}\left
(|\eta\lr\eta|^{A_0A_2}\right)\right)\right]
\end{align}
where the maximum is over all POVMs $\{P_x\}_{x=1}^{d_{B_1}}$ on system $A_1A_2$, and over all bipartite pure states $|\eta\ra^{A_0A_2}$ on system $A_0A_2$.
Note that replacing the optimization over $|\eta\ra^{A_0A_2}$ with optimization over all mixed state will not increase the optimal guessing probability, and furthermore,  we can assume w.l.o.g. that $\dim(\mH^{A_2})\leq\dim(\mH^{A_0})$.
Finally, we define the guessing probability of a quantum-classical channel as:
$$
P_{\text{guess}}(\Omega^{AB})\equiv\frac{1}{d_{B_0}}\sum_{y=1}^{d_{B_0}}P_{\text{guess}}^{(y)}(\Omega^{AB})\;.
$$ 
The above expression can be interpreted as the maximum probability that Alice can guess correctly the value
$x$ of Bob's system $B_1$ if Bob's input $y$ (which is known to Alice) is chosen at random according to a uniform distribution.

\begin{theorem}\label{interp}
Let $\Omega^{AB}$ be a quantum channel as above with a classical system $B$. Then,
$$
P_{\text{guess}}(\Omega^{AB})=2^{-H_{\min}^{\rm ext}\left(B|A\right)_{\Omega}}
$$
\end{theorem}

{\it{Proof.}}
Following the same notations as in~\eqref{om1} and~\eqref{om2}, since $B$ is classical, the (normalized) Choi matrix of $\Omega^{AB}$ can be expressed as
$$
\omega^{AB}=\sum_{y=1}^{d_{B_0}}\sum_{x=1}^{d_{B_1}}\omega^{A}_{x|y}\otimes |y\lr y|^{B_0}\otimes|x\lr x|^{B_1}\;,
$$
with 
\be\label{ome1}
\omega^{A}_{x|y}\equiv\frac{1}{d_{A_0}d_{B_0}}\id^{A_0}\otimes\Omega_{x|y}^{\tA_0\to A_1}\left
(\phi_{+}^{A_0\tA_0}\right)\;.
\ee
Consequently, from~\eqref{maind2}, with $\omega^{AB}$ as above and $\alpha^{AB}\equiv\sum_{x,y}\alpha^{A}_{x|y}\otimes |y\lr y|^{B_0}\otimes |x\lr x|^{B_1}$, we get
\begin{align}
2^{-H_{\min}^{\rm ext}\left(B|A\right)_{\Omega}}&=d_{A_0}\max \sum_{y=1}^{d_{B_0}}\sum_{x=1}^{d_{B_1}}\tr[\alpha^{A}_{x|y}\omega_{x|y}^{A}]\nonumber\\
\text{subject to: }\; &\sum_{x=1}^{d_{B_1}}\alpha^{A}_{x|y}=\zeta^{A_0}_{y}\otimes I^{A_1}\;\;,\;\;\tr[\zeta^{A_0}_y]=1\nonumber\\
&\alpha^{A}_{x|y}\geq 0\nonumber\\
&\text{for all }1\leq x\leq d_{B_1}\text{ and }1\leq y\leq d_{B_0}
\end{align}
where we denoted by 
$$
\zeta^{A_0}_y\equiv \frac{1}{d_{A_1}}\sum_{x=1}^{d_{B_1}}\alpha^{A_0}_{x|y}\;.
$$
Note that we can assume w.l.o.g. that $\zeta^{A_0}_y$ is full rank. Hence, we can define for each $y$ the following POVM on system $A$:
$$
P_{x|y}^{A}\equiv \left(\left(\zeta^{A_0}_y\right)^{-\frac{1}{2}}\otimes I^{A_1}\right)\alpha_{x|y}^{A}\left(\left(\zeta^{A_0}_y\right)^{-\frac{1}{2}}\otimes I^{A_1}\right)
$$
Note that $P_{x|y}^{A}\geq 0$ and $\sum_x P_{x|y}^{A}=I^{A}$. With this notation 
\begin{align}
&d_{A_0} \sum_{y=1}^{d_{B_0}}\sum_{x=1}^{d_{B_1}}\tr[\alpha^{A}_{x|y}\omega_{x|y}^{A}]\nonumber\\
&=d_{A_0}\sum_{x,y}
\tr\left[P^{A}_{x|y}\left(\sqrt{\zeta^{A_0}_y}\otimes I^{A_1}\right)\omega_{x|y}^{A}\left(\sqrt{\zeta^{A_0}_y}\otimes I^{A_1}\right)\right]\nonumber\\
&=\frac{1}{d_{B_0}}\sum_{x,y}\tr\left[P_{x|y}^{A}\left(\id^{A_0}\otimes\Omega_{x|y}^{\tA_0\to A_1}\left
(|\zeta_y\lr\zeta_y|^{A_0\tA_0}\right)\right)\right]
\end{align}
where we used~\eqref{ome1}, and the state $|\zeta_y\ra^{A_0\tA_0}$ is the purification of the (normalized) state $\zeta^{A_0}_y$. That is, 
$$
|\zeta_y\ra^{A_0\tA_0}\equiv \left(\sqrt{\zeta^{A_0}_y}\otimes I^{\tA_0}\right)|\phi_{+}^{A_0\tA_0}\ra\;.
$$
We therefore conclude that 
\begin{align}
2^{-H_{\min}^{\rm ext}\left(B|A\right)_{\Omega}}&=\frac{1}{d_{B_0}}\max\sum_{y=1}^{d_{B_0}}\sum_{x=1}^{d_{B_1}}\nonumber\\
&\tr\left[P_{x|y}^{A}\left(\id^{A_0}\otimes\Omega_{x|y}^{\tA_0\to A_1}\left
(|\zeta_y\lr\zeta_y|^{A_0\tA_0}\right)\right)\right]
\nonumber\\
\text{subject to: }&\;P_{x|y}^{A}\geq 0\;\;,\;\;\sum_{x=1}^{d_{B_1}}P_{x|y}^{A}=I^A\;\;,\la\zeta_y|\zeta_y\ra=1\nonumber\\
&\forall\;x=1,...,d_{B_1}\text{ and }y=1,...,d_{B_0}\;.
\end{align}
This completes the proof.
\QED

The theorem above provides the first operational interpretation for the conditional min-entropy of a bipartite quantum channel when $B$ is classical.
Consider now the special case in which the CP and trace non-increasing maps $\{\Omega^{A_0B_0\to A_1}_x\}$, as defined above, are of the form
$$
\Omega^{A_0B_0\to A_1}_x=p_x\Lambda^{A_0B_0\to A_1}_{(x)}
$$ 
where $\{p_x\}$ form a probability distribution, and each $\Lambda^{A_0B_0\to A_1}_{(x)}$ is a CPTP. For a given fixed value 
$y$ at the input system of $B_0$, we denote by $\Lambda^{A_0\to A_1}_{(x|y)}$ the corresponding CPTP map on Alice's side.
In this special case, the guessing probability~\eqref{guessy}, can be interpreted as the maximum possible probability to guess
which channel $\left\{\Lambda^{A_0\to A_1}_{(x|y)}\right\}_x$ Alice holds. Hence, the theorem above, when applied to this case, implies that $2^{-H_{\min}^{\rm ext}\left(B|A\right)_{\Omega}}$
can be interpreted as the maximum probability to guess which channel Alice holds, out of $\left\{\Lambda^{A_0\to A_1}_{(x|y)}\right\}_x$, where $y$ is chosen from a uniform distribution.

\begin{figure}[h]
    \includegraphics[width=0.3\textwidth]{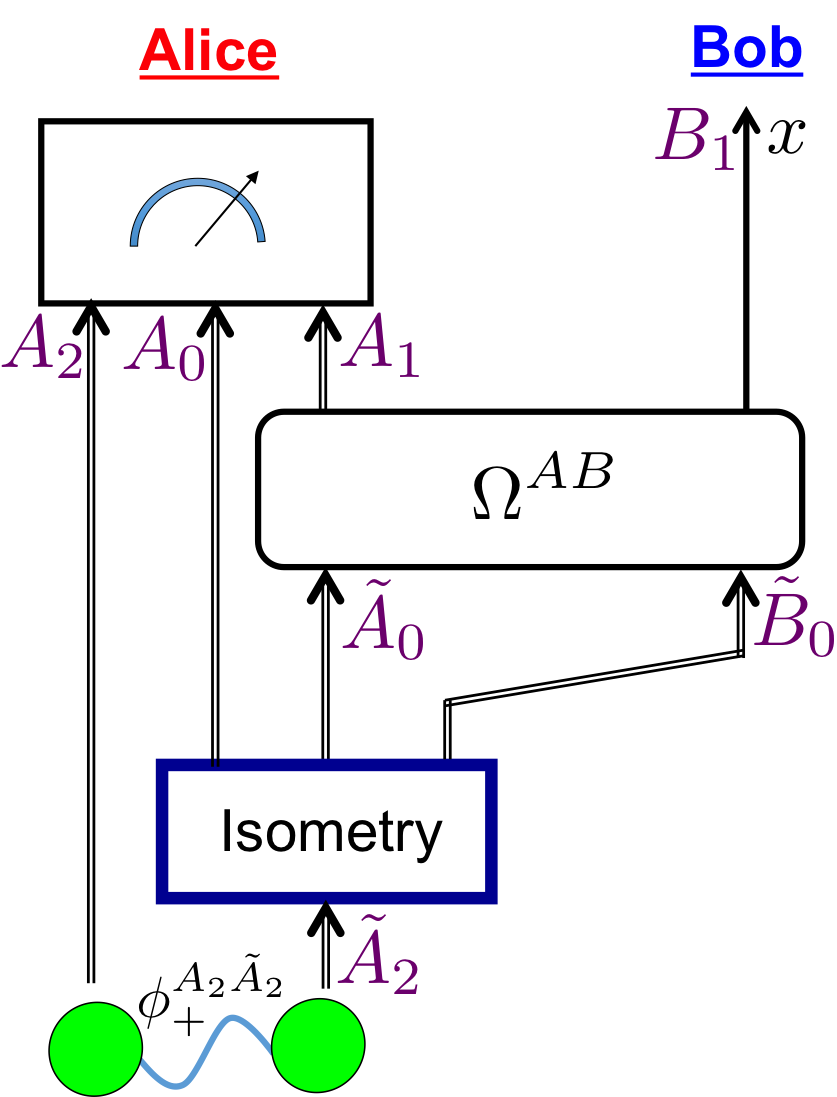}
  \caption{\linespread{1}\selectfont{\small A restrictive strategy for Alice to guess Bob's output. Alice has access to a maximally entangled state $\Phi^{A_2\tA_2}_{+}/d_{A_2}$ (with $d_{A_2}=d_{B_0}$). Alice uses the maximally entangled state and send her system $\tilde{A}_2$ through any (non-local) quantum channel with outputs $A_0$, $\tA_0$, and $\tB_0$ (it can be shown that an isometry channel always achieves the maximal guessing probability). Alice then performs a POVM on her systems $A_0$, $A_1$, and $A_2$ and guesses the output value $x$ of $B_1$. The expression $2^{-H_{\min}^{\rm ext}\left(B|A\right)_{\Omega}}$ is the maximum probability for Alice to guess correctly Bob's output, obtained by optimizing over all possible isometries and POVMs.}} 
  \label{guessing3}
\end{figure}

In Appendix~\ref{fc} we provide another operational interpretation (for the extended conditional min-entropy)  in the case that only system $B_1$ is classical, while systems $A_0,A_1$ and $B_0$ are all quantum. In this case,  the extended conditional min-entropy still can be expressed as the optimal probability to guess correctly the value of $B_1$. In this case, however,
system $B_0$ can be entangled with Alice's systems as described in Fig.~\ref{guessing3}.

\section{Comparison of Quantum Channels}\label{discrim}

Here we consider one of the main problems discussed in the introduction. Given a collection of channels $\Psi^{A}_{j}\in\mC^A$ and $\Phi^{B }_{j}\in\mC^{B}$, with $j=1,...,n$, is there a superchannel $\Theta:\mL^A\to\mL^{B}$ such that for all $j=1,...,n$
\be\label{main}
\Phi^{B}_{j}=\Theta\left[\Psi^{A}_{j}\right]\;?
\ee
These $n$-conditions can be expressed as a single condition given by
$$
\sum_{j=1}^{n} |j\lr j|^{R}\otimes\Phi^{B}_{j}=\sum_{j=1}^{n}|j\lr j|^{R}\otimes\Theta\left[\Psi^{A}_{j}\right] 
$$
where $|j\lr j|^{R}$ can be viewed as a channel from the 1-dimensional system $R_0$ to the $n$-dimensional classical system $R_1$. Therefore, the problem~\eqref{main} is a special case of the following problem.

\begin{figure}[h]
    \includegraphics[width=0.4\textwidth]{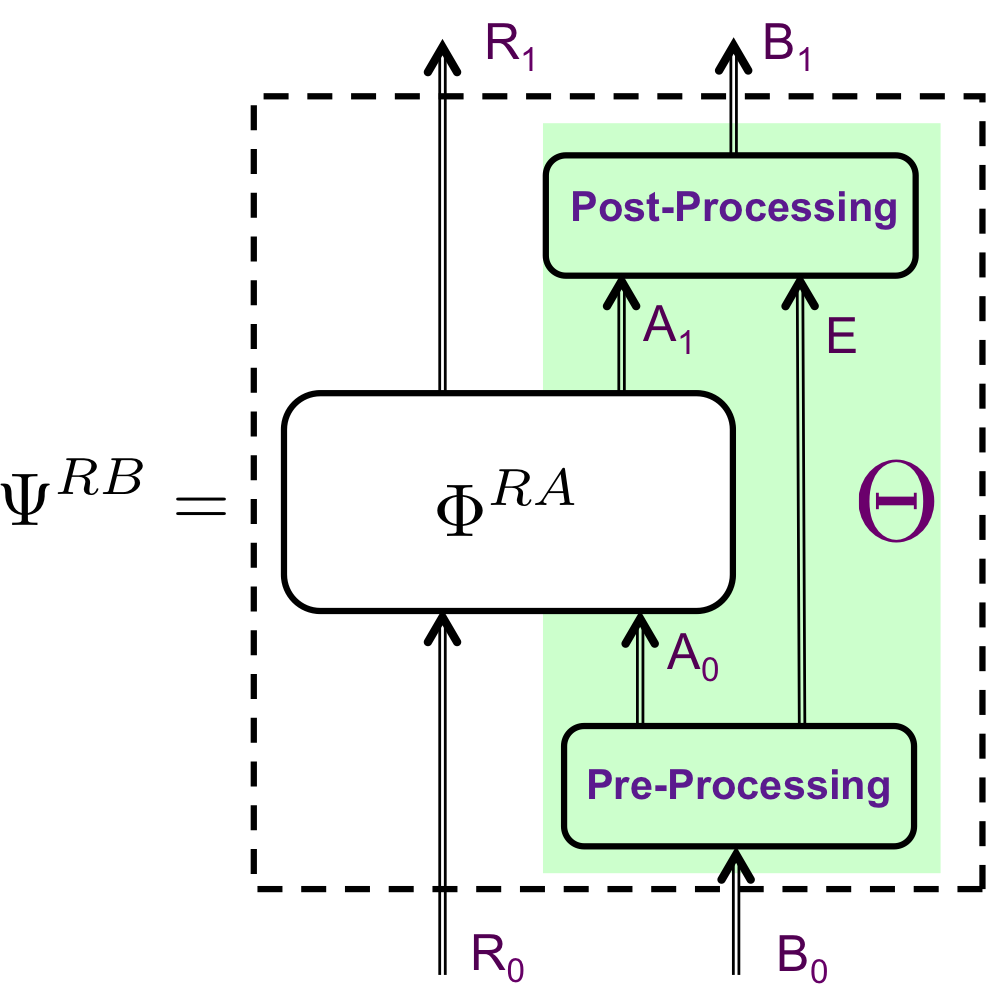}
  \caption{\linespread{1}\selectfont{\small The action of a superchannel $\Theta\in\mbb{L}^{AB}$ on a a bipartite channel $\Phi^{RA}$ yielding the bipartite state $\Psi^{RB}$}}
  \label{super}
\end{figure}

Consider three physical systems $A$, $B$, and $R$, and two bipartite quantum channels $\Phi^{RA}\in\mL^{RA}$ and $\Psi^{RB}\in\mL^{RB}$. Is there  a superchannel $\Theta\in\mbb{L}^{AB}$ such that (see Fig.~\ref{super})
\be\label{supermajor}
\Psi^{RB}=\1^R\otimes\Theta\left[\Phi^{RA}\right]\;?
\ee
If such a superchannel exists, we will say that $\Phi^{RA}$ \emph{quantum majorizes} $\Psi^{RB}$ and denote this relation by
$$
\Psi^{RB}\prec_q\Phi^{RA}\quad\text{or equivalently}\quad \Phi^{RA}\succ_q\Psi^{RB}\;.
$$
The preorder $\prec_q$ was studied in~\cite{Gour2018} for the case of quantum states (i.e. $d_{A_0}=d_{B_0}=d_{R_0}=1$). 

This section is organized as follows. We start with discussion on channel divergences and show that they can be used to provide necessary conditions for the relation~\eqref{main} with $n=2$.
We then move to provide a full characterization of the preorder $\prec_q$ above (and consequently of~\eqref{main} as well). Our characterization is given in terms of the extended conditional min-entropy, and we also show that the problem can be completely solved with  semidefinite programming. We end the section with an application of our results to the resource theory of thermodynamics.

\subsection{Channel Divergences}

Consider~\eqref{main} with $n=2$. As discussed in the introduction, a measure of distinguishability $D(\cdot\|\cdot)$ of two quantum states must satisfy the following condition:
$$
D\left(\Phi(\rho_1)\|\Phi(\rho_2)\right)\leq D\left(\rho_1\|\rho_2\right)\;.
$$
Any such function  provides necessary conditions that a pair of quantum states $(\rho_1,\rho_2)$ is related to another pair of states $(\sigma_1,\sigma_2)$ via some quantum channel $\Phi$. Examples of such functions are the trace norm, the relative entropy, and many R\'enyi divergences (see~\cite{Datta2015} and reference therein for a large class of such functions). In~\cite{Cooney2016,Wilde2018} a method to extend any such divergence from states to channels was introduced. This method is very similar to the extension of the trace norm into the diamond norm~\cite{Kit97}. 

The completely bounded trace norm, which is commonly called the \emph{diamond norm}, is defined as follows (we focus here only on CP maps).
Let $\Phi_1^{A},\Phi_2^{A}:\mB(\mH^{A_0})\to\mB(\mH^{A_1})$ be two CP maps, and let $R$ be a reference system. 
Then, the diamond norm distance between $\Phi_1^A$ and $\Phi_2^A$ is given by
$$
\|\Phi_1^A-\Phi_2^A\|_{\diamond}\equiv\max_{\rho^{A_0R}}\|\Phi_1^A\otimes\id^{R}(\rho^{A_0R})-\Phi_2^A\otimes\id^{R}(\rho^{A_0R})\|_1
$$
where the maximum can be taken over all pure bipartite states $\rho^{A_0R}$ with reference system $R$ of the same dimension as $A_0$. 

Following the above extension of the trace norm, for any divergence $D(\cdot\|\cdot)$, we define now the following induced contraction, $C_D(\cdot\|\cdot)$, that acts on CPTP maps (this induced contraction was first introduced 
in~\cite{Cooney2016,Wilde2018}). For any two CPTP maps $\Phi_1,\Phi_2:\mB(\mH^A)\to\mB(\mH^B)$  define
\ba\label{defc}
&C_D\left(\Phi_1^A\big\|\Phi_2^A\right)\\
&\equiv\sup_{\sigma^{A_0R}}D\left(\Phi_1^A\otimes\id^{R}(\sigma^{A_0R})\big\|\Phi_2^A\otimes\id^{R}(\sigma^{A_0R})\right)
\ea
where the supremum is taken over all density matrices $\sigma^{A_0R}\in\mB(\mH^{A_0}\otimes\mH^{R})$ and over all dimensions of system $R$. Note, however, that we can assume w.l.o.g. that the supremum above is over pure states since since we can always purify $\sigma^{A_0R}$, and $D$ is contractive under a partial trace. Moreover, since 
$D$ is invariant under a local unitary on $R$, we conclude that w.l.o.g. the dimension of $\mH^{R}$ can be taken to be the same as that of $\mH^{A_0}$.  We now show that the above generalized divergence is indeed a contraction under superchannels.

Let $\Theta\in\mbb{L}^{AB}$ be a superchannel, and $\Phi^{A}_1,\Phi^{A}_{2}\in\mL^A$ be two quantum channels.
First note that 
\begin{align}
& C_{D}\left(\Theta\left[\Phi_1^A\right]\big\|\Theta\left[\Phi_2^A\right]\right)\nonumber\\
&=
\sup_{\sigma^{B_0R}}D\left(\Theta[\Phi_1^A]\otimes\id^{R}(\sigma^{B_0R})\Big\|\Theta[\Phi_2^A]\otimes\id^{R}(\sigma^{B_0R})\right)
\end{align}
where the supremum is over all density matrices $\sigma^{B_0R}\in\mB(\mH^{B_0R})$ and over all systems $R$. Although this optimization can be taken over pure states with $R$ having the same dimension as $B_0$, it will be convenient not to assume this at the moment. Now, since $\Theta$ is a superchannel, it can be realized in terms of pre and post processing as in~\eqref{realization}.
Denoting by $\tau^{A_0ER}\equiv\Gamma_{\text{pre}}^{B_0\to A_0E}\otimes\id^{R}(\sigma^{B_0R})$ we get
\begin{align}
&D\left(\Theta[\Phi_1^A]\otimes\id^{R}(\sigma^{B_0R})\Big\|\Theta[\Phi_2^A]\otimes\id^{R}(\sigma^{B_0R})\right)\nonumber\\
&=D\left(\Gamma_{\text{post}}^{A_1E\to B_1}\circ\Phi_{1}^{A}(\tau^{A_0ER})\Big\|\Gamma_{\text{post}}^{A_1E\to B_1}\circ\Phi_{2}^{A}(\tau^{A_0ER})\right)\nonumber\\
&\leq D\left(\Phi_{1}^{A}(\tau^{A_0ER})\Big\|\Phi_{2}^{A}(\tau^{A_0ER})\right)\nonumber\\
&\leq \max_{\gamma^{A_0ER}\in\mD(\mH^{A_0ER})}D\left(\Phi_{1}^{A}(\gamma^{A_0ER})\Big\|\Phi_{2}^{A}(\gamma^{A_0ER})\right)\nonumber\\
&=C_{D}\left(\Phi_1^A\big\|\Phi_2^A\right)\;,
\end{align}
where the first inequality follows from the contractivity of $D$ under $\Gamma_{\text{post}}^{A_1E\to B_1}$, and the last equality follows from the definition. We therefore conclude that
\be
C_{D}\left(\Theta\left[\Phi_1^A\right]\big\|\Theta\left[\Phi_2^A\right]\right)\leq C_D\left(\Phi_1^A\big\|\Phi_2^A\right)\;.
\ee

There are a couple of simple consequences of the above data processing inequality. First, any such contraction under superchannels, $C(\cdot\|\cdot)$, is invariant under unitary superchannels as defined in~\eqref{reversible}. For such reversible superchannel $\Theta$ we have
$$
C\left(\Theta\left[\Phi_1^A\right]\big\|\Theta\left[\Phi_2^A\right]\right)= C\left(\Phi_1^A\big\|\Phi_2^A\right)\;.
$$
Second, consider a superchannel $\Theta_{\Lambda}:\mL^{A}\to\mL^{AB}$  defined by
$$
\Theta_{\Lambda}[\Phi^{A}]=\Phi^{A}\otimes\Lambda^{B}
$$
It is straightforward to show that it is a superchannel. Similarly, consider the linear map
$
\mT:\mL^{AB}\to\mL^{A}
$
acting on a bipartite map $\Omega^{AB}\in\mL^{AB}$ as
$$
\mT\left[\Omega^{AB}\right](\rho^{A_0})=\tr_{B_1}\left[\Omega^{AB}(\rho^{A_0}\otimes u^{B_0})\right]
$$
where $u^{B_0}\equiv\frac{1}{d_{B_0}}I^{B_0}$ is the maximally mixed (uniform) state (note that one can choose to input another state and the choice of maximally mixed state is just a convenient one). Since $\mT$ is also a superchannel, the contractivity of $C(\cdot\|\cdot)$ implies that
$$
C\left(\Phi^{A}\otimes\Lambda^{B}\big\|\Psi^{A}\otimes\Lambda^B\right)=C\left(\Phi^{A}\big\|\Psi^{A}\right)
$$

\subsection{Characterization of Quantum Majorization for Channels}\label{supermaj}

To characterize the preorder $\prec_q$, we first show that we can assume w.l.o.g. that system $R$ is classical, and that $d_{R_0}=1$. For this purpose, we define two sets of CP trace non-increasing maps that we construct from the two bipartite CPTP maps $\Phi^{RA}$ and $\Psi^{RB}$. Let $\{|\varphi_x\lr\varphi_x|^{R_0}\}_{x=1}^{d_{R_0}^2}$ be a normalized rank one basis of $\mB_h(\mH^{R_0})$ and let $\{E_y^{R_1}\}_{y=0}^{d_{R_1}^2}$ be an informationally complete (basis) POVM of $\mB(\mH^{R_1})$. Then, we define for all $\rho^{A_0}\in\mB(\mH^{A_0})$
$$
\Phi_{y|x}^A(\rho^{A_0})\equiv\tr_{R_1}\left[\left(E_y^{R_1}\otimes I^{A_1}\right)\Phi^{RA}\left(|\varphi_x\lr\varphi_x|^{R_0}\otimes\rho^{A_0}\right)\right]
$$ 
and $\Psi_{y|x}^B$ is defined similarly as in the above equation, by replacing $\Phi^{RA}$ with $\Psi^{RB}$.
By definition, for all $x$ and $y$, $\Phi_{y|x}^A$ and $\Psi_{y|x}^B$ are trace non-increasing CP maps, and for any $x$ the maps $\sum_{y=1}^{d_{R_1}^2}\Phi_{y|x}^A$ and $\sum_{y=1}^{d_{R_1}^2}\Psi_{y|x}^B$ are 
CPTP maps.

A key observation with these definitions is that~\eqref{supermajor} holds if and only if
$$
\Psi_{y|x}^B=\Theta\left[\Phi_{y|x}^A\right]
$$
for all $x=1,...,d_{R_0}^2$ and all $y=1,...,d_{R_1}^2$. With this at hand, we can define two classical systems $X$ and $Y$, with 1-dimensional input, and with $d_X\equiv d_{R_0}^2$ and $d_Y\equiv d_{R_1}^2$ dimensional outputs, such that
\be\label{formc}
\Phi^{XYA}\equiv\frac{1}{d_{X}}\sum_{x=1}^{d_{X}}\sum_{y=1}^{d_{Y}}|xy\lr xy|^{XY}\otimes \Phi_{y|x}^A\;.
\ee
Note that $\Phi^{XYA}\in\mC^{XYA}$; i.e. it is a CPTP map. Defining $\Psi^{XYB}\in\mC^{XYB}$ in the same way, we conclude that~\eqref{supermajor} holds if and only if 
\be\label{supmaj}
\Psi^{XYB}=\1^{XY}\otimes\Theta\left[\Phi^{XYA}\right]\;.
\ee
Note further that this relation reduces to~\eqref{main} in the special case that $d_Y=1$.

We are now ready to characterize the preorder $\prec_q$ in terms of the extended conditional min-entropy. 
For this purpose, we will denote by $\mC^{XYA}_\star$ (and similarly $\mC^{XYB}_\star$) a subset of CPTP maps in $\mC^{XYA}$ that has the form~\eqref{formc} with the property that for any $x$ the map $\sum_{y=1}^{d_{R_1}^2}\Phi_{y|x}^A$ is CPTP. 
\begin{theorem}\label{superent}
Let $\Phi^{RA}\in\mC^{RA}$ and $\Psi^{RB}\in\mC^{RB}$ be two bipartite quantum channels, and let $\Phi^{XYA}\in\mC^{XYA}_\star$ and $\Psi^{XYB}\in\mC^{XYB}_{\star}$ be their corresponding classical-quantum channels as described above. For any quantum channel $\Lambda^{XYB}\in\mC^{XYB}_{\star}$ define the two bipartite CP maps
\ba
&\Lambda^{AB}_{\Phi}\equiv\frac{1}{d_X}\sum_{x=1}^{d_{X}}\sum_{y=1}^{d_{Y}}\Phi_{y|x}^A\otimes\Lambda_{y|x}^B\\
&\Lambda^{\tB B}_{\Psi}\equiv\frac{1}{d_X}\sum_{x=1}^{d_{X}}\sum_{y=1}^{d_{Y}}\Psi_{y|x}^{\tB}\otimes\Lambda_{y|x}^B\nonumber\;.
\ea  
Then, 
$
\Psi^{RB}\prec_q\Phi^{RA}
$
if and only if for all $\Lambda^{XYB}\in\mC^{XYB}_{\star}$
$$
H_{\min}^{\rm ext}(B|A)_{\Lambda_\Phi}\leq H_{\min}^{\rm ext}(B|\tB)_{\Lambda_\Psi}\;.
$$
\end{theorem}
\begin{remark}
Note that the bipartite CP maps $\Lambda^{AB}_{\Phi}$ and $\Lambda^{\tB B}_{\Psi}$ are in general not trace preserving. They can be expressed as
\ba
&\Lambda^{AB}_{\Phi}=\Theta_{\Lambda}^{XY\to B}\otimes\1^A\left[\Phi^{XYA}\right]\\
&\Lambda^{\tB B}_{\Psi}=\Theta_{\Lambda}^{XY\to B}\otimes\1^{\tB}\left[\Psi^{XY\tB}\right]\nonumber\;.
\ea 
where  $\Theta_{\Lambda}$ is a CPP map from the classical variables space $\mL^{XY}$ to $\mL^B$. It is defined by its action on the elements of $\mL^{XY}$: for any $x\in\{1,...,d_X\}$ and $y\in\{1,...,d_Y\}$
$$
\Theta_{\Lambda}\left[|xy\lr xy|^{XY}\right]\equiv\Lambda_{y|x}^{B}
$$
(recall that $|xy\lr xy|^{XY}$ is viewed as a preparation channel; i.e. with 1-dimensional channel input space).
\end{remark}

{\it{Proof.}}
The necessity of the condition follows trivially from the monotonicity condition of the extended conditional min-entropy (see~\eqref{noisyevolution} with $\Omega^{AB}$ replaced by $\Lambda^{AB}_{\Phi}$). To prove sufficiency consider the set 
$$
\mK_\Phi\equiv\left\{\1^{XY}\otimes\Theta\left[\Phi^{XYA}\right]\;\Big|\;\Theta\in\mbb{S}^{AB}\right\}\subset\mL^{XYB}\;.
$$
Note that the space $\mL^{XYB}$ consists of maps of the form $\sum_{x,y}|xy\lr xy|^{XY}\otimes\Omega_{xy}^{B_0\to B_1}$ with each $\Omega_{xy}^{B_0\to B_1}\in\mL^B$.
$\mK_{\Phi}$ is a closed convex (compact) set in $\mL^{XYB}$. Therefore, from the separation theorem it follows that $\Psi^{XYB}\not\in\mK_\Phi$ if and only if there exists a map $\Lambda^{XYB}\in\mL^{XYB}$ such that
\be
\la\Psi^{XYB},\Lambda^{XYB}\ra>\max_{\Theta\in\mbb{S}^{AB}}\la\1^{XY}\otimes\Theta\left[\Phi^{XYA}\right],\Lambda^{XYB}\ra
\ee
Alternatively, this can be expressed as follows. $\Psi^{XYB}\in\mK_{\Phi}$ if and only if for all $\Lambda^{XYB}\in\mL^{XYB}$
\be\label{sim}
\la\Psi^{XYB},\Lambda^{XYB}\ra\leq\max_{\Theta\in\mbb{S}^{AB}}\la\1^{XY}\otimes\Theta\left[\Phi^{XYA}\right],\Lambda^{XYB}\ra\;.
\ee
We first show that~\eqref{sim} holds for all linear maps $\Lambda^{XYB}\in\mL^{XYB}$ if and only if for all $\Lambda^{XYB}\in\mL^{XYB}$
\ba\label{sim2}
&\max_{\Theta'\in\mbb{S}^{\tB B}}\big\la\1^{XY}\otimes\Theta'\left[\Psi^{XY\tB}\right],\Lambda^{XYB}\big\ra\nonumber\\
&\leq\max_{\Theta\in\mbb{S}^{AB}}\left\la\1^{XY}\otimes\Theta\left[\Phi^{XYA}\right],\Lambda^{XYB}\right\ra\;.
\ea
 Indeed, if~\eqref{sim2} holds, then take $\Theta'$ to be the identity superchannel $\1^B$ so that~\eqref{sim} holds. Conversely, suppose~\eqref{sim} holds for all maps
$\Lambda^{XYB}\in\mL^{XYB}$, and let $\Theta'$ be an optimal superchannel in the LHS of~\eqref{sim2}.
We then get
\begin{align*}
&\la\1^{XY}\otimes\Theta'\left[\Psi^{XY\tB}\right],\Lambda^{XYB}\ra=\la\Psi^{XY\tB},\1^{XY}\otimes\Theta^{\prime*}\left[\Lambda^{XYB}\right]\ra\\
&\leq \max_{\Theta\in\mbb{S}^{AB}}\la\1^{XY}\otimes\Theta\left[\Phi^{XYA}\right],\1^{XY}\otimes\Theta^{\prime*}\left[\Lambda^{XYB}\right]\ra\\
&= \max_{\Theta\in\mbb{S}^{AB}}\la\1^{XY}\otimes\left(\Theta'\circ\Theta\right)\left[\Phi^{XYA}\right],\Lambda^{XYB}\ra\\
&\leq \max_{\Theta\in\mbb{S}^{AB}}\la\1^{XY}\otimes\Theta\left[\Phi^{XYA}\right],\Lambda^{XYB}\ra\;,
\end{align*}
where for the first inequality we used~\eqref{sim}, and for the second inequality we used the fact that $\Theta'\circ\Theta$ is itself a superchannel in $\mbb{S}^{AB}$.

We now prove that it is sufficient to consider in~\eqref{sim2} only maps $\Lambda^{XYB}\in\mC^{XYB}_{\star}\subset\mL^{XYB}$.
From its definition $\Lambda^{XYB}\in\mC^{XYB}_{\star}$ if and only if its Choi matrix is positive semidefinite and has marginal
$$
J_{\Lambda}^{XB_0}=u^X\otimes I^{B_0}\;,
$$
(in particular, $J_{\Lambda}^{B_0}= I^{B_0}$ so that $\Lambda^{XYB}$ is trace preserving).
Now, if $\Lambda^{XYB}\in\mL^{XYB}$ does not have this form, then set $T^{XB_0}\equiv u^X\otimes I^{B_0}-J_{\Lambda}^{XB_0}$ and define ${\Lambda'}^{XYB}$ by its Choi matrix
$$
J_{\Lambda'}^{XYB}\equiv (1-\epsilon)u^{XYB_1}\otimes I^{B_0}+\epsilon\left(J_{{\Lambda}}^{XYB}+T^{XB_0}\otimes u^{YB_1}\right)\;,
$$
where $\epsilon>0$ is small enough so that $J_{\Lambda'}^{XYB}\geq 0$. Note also that $J_{{\Lambda'}}^{XB_0}=u^X\otimes I^{B_0}$.
Now, for any channel $\Psi^{XYB}\in\mC_{\star}^{XYB}$
we have
\ba
&\la\Psi^{XYB},{\Lambda'}^{XYB}\ra=\nonumber\\
&(1-\epsilon)\frac{d_{B_0}}{d_{X}d_{Y}d_{B_1}}+\epsilon \la\Psi^{XYB},\Lambda^{XYB}\ra+\frac{\tr[T^{XB_0}]}{d_X}\nonumber
\ea
A key observation is that except for the term $\epsilon\la\Psi^{XYB},\Lambda^{XYB}\ra$ on the RHS of the equation above, all the other terms are independent of $\Psi^{XYB}$. Hence, equations~\eqref{sim} and~\eqref{sim2} holds for all maps $\Lambda^{XYB}\in\mL^{XYB}$ if and only if they hold for all CPTP maps ${\Lambda'}^{XYB}\in\mC_{\star}^{XYB}$. 

With this form of $\Lambda^{XYB}$ we have
\ba
\big\la\1^{XY}\otimes\Theta\left[\Phi^{XYA}\right]&,\Lambda^{XYB}\big\ra\\
&=\frac{1}{d_{X}^2}\sum_{x=1}^{d_{X}}\sum_{y=1}^{d_{Y}}\left\la\Theta\left[\Phi_{y|x}^A\right],\Lambda_{y|x}^B\right\ra\nonumber
\ea
Expressing the inner product in terms of the Choi matrices and using~\eqref{trans} we get
\ba
\left\la\Theta\left[\Phi_{y|x}^A\right],\Lambda_{y|x}^B\right\ra&=\tr\left[\J_{\Theta}^{AB}\left(\left(J_{\Phi_{y|x}}^A\right)^T\otimes J_{\Lambda_{y|x}}^B\right)\right]\\
&=\tr\left[\left(\J_{\Theta}^{AB}\right)^T\left(J_{\Phi_{y|x}}^A\otimes \left(J_{\Lambda_{y|x}}^B\right)^T\right)\right]\nonumber
\ea
Denoting the CPTP map $\tilde{\Lambda}^{XYB}$ to be 
$$
J_{\tilde{\Lambda}}^{XYB}\equiv\frac{1}{d_{X}}\sum_{x=1}^{d_{X}}\sum_{y=1}^{d_{Y}}|xy\lr xy|^{XY}\otimes\left(J_{\Lambda_{y|x}}^B\right)^T\;,
$$
we conclude that 
\ba
&\max_{\Theta\in\mbb{S}^{AB}}\la\1^{XY}\otimes\Theta\left[\Phi^{XYA}\right],\Lambda^{XYB}\ra\\
&=\max_{\Theta\in\mbb{S}^{AB}}\frac{1}{d_{X}^2}\sum_{x=1}^{d_{X}}\sum_{y=1}^{d_{Y}}\tr\left[\left(\J_{\Theta}^{AB}\right)^T\left(J_{\Phi_{y|x}}^A\otimes J_{\tilde{\Lambda}_{y|x}}^B\right)\right]\\
&=\max_{\Theta\in\mbb{S}^{AB}}\frac{1}{d_{X}^2}\sum_{x=1}^{d_{X}}\sum_{y=1}^{d_{Y}}
\tr\left[\J_{\Theta}^{AB}J_{\Phi_{y|x}\otimes\tilde{\Lambda}_{y|x}}^{AB}\right]\\
&=\frac{1}{d_X}\max_{\Theta\in\mbb{S}^{AB}}
\tr\left[\J_{\Theta}^{AB}J_{\tilde{\Lambda}_{\Phi}}^{AB}\right]\\
&=\frac{d_{B_0}}{d_X}2^{-H_{\min}^{\rm ext}(B|A)_{\tilde{\Lambda}_{\Phi}}}\label{95}
\ea
where the last equality follows from~\eqref{sf}, and
\be\label{ltilde}
\tilde{\Lambda}^{AB}_\Phi\equiv\frac{1}{d_X}\sum_{x=1}^{d_{X}}\sum_{y=1}^{d_{Y}}\Phi_{y|x}^A\otimes\tilde{\Lambda}_{y|x}^B
\ee
Hence, Eq.\eqref{sim2} is equivalent to
$$
H_{\min}^{\rm ext}(B|A)_{\tilde{\Lambda}_\Phi}\leq H_{\min}^{\rm ext}(B|\tB)_{\tilde{\Lambda}_\Psi}\;.
$$
Since $\tilde{\Lambda}^{XYB}\in\mC^{XYB}_{\star}$ if and only if $\Lambda^{XYB}\in\mC^{XYB}_{\star}$
we can remove the tilde from $\Lambda^{XYB}$. This completes the proof.
\QED

\subsection{Comparison of channels with the extended conditional min-entropy}

The family of generalized divergences discussed earlier (see also~\cite{Kun18}), along with other methods to extend divergences~\cite{Wil18b}, provide only necessary conditions for the existence of a superchannel that converts one pair of channels to another.
We now focus on yet another family of channel divergences that provides both necessary and \emph{sufficient} conditions for the existence of such a superchannel. 

By taking the special case of $d_{Y}=1$ in the Theorem~\ref{superent} above we get the following corollary.
\begin{corollary}\label{maincor}
For each $j=1,...,n$, let $\Psi^{A}_{j}\in\mC^A$ and $\Phi^{B }_{j}\in\mC^{B}$ be quantum channels, and let $R$ denote a third system of the same input and output dimensions as $B$.
Then, there exists a superchannel satisfying~\eqref{main} for all $j=1,...,n$ if and only if
for any set of $n$ quantum channels $\Lambda_{j}^{R}\in\mC^{R}$, the tripartite quantum channel
$$
\Lambda^{ABR}\equiv\frac{1}{n}\sum_{j=1}^{n}\Psi^{A}_{j}\otimes \Phi^{B }_{j}\otimes \Lambda_{j}^{R}
$$
satisfies
$$
H_{\min}^{\rm ext}(R|A)_\Lambda \leq H_{\min}^{\rm ext}(R|B)_\Lambda\;.
$$
\end{corollary}

\begin{remark}
In the case that $n=2$, we can define for any two channels $\Lambda_1^R$ and $\Lambda_2^R$, the channel contraction (i.e. divergence) as
$$
C_\Lambda(\Psi^A_1\|\Psi^A_2)\equiv H_{\min}^{\rm ext}(R|A)_{\Lambda^{AR}}\;,
$$
where 
$$
\Lambda^{AR}\equiv\frac{1}{2}\left(\Psi^{A}_{1}\otimes \Lambda_{1}^{R}+\Psi^{A}_{2}\otimes \Lambda_{2}^{R}\right)\;.
$$
The corollary above states that this family of divergences $\{C_\Lambda\}$ provides both necessary and sufficient conditions for~\eqref{main} to hold in the case that $n=2$.
\end{remark}

\subsection{Characterization with semidefinite programming}\label{sdp}

A caveat of the characterization of quantum majorization with the extended conditional min-entropy is that it involves an infinite number of conditions, and therefore it is not practical 
for determining whether there exists a superchannel that satisfies~\eqref{supermajor} or~\eqref{main}. Nonetheless,
here we show that the problems in~\eqref{supermajor} and~\eqref{main} can be solved with semidefinite programming (SDP).

From the proof of Theorem~\ref{superent} it follows (see particularly~\eqref{sim}) that $\Psi^{XYB}\prec_q\Phi^{XYA}$ if and only if the function
\ba\label{sim3}
f(\Phi,\Psi)&\equiv\min_{\Lambda^{XYB}\in\mC^{XYB}_{\star}}\;\max_{\Theta\in\mbb{S}^{AB}}\\
&\la\1^{XY}\otimes\Theta\left[\Phi^{XYA}\right]-\Psi^{XYB},\Lambda^{XYB}\ra
\ea
is not negative. We now show that the above optimization problem is an SDP. 

Recall that in~\eqref{95} we showed that
$$
\max_{\Theta\in\mbb{S}^{AB}}\la\1^{XY}\otimes\Theta\left[\Phi^{XYA}\right],\Lambda^{XYB}\ra=\frac{d_{B_0}}{d_X}2^{-H_{\min}^{\rm ext}(B|A)_{\tilde{\Lambda}_{\Phi}}}
$$
where $\tilde{\Lambda}^{AB}$ is defined in~\eqref{ltilde}. Combining the expression for $f(\Phi,\Psi)$ with the definition in~\eqref{maind} for the extended conditional min-entropy gives
$$
f(\Phi,\Psi)=\min\left\{\tr[\gamma^{AB_0}]-\tr\left[J^{XYB}_{\Psi}J^{XYB}_\Lambda\right]\right\}
$$
subject to: 
\begin{align}
&{\it 1.}\;\;\gamma^{AB_0}\otimes I^{B_1}\geq\frac{1}{d_{A_0}d_X^2}\sum_{x=1}^{d_{X}}\sum_{y=1}^{d_{Y}}\left(J_{\Phi_{y|x}}^A\right)^T\otimes J_{\Lambda_{y|x}}^B\label{98}\\
&{\it 2.}\;\;\gamma^{A_0B_0}=u^{A_0}\otimes \gamma^{B_0}\label{99}\\
&{\it 3.}\;\; J^{XB_0}_\Lambda=u^X\otimes I^{B_0}\quad;\quad J^{XYB}_\Lambda\geq 0\label{100}
\end{align}
where we absorbed the factor $\frac{d_{B_0}}{d_X}$ into $\gamma^{AB_0}$. To bring the above optimization to a canonical SDP form define the following (real) vector spaces consisting of the direct sum of three spaces:
\be
V\equiv\mB_h(\mH^{AB_0})\oplus\mB_h(\mH^{XYB})\oplus\mB_h(\mH^{AB})
\ee
We denote the elements of $V$ by 
$$
\xi\equiv(\gamma^{AB_0},\alpha^{XYB},\eta^{AB})\in V\;,
$$
and set $$\mu\equiv\left(I^{AB_0},-J_{\Psi}^{XYB},{0}^{AB}\right)\;,$$
where ${0}^{AB}$ is the zero matrix in $\mB_h(\mH^{AB})$. The inner product between elements in $V$ is defined as the sum of the inner products among the three components.
We also define the linear map $\mT:V\to \mB_h(\mH^{AB})$ by
\ba
\mT(\xi)&=\gamma^{AB_0}\otimes I^{B_1}\nonumber\\
&-\tr_{XY}\left[\left(\left(J^{XYA}_{\Phi}\right)^{T}\otimes I^B\right)\left(\alpha^{XYB}\otimes I^A\right)\right]-\eta^{AB}
\ea
Note that $\mT$ is indeed linear, and~\eqref{98} is equivalent to $\mT(\xi)+\eta^{AB}\geq 0$ when we identify $\alpha^{XYB}$ with $J_{\Lambda}^{XYB}$. 
Finally, set
\ba
& G_{x,k}\equiv ({0}^{AB_0},|x\lr x|^X\otimes |\psi_k\lr\psi_k|^{B_0}\otimes I^{YB_1},{0}^{AB})\in V\;,\nonumber\\
& F_{\ell}\equiv \left(M_{\ell}^{A_0B_0}\otimes I^{A_1},{0}^{XYB},{0}^{AB}\right)\in V,
\ea
where $\left\{|\psi_k\lr\psi_k|^{B_0}\right\}_{k=1}^{d_{B_0}^2}$ is a pure state basis of $\mB_h(\mH^{B_0})$, and $\{M^{A_0B_0}_{\ell}\}_{\ell=1}^{(d_{A_0}^2-1)d_{B_0}^2}$ is a basis of the subspace of $\mB_h(\mH^{A_0B_0})$ consisting of all matrices with zero marginal on $B_0$. Note that~\eqref{99} holds if and only if $\tr[\gamma^{A_0B_0}M^{A_0B_0}_{\ell}]=0$ for all $\ell$, and similarly the first equality of~\eqref{100} holds if and only if $\tr\left[J^{XB_0}_\Lambda\left(|x\lr x|^X\otimes|\psi_k\lr\psi_k|^{B_0}\right)\right]=1/d_{X}$ for all $x$ and $k$. Hence, with
these notations we get
$$
f(\Phi,\Psi)=\min\tr[\xi \mu]
$$
subject to: for all $x=1,...,d_X$, $k=1,...,d_{B_0}^2$, and $\ell=1,...,(d_{A_0}^2-1)d_{B_0}^2$
$$
\mT(\xi)=0\;\;;\;\;\tr[\xi G_{k,x}]=\frac{1}{d_X}\;\;;\;\;\tr\left[\xi F_{\ell}\right]=0\;\;;\;\;\xi\geq 0
$$
Finally, note that the condition $\mT(\xi)=0$ can also be expressed in terms of inner products. That is,
 let $\{E_j\}_{j=1}^{d}$ be a basis of $\mB_h(\mH^{AB})$, 
and for each $j=1,...,d$, let $K_j\equiv\mT^*(E_j)$.  Then, with these notations, we can replace the condition $\mT(\xi)=0$ above with 
$$
\tr[\xi K_j]=0\;\;\forall\;j=1,...,d\;.
$$
We therefore obtained a canonical form of a SDP optimization problem that can be plugged into standard packages such as CVX. Note that the number of all the constraints is polynomial in the dimensions.

\subsection{An application to thermodynamics}\label{thermo}

Recall that if $d_{A_0}=d_{B_0}=1$ then all the channels involved in Theorems~\ref{superent}, and the semidefinite programming above, become states and the extended conditional min-entropy reduces to the standard conditional min-entropy of states. In this case, the Theorem~\ref{superent} above reduces to the state analog that was proved in~\cite{Gour2018}. As was shown in~\cite{Gour2018}, the state version of the theorem above has many applications particularly in state transformations of quantum resources theories of thermodynamics and asymmetry. We expect that the theorem above will also have applications in the simulation of channels in various resource theories of quantum processes (see some very recent work on the subject~\cite{FBB18,Bus18,Diaz18,Wint2}). We give here a very brief discussion on one such application in the quantum resource theory of athermality in thermodynamics.

In the resource theory of quantum thermodynamics the Gibbs state, $\gamma$, is known to be the only free state of the model.  In this model, a replacement channel that outputs the Gibbs state irrespective of the input state is a free channel. Denote such a channel by $\Pi^A_\gamma\in\mC^A$; that is,
$$
\Pi^{A}_\gamma(\rho^{A_0})\equiv\tr\left[\rho^{A_0}\right]\gamma^{A_1}\quad\forall\;\rho^{A_0}\in\mB(\mH^{A_0})\;.
$$
Now, consider a superchannel $\Theta:\mL^A\to\mL^B$ in which system $A_1$ is associated with a Gibbs state $\gamma^{A_1}$ and system $B_1$ with Gibbs state $\gamma^{B_1}$. Then, if $\Theta$ is a free superchannel it must take the Gibbs channel of system $A$ to the Gibbs channel of system B. That is, it satisfies
$$
\Theta\left[\Pi^{A}_\gamma\right]=\Pi^{B}_\gamma\;.
$$
We call such superchannels that preserve the Gibbs channel $\Pi_\gamma$, \emph{Gibbs preserving superchannels}. In this case, Corollary~\ref{maincor} and~Sec.~\ref{sdp} provide the necessary and sufficient conditions that a given channel $\Phi^B$ can be simulated by another channel $\Psi^A$ via Gibbs preserving superchannels.
While Corollary~\ref{maincor} provides a complete family of athermality monotones (of dynamical resources, i.e. channels) in terms of the extended conditional min-entropy, Sec.~\ref{sdp} shows that the problem can be solved efficiently with SDP.
We leave the extensions of these ideas to other, more physical models of thermodynamics (e.g. channel simulations under thermal operations, etc), and other resource theories, for future work.


\section{Summary and Conclusions}\label{outlook}

We discussed in this paper several different noise models for superchannels, and used that to define 
 entropy functions for quantum channels. Our approach was axiomatic and minimalistic in nature, requiring the entropy function to be additive and monotonic only under random unitary superchannels. As an example, we found that the extended min-entropy is an entropy function that is monotonic under a much larger set of operations than the random unitary ones. We called these operations doubly stochastic superchannels since they consist of superchannels whose dual maps are also superchannels. We gave doubly stochastic superchannels a physical interpretation by showing that they have the property that they are completely uniformity preserving (see Sec.~\ref{unipre}).
 
 We then introduced an extension to the conditional min-entropy from bipartite states to bipartite channels. Given our definition of an entropy of a channel, we were able to show that the extended conditional min-entropy has many similar properties to the ones of its state version (i.e. the conditional min-entropy), including an operational interpretation in terms of a guessing probability if one of the subsystems is classical. 
 
 The extended conditional min-entropy turned out to play a key role in our extension of quantum majorization
from bipartite states to bipartite channels. Quantum majorization, as defined originally  in~\cite{Gour2018}, is a pre-order for bipartite states that can be viewed as a generalization of majorization. It has applications to quantum resource theories, degradability of channels, and quantum statistical comparisons. Here we extended this definition from bipartite states to bipartite channels. A special case of this pre-order is the problem of comparison of channels given in~\eqref{main}. In theorem~\ref{superent} we showed that quantum majorization for channels can be fully characterized with a family of functions given in terms of the extended conditional min-entropy. In particular, for the comparison between two channels, Corollary~\ref{maincor} provides a complete set of channel divergences that are both necessary and sufficient to determine if~\eqref{main} holds.
We also showed that determining whether one bipartite channel quantum majorizes another can be solved efficiently with semidefinite programming.

We expect that the results and techniques used here will be useful particularly in resource theories of quantum processes. We gave an indication for that in Sec.~\ref{thermo}. Moreover, some of the definitions given here can be extended further. For example, it is straightforward to define the smoothed version of the extended conditional min-entropy. 
Let $\Psi^{AB}:\mB(\mH^{A_0B_0})\to\mB(\mH^{A_1B_1})$ be a bipartite quantum channel. The $\epsilon$-extended conditional min-entropy is defined by:
$$
H^{{\rm ext},\epsilon}_{\min}(B|A)_{\Psi}\equiv\sup_{\|\Phi-\Psi\|_{\diamond}\leq\epsilon}H^{{\rm ext}}_{\min}(B|A)_{\Phi}\;,
$$
where the supremum is over all bipartite channels $\Phi^{AB}$ that are $\epsilon$-close (in the diamond norm) to the channel $\Psi^{AB}$. With this definition we can also define
$$
S^{\rm ext}(B|A)_\Psi\equiv\lim_{\epsilon\to 0}\lim_{n\to\infty}\frac{1}{n}H^{{\rm ext},\epsilon}_{\min}(B^n|A^n)_{\Psi^{\otimes n}}
$$
In~\cite{Tomamichel-2009a} the asymptotic equipartition property was proved for states. This means that the function above becomes the von-Neumann conditional entropy whenever $\Psi^{AB}$ is a replacement map; i.e. for any input density matrix $\sigma^{A_0B_0}$ we have $\Psi^{AB}(\sigma^{A_0B_0})=\rho^{A_1B_1}$, where $\rho^{A_1B_1}$ is a fixed output density matrix. In this case,
$$
S^{\rm ext}(B|A)_\Psi=S(B_1|A_1)_{\rho}\;,
$$
where $S(B_1|A_1)_{\rho}$ is the conditional von-Neumann entropy. We therefore expect that the above quantity will have an interesting operational interpretation and leave its investigation for future work.

\appendices
\section{Strong Duality in Conic Linear programming}\label{AppA}

There were several places in the paper that we were using the strong duality of SDP or conic linear programming. We present here the strong duality relation as given in~\cite{barvinok2002course}, and use it in the following subsections to prove the various statements made in the paper for its specific applications.
  
Let $V_1$ and $V_2$ be two (real) vector spaces (here will will assume that they consists of Hermitian matrices) and let $\Gamma:V_1\to V_2$ be a linear map.  Let $\mK_1\subset V_1$ and $\mK_2\subset V_2$ be two convex cones. Moreover, let $H_1\in V_1$ and $H_2\in V_2$ be two (fixed) elements.
\begin{enumerate}
\item The Primal Problem:
\begin{align}
\text{Find}\quad & \alpha\equiv\inf\tr\left[XH_1\right]\nonumber\\
\text{Subject to}\quad &\Gamma(X)-H_2\in\mK_2\quad\text{and}\nonumber\\
& X\in\mK_1\label{primal}
\end{align}
\item The Dual Problem:
\begin{align}
\text{Find}\quad & \beta\equiv\sup\tr\left[YH_2\right]\nonumber\\
\text{Subject to}\quad &H_1-\Gamma^*(Y)\in\mK_1^*\quad\text{and}\nonumber\\
& Y\in\mK_2^*\label{dual}
\end{align}
\end{enumerate}
Here $\Gamma^*:V_2\to V_1$ is the dual map of $\Gamma$, and $\mK_1^*$ and $\mK_2^*$ are the dual cones, respectively, of $\mK_1$ and $\mK_2$.\\ 

\noindent\emph{Weak duality:} 

For any feasible plan $X$ (i.e. $X$ satisfies $\Gamma(X)-H_2\in\mK_2$ and $X\in\mK_1$) and a dual feasible plan $Y$
(i.e. $Y$ satisfies $H_1-\Gamma^*(Y)\in\mK_1^*$ and $Y\in\mK_2^*$), we have
$$
\tr\left[XH_1\right]\geq  \tr\left[YH_2\right]\quad\text{and, in particular}\quad\alpha\geq\beta.
$$

\noindent\emph{Strong Duality:}
\begin{enumerate}
\item  Consider the cone $\mK\subset V_2\oplus\mbb{R}$ defined by
$$
\mK\equiv\left\{\Big(\Gamma(X)-Y\;,\;\tr[XH_1]\Big)\;:\; X\in\mK_1\;,\;Y\in\mK_2\right\}.
$$
If $\mK$ is closed in $V_2\oplus\mbb{R}$ and there exists a primal feasible plan then $\alpha=\beta$. Moreover, if $\alpha>-\infty$ then there exists a primal optimal plan (i.e. a feasible $X$ such that $\alpha=\tr[XH_1]$).
\item The Slater's condition: Suppose that there is a primal feasible plan $X_0\in{\rm int}\left(\mK_1\right)$ such that $\Gamma(X_0)-H_2\in{\rm int}\left(\mK_2\right)$. Suppose also that there exists a primal optimal plan. Then, there is no duality gap; i.e. $\alpha=\beta$.
\end{enumerate}
In our cases, the strong duality will always hold.

\subsection{Proof of the Equivalence of~\eqref{maind} and~\eqref{maind2} }

Consider now the primal problem in~\eqref{maind}:
\begin{align}
&\min\tr[\gamma^{AB_0}]\nonumber\\
\text{subject to: }\; &{\it 1.}\;\;\gamma^{AB_0}\otimes I^{B_1}\geq\omega^{AB}\nonumber\\
&{\it 2.}\;\;\gamma^{A_0B_0}=u^{A_0}\otimes \gamma^{B_0}
\label{amaind}
\end{align}
We can identify it with the primal problem of the above conic programming in which $V_1\equiv \mB_h(\mH^{AB_0})$, $V_2\equiv \mB_h(\mH^{AB})\oplus\mB_h(\mH^{A_0B_0})$, $H_1\equiv I^{AB_0}\in V_1$, $H_2\equiv (\omega^{AB},0^{A_0B_0})\in V_2$, $\mK_1\equiv V_1$ (hence, $\mK_1^*= \{0^{AB_0}\}$), and 
$$
\mK_2\equiv\left\{(\eta^{AB},0^{A_0B_0})\;:\;\eta^{AB}\geq 0\right\}\subset V_2
$$ 
 and $\Gamma: V_1\to V_2$ defined by: for all $\gamma^{AB_0}\in V_1$
$$
\Gamma(\gamma^{AB_0})\equiv \left(\gamma^{AB_0}\otimes I^{B_1},\gamma^{A_0B_0}-u^{A_0}\otimes\gamma^{B_0}\right)\;.
$$
With these identifications, we get that the problem~\label{amaind} is identical to~\eqref{primal}.
Hence, to get its dual, observe that $\mK_2^*=\mB_+(\mH^{AB})\oplus\mB_h(\mH^{A_0B_0})$, and $\Gamma^*: V_2\to V_1$ satisfies for any 
$(\eta^{AB},\zeta^{A_0B_0})\in V_2$ 
\be
\Gamma^*(\eta^{AB},\zeta^{A_0B_0})
=\eta^{AB_0}+\left(\zeta^{A_0B_0}-u^{A_0}\otimes\zeta^{B_0}\right)\otimes I^{A_1}
\ee
With these identifications at hand, we get that the dual problem in~\eqref{dual} is given by:
\begin{align*}
& \max\tr\left[\eta^{AB}\omega^{AB}\right]\\
\text{Subject to}\quad &I^{AB_0}=\eta^{AB_0}+\left(\zeta^{A_0B_0}-u^{A_0}\otimes\zeta^{B_0}\right)\otimes I^{A_1}\\
& \eta^{AB}\geq 0\quad;\quad\zeta^{A_0B_0}\in \mB_h(\mH^{A_0B_0})\;.
\end{align*}
Finally, note that the condition
$$
\eta^{AB_0}=I^{AB_0}-\left(\zeta^{A_0B_0}-u^{A_0}\otimes\zeta^{B_0}\right)\otimes I^{A_1}=\eta^{A_0B_0}\otimes u^{A_1}
$$
where $\eta^{A_0B_0}=d_{A_1}\left(I^{A_0B_0}-\zeta^{A_0B_0}+u^{A_0}\otimes\zeta^{B_0}\right)$. 
Furthermore, $\eta^{B_0}=d_{A_0}d_{A_1}I^{B_0}$. Finally, denoting by $\alpha^{AB}\equiv\frac{1}{d_{A_0}}\eta^{AB}$ we conclude that the above dual problem can be expressed as:
\begin{align*}
& d_{A_0}\max\tr\left[\alpha^{AB}\omega^{AB}\right]\\
\text{Subject to}\quad &\alpha^{AB_0}=\alpha^{A_0B_0}\otimes u^{A_1}\\
& \alpha^{AB}\geq 0\quad;\quad\alpha^{B_0}=d_{A_1}I^{B_0}\;.
\end{align*}
This expression is equivalent to~\eqref{maind2}.

\section{Operational interpretation of $H_{\min}^{\rm ext}\left(B|A\right)_{\Omega}$ in the case that only system $B_1$ is classical}\label{fc}

Since system $B_1$ is classical, for all $\rho^{A_0B_0}\in\mB(\mH^{A_0B_0})$
$$
\Omega^{A_0B_0\to A_1B_1}(\rho^{A_0B_0})=\sum_{x=1}^{d_{B_1}}\Omega_{x}^{A_0B_0\to A_1}(\rho^{A_0B_0})\otimes |x\lr x|^{B_1}\;,
$$
where $\{\Omega_{x}^{A_0B_0\to A_1}\}$ form a quantum instrument, and
the Choi matrix of $\Omega^{AB}$ can be expressed as:
$$
\omega^{AB}=\sum_{x=1}^{d_{B_1}}\omega^{AB_0}_{x}\otimes |x\lr x|^{B_1}\;,
$$
with 
\be\label{ome}
\omega^{AB_0}_{x}=\frac{1}{d_{A_0}d_{B_0}}\id^{A_0B_0}\otimes\Omega_{x}^{\tA_0\tB_0\to A_1}\left
(\phi_{+}^{A_0\tA_0}\otimes\phi_{+}^{B_0\tB_0}\right)
\ee
Consequently, from~\eqref{maind2}, with $\omega^{AB}$ as above and $\alpha^{AB}\equiv\sum_{x=1}^{d_{B_1}}\alpha^{AB_0}_{x}\otimes |x\lr x|^{B_1}$, we get
\begin{align}
2^{-H_{\min}^{\rm ext}\left(B|A\right)_{\Omega}}&=d_{A_0}\max \sum_{x=1}^{d_{B_1}}\tr[\alpha^{AB_0}_x\omega_{x}^{AB_0}]\nonumber\\
\text{subject to: }\; &\sum_{x=1}^{d_{B_1}}\alpha_{x}^{AB_0}=\eta^{A_0B_0}\otimes I^{A_1}\;\;,\;\;\eta^{B_0}=I^{B_0}\nonumber\\
&\alpha_{x}^{AB_0}\geq 0\quad\forall\;x\in\{1,...,d_{B_1}\}
\end{align}
where we denoted by 
$$
\eta^{A_0B_0}\equiv \frac{1}{d_{A_1}}\sum_{x=1}^{d_{B_1}}\alpha^{A_0B_0}_x\;.
$$
Note that we can assume w.l.o.g. that $\eta^{A_0B_0}$ is full rank. Hence, we can define the following POVM on system $AB_0$:
$$
P_{x}^{AB_0}\equiv \left(\left(\eta^{A_0B_0}\right)^{-\frac{1}{2}}\otimes I^{A_1}\right)\alpha_{x}^{AB_0}\left(\left(\eta^{A_0B_0}\right)^{-\frac{1}{2}}\otimes I^{A_1}\right)
$$
Note that $P_x^{AB_0}\geq 0$ and $\sum_x P_x^{AB_0}=I^{AB_0}$. With this notation 
\begin{align}
 &d_{A_0}\sum_{x=1}^{d_{B_1}}\tr[\alpha^{AB_0}_x\omega_{x}^{AB_0}]=d_{A_0}\sum_{x=1}^{d_{B_1}}\nonumber\\
& \tr\left[P^{AB_0}_x\left(\sqrt{\eta^{A_0B_0}}\otimes I^{A_1}\right)\omega_{x}^{AB_0}\left(\sqrt{\eta^{A_0B_0}}\otimes I^{A_1}\right)\right]\nonumber\\
&=\sum_{x=1}^{d_{B_1}}\tr\left[P_{x}^{AB_0}\left(\id^{A_0B_0}\otimes\Omega_{x}^{\tA_0\tB_0\to A_1}\left
(|\eta\lr\eta|^{A_0B_0\tA_0\tB_0}\right)\right)\right]
\end{align}
where we used~\eqref{ome}, and the state $|\eta\ra^{A_0B_0\tA_0\tB_0}$ is the purification of the normalized state $\frac{1}{d_{B_0}}\eta^{A_0B_0}$. That is, 
$$
|\eta\ra^{A_0B_0\tA_0\tB_0}\equiv \frac{1}{\sqrt{d_{B_0}}}\left(\sqrt{\eta^{A_0B_0}}\otimes I^{\tA_0\tB_0}\right)|\phi_{+}^{A_0\tA_0}\ra|\phi_{+}^{B_0\tB_0}\ra\;.
$$
We therefore conclude that 
\begin{align}
&2^{-H_{\min}^{\rm ext}\left(B|A\right)_{\Omega}}=\max \sum_{x=1}^{d_{B_1}}\nonumber\\
&\tr\left[P_{x}^{AB_0}\left(\id^{A_0B_0}\otimes\Omega_{x}^{\tA_0\tB_0\to A_1}\left
(|\eta\lr\eta|^{A_0B_0\tA_0\tB_0}\right)\right)\right]
\nonumber\\
&\text{subject to: }\;\tr_{A_0\tA_0\tB_0}\left
[|\eta\lr\eta|^{A_0B_0\tA_0\tB_0}\right]=u^{B_0}\nonumber\\
&\quad\quad\quad\quad\quad P_{x}^{AB_0}\geq 0\quad\forall\;x=1,...,d_{B_1}
\end{align}
Note that we can think of system $B_0$ above in Alice's system. Denoting it by $A_2$ (hence $d_{A_2}=d_{B_0}$), and expressing
$$
|\eta\lr\eta|^{A_0A_2\tA_0\tB_0}=\id^{A_2}\otimes\mV^{\tA_2\to A_0\tA_0\tB_0}\left(\frac{1}{d_{A_2}}\phi_{+}^{A_2\tA_2}\right)
$$
with $\mV^{\tA_2\to A_0\tA_0\tB_0}$ is an isometry, we get
\begin{align}
&2^{-H_{\min}^{\rm ext}\left(B|A\right)_{\Omega}}=\max \sum_{x=1}^{d_{B_1}}
\tr\Big[P_{x}^{A_0A_1A_2}\left(\id^{A_0A_2}\otimes\Omega_{x}^{\tA_0\tB_0\to A_1}\right)\nonumber\\
&\circ\left(\id^{A_2}\otimes\mV^{\tA_2\to A_0\tA_0\tB_0}\right)\left
(\phi_{+}^{A_2\tA_2}/d_{A_2}\right)\Big]
\end{align}
Subject to: $\mV^{\tA_2\to A_0\tA_0\tB_0}$ being an isometry, and  $P_{x}^{A_0A_1A_2}\geq 0$ for all $x=1,...,d_{B_1}$\;.
This optimization problem is illustrated in Fig.~\ref{guessing3}.

\section*{Acknowledgment}
The author would like to thank Francesco Buscemi, Giulio Chiribella, Eric Chitambar, Nilanjana Datta, Teiko Heinosaari, Barbara Kraus, Rob Spekkens, Mark Wilde, and Andreas Winter for useful discussions related to the topic of this paper. 
Particularly, the author appreciates the many useful comments by Mark Wilde on the first draft. The author acknowledges support from the Natural Sciences and Engineering Research Council of Canada (NSERC).

\ifCLASSOPTIONcaptionsoff
  \newpage
\fi



%

%

\begin{IEEEbiographynophoto}{Gilad Gour}
Dr. Gour is a Professor at the Department of Mathematics and Statistics at the University of Calgary, and an affiliated member of the Perimeter Institute in Waterloo, Canada. Dr. Gour is a recipient of the faculty of science early career research excellence award (2015), of the Katzir Prize (2003) from the Hebrew University, and the Teaching Excellence Award from the Schulich School of Engineering (2017). Dr. Gour is especially well known for his contributions to quantum resource theories.  Dr. Gour current research interests include quantum Shannon theory, resource theories of quantum processes, entanglement theory, thermodynamics at the nanoscale, quantum cryptography, and foundations of physics and science. 
\end{IEEEbiographynophoto}





\end{document}